\documentclass[preprintnumbers,amsmath,amssymb,twocolumn,groupedaddress,floatfix]{revtex4}

\usepackage{graphicx}
\usepackage{bm}
\usepackage{slashed}
\usepackage{amsfonts}
\usepackage{graphicx,color}
\usepackage[colorlinks,linkcolor=blue,citecolor=green]{hyperref}
\usepackage{wrapfig}
\newcommand{\dd}{\textrm{\,d}}
\newcommand{\ee}{\textrm{\,e}}

\makeatletter
\@addtoreset{figure}{section}
\makeatother

\begin{document}

\title{Hadronization within the Non-Extensive Approach and the Evolution of the Parameters}

\author{Keming Shen}
\email{shen.keming@wigner.mta.hu}

\author{Gergely G\'abor Barnaf\"oldi}

\author{Tam\'as S\'andor Bir\'o}
\affiliation{Wigner Research Center for Physics of the HAS, 29-33 Konkoly-Thege Mikl\'os Street, 1121 Budapest, Hungary}

\date{\today}

\begin{abstract}

We review transverse momentum distributions of various identified charged particles stemming from high energy collisions fitted by various non-extensive distributions as well as by the usual Boltzmann-Gibbs statistics.
We investigate the best-fit formula with the obtained $\chi^2/ndf$ values.
We find that the physical mass and $\sqrt{s}$ scaling become more explicit with heavier produced hadrons in both proton-proton and heavy-ion collisions. 
The spectral shape parameters, in particular the temperature $T$ and the non-extensive Tsallis parameter $q$, do exhibit an almost linear dependence with the centrality-dependence in heavy-ion collisions.

\end{abstract}

\maketitle

\section{Introduction}\label{sec:sec1}

More and more attention has been paid recently to the analysis of transverse momentum ($p_T$) spectra in heavy ion collisions due to its importance in novel statistical approaches~\cite{app-p1}. 
The $p_T$ spectrum reveals information on the early thermal or close to thermal properties of the hottest state of such collisions.
It has been realized that data on single-inclusive-particle distributions show a power-law behavior in the high-$p_T$ region.
This cannot follow from traditional statistical models based on Boltzmann-Gibbs (BG) statistics.

Thanks to the high multiplicities produced at high energies one may utilize statistical models to study the production mechanism even in small systems like the proton-proton ($pp$) \cite{pp-Ortiz}.
In these collisions identified particle spectra at RHIC and LHC energies do not follow a BG distribution, especially not at the high-$p_T$ values. 
In the past decades the Tsallis-Pareto-Hagedorn distribution has been frequently applied \cite{app-p1}. 
Its precise form is based on the generalised $q$-exponential function:
\begin{eqnarray}
\ee_q(x):=\left[1+(1-q)x\right]^{\frac{1}{1-q}}~,
\label{qexp}
\end{eqnarray}
where $1+(1-q)x \ge0$ for $q>1$.
For allowed ranges in $x$ and more discussions, see ref.~\cite{Hasegawa-2009}.
Specifically, several different shape functions are in use to fit the spectra in the study of high energy collisions both experimentally and theoretically~\cite{all-1, all-2, all-3, all-4, all-5, all-6, all-7, all-8, all-9, all-10, all-10-1-1, all-10-1, all-11, all-12, all-13, all-14, all-15, all-16, all-17, all-17-1, all-18, all-19, all-20, all-21, all-22, all-23, all-24, all-25, all-25-1, all-26, all-27, all-28, all-29, all-30, all-30-1, all-31, all-32, all-33, all-34}.

In this work we review momentum distributions within the non-extensive Tsallis statistical approach, both in proton-proton and in nucleus-nucleus collisions. 
Various momentum distribution formulas are investigated and compared to each other while seizing for a physical interpretation of their parameters. 
Our aim is to find the natural situation for the best data-fit formulas with the possibly strongest physical content.

\section{Modeling hadron spectra}\label{sec:sec2}

In order to describe hadron spectra measured in various hadron collisions, one tries to disentangle hard QCD and soft collective effects, and a possible transverse flow.
Then one tests whether the result complies with the thermal assumption, i.e. that the dependence on momenta is solely through a dependence on the kinetic energy.
In statistical distributions, like the BG distribution, the independent variable is $E-\mu$, where $E$ denotes the energy of a single particle state and $\mu$ is the chemical potential. At low density and high temperature, $T\gg (\mu -m)$, this variable can be approximated by the pure kinetic energy, $E-m$, where $m$ is the rest mass. 

In general it is assumed that the source, emitting the detected hadrons, is flowing in an arbitrary direction. 
Here the 4-velocity ($u_\mu$) of the source and the actual 4-momentum ($p_\mu$) of the particle are parameterized by rapidity and angle variables respectively\cite{Biro-p1}:
\begin{eqnarray}
u_\mu&&=(\gamma_T\cosh \xi,\gamma_T\sinh \xi, \gamma_T v_T\cos\phi, \gamma_Tv_T\sin\phi), \nonumber \\
p_\mu&&=(m_T\cosh y, m_T\sinh y, p_T\cos\psi, p_T\sin\psi)
\end{eqnarray}
leading to the covariant J\"uttner expression for the energy of the particle in the co-moving fluid element:
\begin{eqnarray}
E=u_\mu p^\mu=\gamma_Tm_T\cosh(y-\xi)-\gamma_Tv_Tp_T\cos(\psi-\phi) .
\label{E}
\end{eqnarray}
Further details are displayed in Appendix~\ref{sec:secA}.

In high energy physics one investigates the Lorentz-invariant particle spectrum.
In order to fit the hadron spectra, the following approximations to Eq.($\ref{E}$) are utilized:
\begin{enumerate}
\item For large $m_T$ and small longitudinal momentum $p_L\ll m_T$, $E=\sqrt{m_T^2+p_L^2}\approx m_T+\frac{p_L^2}{2m_T}$~\cite{all-23}.
\item For the midrapidity plateau calculations, $y-\xi\approx 0$ and $\psi-\phi=0$, $E=\gamma_Tm_T-\gamma_Tv_Tp_T$. 
In case of light particles, like pions, in the observed $p_T$ range the assumption $p_T\gg m$ can safely be made.
Thus we can rewrite the thermal ratio $E/T$ in spectra as $E/T=\gamma(1-v)p_T/T=p_T/T_D$, with the so-called Doppler-blue-shifted temperature $T_D:=\sqrt{\frac{1+v_T}{1-v_T}}T$  \cite{doppler-p1,doppler-p2}. 
In such a case, the transverse flow velocity can be absorbed into the fitting temperature. 
Note that we will follow this assumption in the present paper with respect to the flow effects in high energy collisions.
\item Typically in small systems, like $ee$, $ep$ and $pp$ collisions~\cite{Urmossy-epp, Urmossy-epp-1}, no transverse flow effects are considered: $v_T\to 0$, hereafter $E= m_T\cosh(y-\xi)$, see Refs.~\cite{all-1, all-2, all-3}. 

\item Taking both $v_T= 0$ (no transverse flow) and $y-\xi =0$ (midrapidity plateau), one arrives at $E= m_T$, hence  $E-\mu\approx m_T-m$. 
In order to study the fitting functions on hadron spectra in high energy collisions, this proxy is widely used and cited as ``$m_T$-scaling". 

\end{enumerate}
It is worthwhile to mention that another non-extensive approach exists, inspired by the superstatistics named in Ref.~\cite{all-28}. 
We adopt 
\begin{eqnarray}
f_q(x):=\left[\ee_q(x)\right ]^q=\left[1+(1-q)x \right ]^{\frac{q}{1-q}} ,
\label{qqexp}
\end{eqnarray}
in phenomenological investigations on $p_T$ spectra in high energy collisions, where $x$ is the corresponding physical variable as described above. 
Here the following remark is in order: a change in the power index $n$ to,
\begin{eqnarray}
n\cdot q=n+1
\end{eqnarray}
where $q=1+1/n$ is the non-extensive parameter in our fittings, into $n+1$,  since $\frac{q}{q-1}=\frac{1}{q-1}+1$.

As for describing the shape of spectra, different groups use various kinds of definitions:

\begin{itemize}

\item Experiments usually present the high-energy multiparticle production spectra fitted by a Tsallis distribution, well described by the formula
\begin{eqnarray}
E\frac{\dd^3N}{\dd p^3}\propto \ee_q\left(-\frac{E}{T}\right) ,
\end{eqnarray}
where the shape factor $\ee_q(x)$, given in Eq.(\ref{qexp}), is a generalized distribution function of the BG one.
The latter is reconstructed as a special case for the particular value $q=1$.

\item 
Some works use a normalization constant, $C_q$, which also depends on the particle mass, $m$, fitting temperature, $T$, and the non-extensive Tsallis parameter, $q$ (or  $n\equiv1/(q-1$)) to redefine the hadron spectra 
\begin{eqnarray}
E\frac{\dd^3N}{\dd p^3}\propto C_q(n, m,T)\cdot \ee_q\left(-\frac{E}{T}\right) .
\end{eqnarray}
The normalization factor $C_q$ can be fixed, for example, by normalizing the transverse momentum spectrum with the rapidity distribution
\begin{eqnarray}
E\frac{\dd ^3 N}{\dd p^3}&&=\frac{1}{2\pi p_T}\frac{\dd ^2N}{\dd y\dd p_T}
\nonumber \\
&&=\frac{\dd N}{\dd y}\cdot C_q\cdot \left(1+\frac{m_T-m}{nT}\right)^{-n} ~.
\end{eqnarray}
With the condition 
\begin{eqnarray}
\int^\infty_0 \frac{1}{2\pi p_T}\frac{\dd ^2N}{\dd y\dd p_T} 2\pi p_T\dd p_T =\frac{\dd N}{\dd y} ,
\end{eqnarray} 
one obtains the normalization constant $C_q$ in the $m_T-m$ scaling case as
\begin{eqnarray}
C_q=\frac{(n-1)(n-2)}{2\pi nT[nT+m(n-2)]}.
\end{eqnarray}
Here
$\dd N/\dd y$ is the particle yield per unit rapidity,
determined by integrating the $p_T$ spectrum for given particles.

\item  
Based on Ref.~\cite{Lavagno-2002},  the total number of particles is
\begin{eqnarray}
N=gV\int \frac{\dd^3p}{(2\pi)^3}\Big[\ee_q\left(-\frac{E}{T}\right)\Big ] ,
\end{eqnarray} 
so one can interpret the spectra also as being proportional to
\begin{eqnarray}
E\frac{\dd^3N}{\dd p^3}=\frac{gV}{(2\pi)^3}\cdot E\cdot \ee_q\left(-\frac{E}{T}\right)\propto E\cdot \ee_q\left(-\frac{E}{T}\right) .
\end{eqnarray}
Here $V$ denotes a 3-space volume absorbed in the normalized constant.
Note that there are two equivalent ways of expressing the normalization, either in terms of $V$ or in terms of $\dd N/\dd y$. 
The general choice for normalization, however, is $\dd N/\dd y$ since it is experimentally available, while $V$ is not.

\end{itemize}

In the followings we focus on some of the formulas used for approximating the identified particle spectra in various collisions.
This work compares the differences of ($m_T-m$) and $m_T$-dependent {\it ansatz}, as well as the simple $p_T$ function in the full azimuthal integrated space:
\begin{eqnarray}
E\frac{\dd N}{\dd^3p}=\frac{1}{2\pi p_T}\frac{\dd^2N}{\dd y\dd p_T}=f(p_T) .
\end{eqnarray}

Next we will investigate different formulas with respect to fitting experimental data. 
There are variances on using the non-extensive function, $\ee_q(x)$, based on different assumptions on the physical level of their respective cause.
For example, a power of $q/(1-q)$ is derived from thermodynamical derivative with respect to the chemical potential, while $n=1/(q-1)$ is just a direct replacement. 
$m_T$ and $p_T$ as variables again are more direct assumptions: at high $p_T$, $p_T\gg m$, these should coincide.
See our items 1-4 above.
Further differences root in various physical pictures on the proper one-particle variable to be used in the fits: $(m_T-m)$ is motivated by a leading-order estimate of $\mu\approx m$ and can be falsified by investigating hadron spectra with various masses.

Without loss of generality, here we investigate the shape functions listed in Eqs.(\ref{functions}).
In all the following figures, these different spectra-fitting formulas correspond in order to, 
\begin{align}
f_0&=f_{BG}=A_0\cdot \exp\left(\textcolor{blue}{-}\frac{m_T-m}{T_0} \right), \nonumber \\
f_1&=A_1\cdot \left(1+\frac{m_T-m}{n_1T_1}\right)^{-n_1}, \nonumber \\
f_2&=A_2\cdot\frac{(n_2-1)(n_2-2)}{2\pi n_2T_2[n_2T_2+m(n_2-2)]}\cdot \nonumber \\
&~~~~~\times\left(1+\frac{m_T-m}{n_2T_2}\right)^{-n_2}, \nonumber \\
f_3&=A_3\cdot m_T\left(1+\frac{m_T-m}{n_3T_3}\right)^{-n_3}, \nonumber \\ 
f_4&=A_4\cdot \left(1+\frac{m_T}{n_4T_4} \right)^{-n_4}, \nonumber \\
f_5&= A_5\cdot \left(1+\frac{p_T}{n_5T_5}\right)^{-n_5}.
\label{functions}
\end{align} 
Our aim is to find out the best fit with all of these functions, in parallel, obtaining physical information from these parameters.

\section{Analysis of $pp$ Results}\label{sec:sec3}

In this section we compare the transverse momentum distributions in the elementary ($pp$) collisions and fittings by all the functions listed in Eqs.(\ref{functions}). 
Since there still exist fierce differences over the proper functional shape of $p_T$ spectra within the non-extensive approaches, it is also worthy to compare the $\chi^2/ndf$ values over different fitting functions. 
All the corresponding parameters are then analyzed in order to investigate further the non-extensive physics behind.

\subsection{The hadron spectra}

In $pp$ collisions, the flow effects seem to contribute less than in heavy-ion collisions.
However, ``elementary" $pp$ collisions are rather complicated processes: there exist two different regimes of particle production.
One is a soft multiparticle production, dominant at low transverse momenta, here the spectra reveal an almost exponential behavior close to the BG statistics~\cite{Hagedorn-1995}, cf. in Fig.\ref{figspe1} in Appendix~\ref{sec:secB}.
At high $p_T$, on the other hand, they display power law tails, shown in Fig.\ref{figspe1} and Fig.\ref{figspe2} in Appendix~\ref{sec:secB}.
They are well described by perturbative QCD owing to the hard scattering of current quarks.
In a number of publications \cite{all-4,all-10, all-10-1-1, all-10-1,all-16,all-32,all-34} the Tsallis statistical distribution was successfully applied to describe data for $pp$ collisions over a wide range of the transverse momenta because of interpolating between low- and high- $p_T$ limit, first proposed by Hagedorn~\cite{Hagedorn-1995}:
\begin{eqnarray}
\ee_q(-\frac{p_T}{T}) \longrightarrow \left\{ \begin{array}{l}  
		~\ee^{-p_T/T} \qquad  ~~~~~~~~~~ {\bf low}~p_T \\ 
		\\	
		\Big( (q-1)\frac{p_T}{T}\Big)^{\frac{1}{1-q}}  \qquad  {\bf high}~ p_T 	
\end{array} \right. ~.
\end{eqnarray}
We henceforth focus on the fittings of the produced charged particle spectra in elementary collisions at different colliding energies.
Data are taken for pions, kaons and protons in $pp$ collisions at $\sqrt{s}=62.4$ GeV, $200$ GeV from PHENIX Collaboration~\cite{all-16} and at 900 GeV~\cite{all-10, all-10-1-1}, 2.76 TeV~\cite{all-32}, 5.02 TeV and 7 TeV~\cite{all-10-1} from ALICE Collaboration. 
We apply all fitting functions listed in Eqs.(\ref{functions}) to the data within as wide $p_T$ range as possible. 
The exact fitting ranges are shown in Table \ref{tab2}, while we restrict our analysis to the mid-rapidity region $|y|<0.5$ in which region both the temperature $T$ and non-extensive parameter $q$ are $y$-independent based on the work in Ref.~\cite{Rybc-2014}. 
Note that in the followings $\pi$, $K$ and $p$ mark the spectra of $\frac{\pi^++\pi^-}{2}$, $\frac{K^++K^-}{2}$ and $\frac{p+\bar{p}}{2}$ respectively.

Fig.\ref{figspe1} and Fig.\ref{figspe2} in Appendix~\ref{sec:secB} show the $p_T$ spectra fitted by all the functions in Eqs.(\ref{functions}) for pions and protons in $pp$ collisions for various kinds of beam energies at midrapidity as examples. We could see that the classical BG distribution, $f_{BG}$, no longer fits the spectra, especially in larger $p_T$ regions or higher colliding energies. 
On the other hand, it seems that all the other five non-extensive fitting functions can fit the $p_T$ spectra in the elementary $pp$ collisions nicely.
There appear some differences for the proton spectra than for the pion ones: typically in the high $p_T$ part.
This is due to the fact that all these non-extensive formulas become similar when the particle mass is getting negligible in comparison to $p_T$.

\begin{table}[htb]
\caption{Fitting $p_T$ ranges of spectra for different charged particles in $pp$ collisions~\cite{all-10, all-10-1-1, all-10-1,all-16,all-32}:}
\scalebox{1.2}[1.2]{
\begin{tabular}{c|c|c|c}
\hline
\hline
$\sqrt{s}$ & $\pi$ [GeV/c] & $K$ [GeV/c] & $p$ [GeV/c] \\
\hline
62.4 GeV & 0.3-2.9 & 0.4-2 & 0.6-3.6 \\
\hline
200 GeV  &  0.3-3  & 0.4-2 &  0.5-4.6 \\
\hline 
900 GeV  & 0.1-2.6 & 0.2-2.4 & 0.35-2.4 \\
\hline
2.76 TeV  & 0.1-20 & 0.2-20 & 0.3-20 \\ \hline
5.02 TeV   &  0.1-20  & 0.2-20   &  0.3-20 \\ \hline
7 TeV  & 0.1-20 & 0.2-20 & 0.3-20 \\ 
\hline
\hline
\end{tabular}
}
\label{tab2}
\end{table}

More details could be found when we focus on the values of $\chi^2/ndf$ cf. Fig.\ref{figchi}.
One realizes that for all different colliding energies their values are around or smaller than 1, except for the formula $f_0$ ($f_{BG}$) which gives the worst fitting results of all. 
The BG statistics fails describing $p_T$ spectra in $pp$ collisions. 
This can be understood from the feature that in the elementary collisions over-estimated values of the relative variances $\sigma$ occur in the experimental data and result in smaller $\chi^2/ndf$.
Specifically, the first two distributions ($f_1$ and $f_2$) of $m_T-m$ and $f_4$ of $m_T$ show close goodness.
The distribution, $f_3$, derived thermodynamically, also does not display big differences for the fit goodness.

Checking the fitting parameters, $A$, $T$ and $q=1+1/n$, we observe that all these share the same Tsallis parameter $n$, but the two $m_T-m$ functions ($f_1$ and $f_2$) conclude to values of the fitting temperature $T$ different from the pure $m_T$ one ($f_4$). 
This indicates that the normalization constant does not affect the fitted $T$ and $q$ parameters but the integrated yield $\dd N/\dd y$, namely, $A_2$.

As for the distribution of $f_5(p_T)$, the results of $\chi^2/ndf$ turn out to be the best of all in the spectra for pions, because of its small mass, cf. Fig.\ref{figchi}.

\subsection{Connections between the fitting functions}\label{sec:secC}

Some universality discussed in the previous text is reasonably expected on inspecting Eqs. (\ref{functions}). 
Comparing the formulas for $f_1$ and $f_2$ it is easy to realize that they coincide whenever their amplitudes satisfy the relation
\begin{eqnarray}
A_1=A_2\cdot\frac{(n_2-1)(n_2-2)}{2\pi n_2T_2[n_2T_2+m(n_2-2)]}=A_2\cdot C_q.
\end{eqnarray}

The first panel in Fig.\ref{figAT} in Appendix~\ref{sec:secB} demonstrates this relation.
Here $T_2$ stands for the fitting temperature with the fitting function $f_2$ and then the values of all the other fitting parameters.
Accounting for the differences between ($m_T-m$) and $m_T$ scalings, {\it i.e.}, comparing $f_1$ with $f_4$, we re-cast $f_1$ given in Eqs. (\ref{functions}) as follows
\begin{eqnarray}
f_1=A_1\cdot \left(1-\frac{m}{n_1T_1}\right)^{-n_1}\cdot \left(1+\frac{m_T}{n_1T_1-m}\right)^{-n_1} .
\end{eqnarray}
Comparing it with $f_4$, it is straightforward to arrive at the relations:
\begin{eqnarray}
A_1\cdot \left(1-\frac{m}{n_1T_1}\right)^{-n_1} &&=A_4, \\
n_1&&=n_4, \\
n_1T_1-m&&=n_4T_4.
\label{mmT}
\end{eqnarray}
The lower panel in Fig.\ref{figAT} points out this clearly.
This in turn explains the similarity of  $\chi^2/ndf$ values over these fitting functions. This in fact was a triviality check, useful to demonstrate that no inconsistency error occur by applying different fit formulas of $m_T-m$ or $m_T$ and whether it is normalized properly.

\subsection{The scaling of fitting temperature $T$}

\begin{figure}[!htb]
\includegraphics[width=0.95\linewidth]{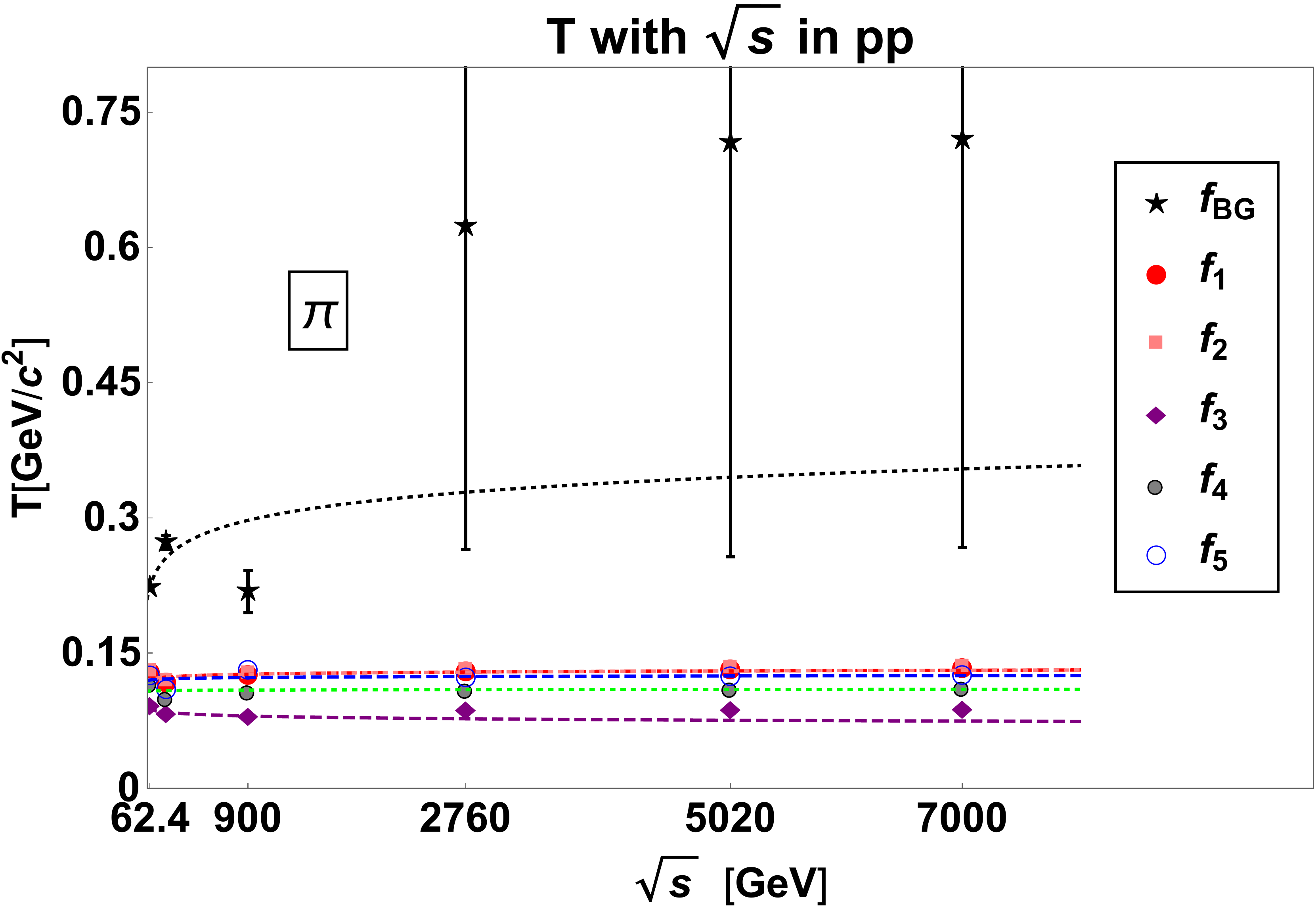}
\includegraphics[width=0.95\linewidth]{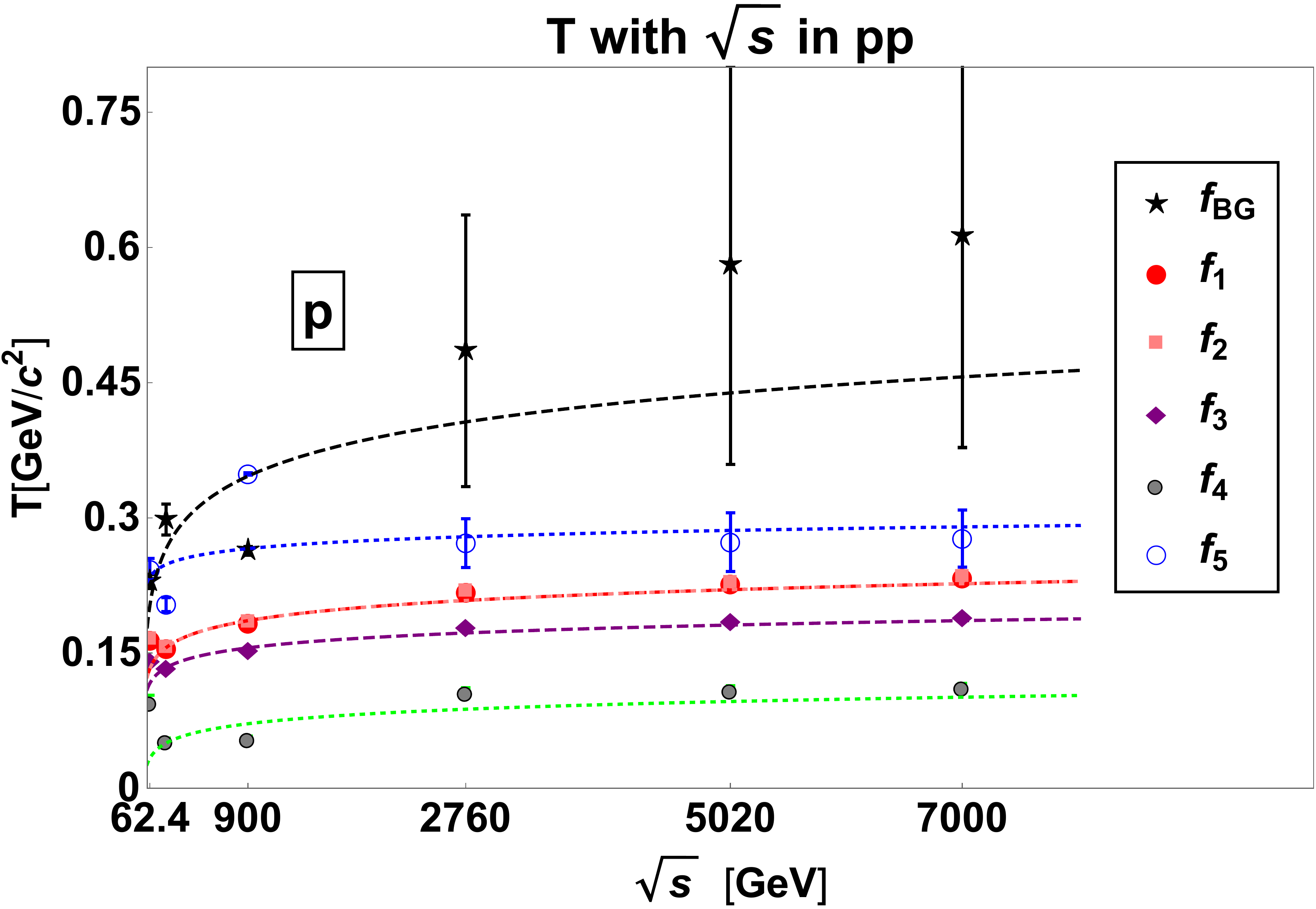}
\caption{The center-of-mass energy $\sqrt{s}$-dependence of the inverse slope parameter $T$ from $pp$. Results of $\pi$ spectra are shown in the upper panel and the lower is of protons. All dashed lines are fitted by: $T=T_0+T_1\cdot\ln\sqrt{s}$.}
\label{fig3}
\end{figure}

\begin{figure}[!htb]
\includegraphics[width=0.95\linewidth]{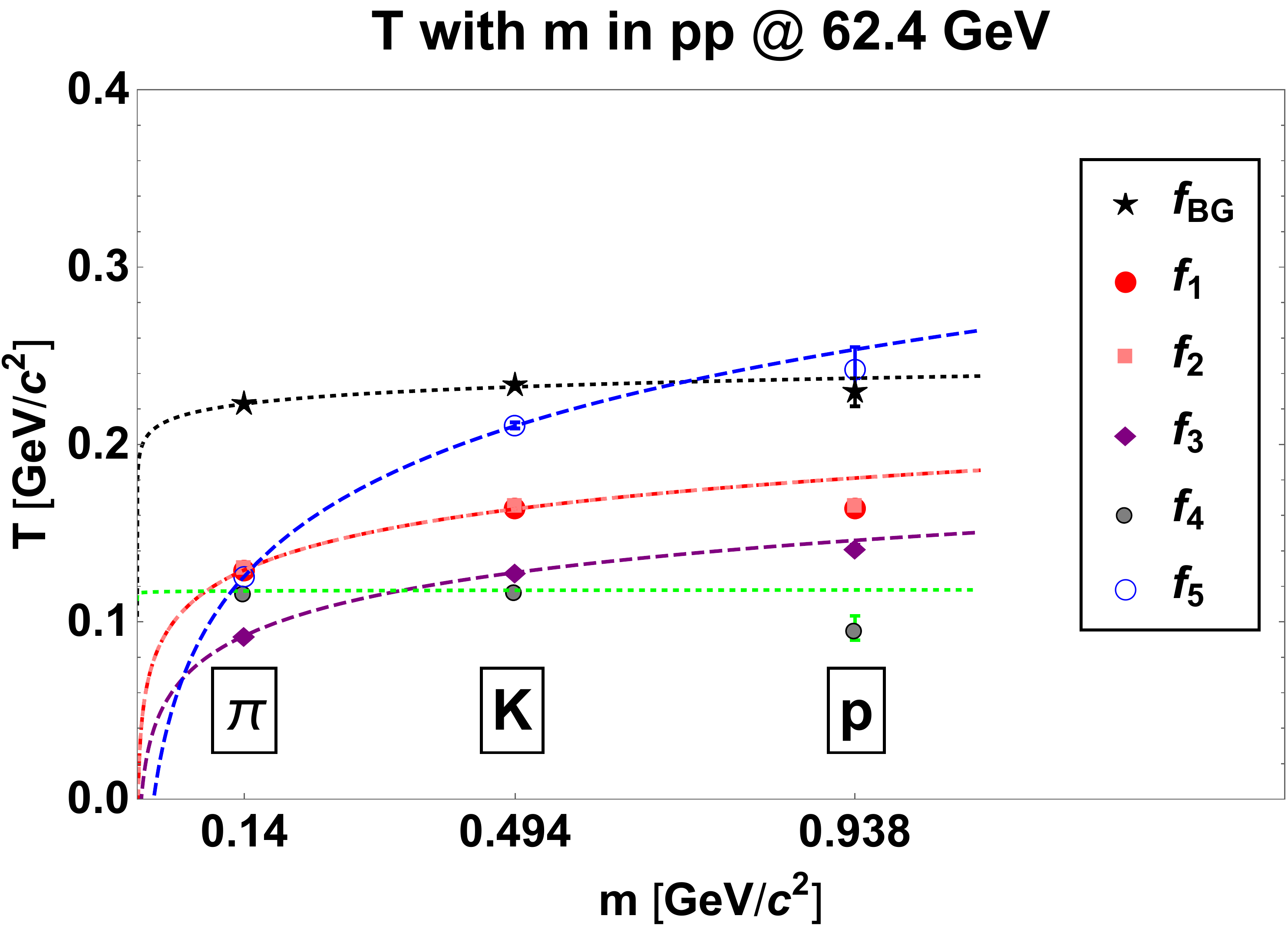}
\includegraphics[width=0.95\linewidth]{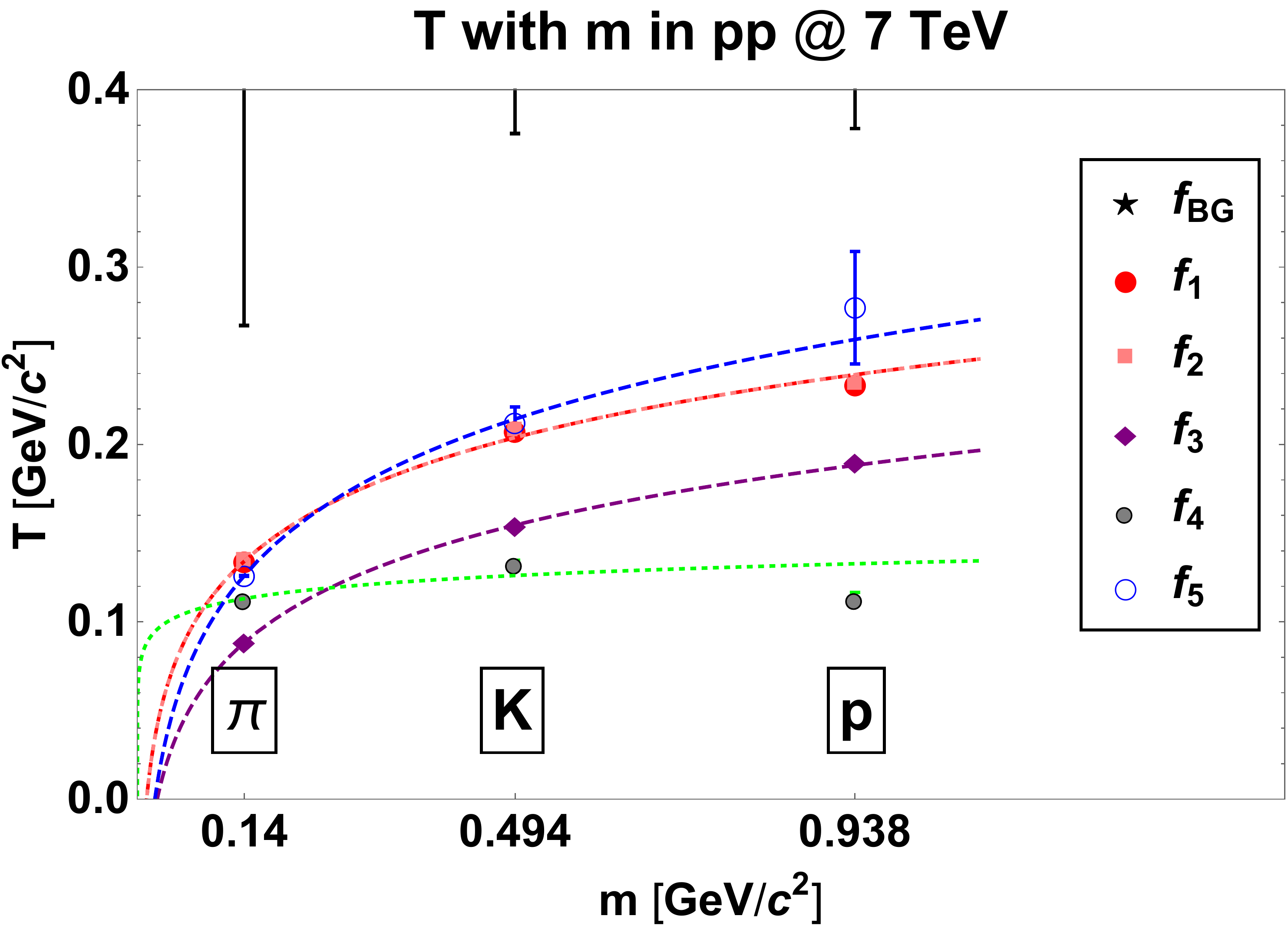}
\caption{The particle mass $m$-dependence analysis of the inverse slope parameter $T$ from $pp$. We list the analysis in $pp$ at 62.4 GeV and 7 TeV as examples. All the dashed lines are obtained by the relationship: $T=T'_0+T'_1\cdot\ln m$.}
\label{fig3-1}
\end{figure}

\begin{table}[htb]
\caption{Fitting parameters of $T_0$ and $T_1$ of Eq.(\ref{Ts}) at different hadron masses:}
\scalebox{1.1}[1.1]{
\begin{tabular}{c|c|c|c}
\hline
\hline
($T_0,T_1$) & $m_\pi$ & $m_K$ & $m_p$ \\
\hline
$f_{BG}$ & (0.108,0.028) & (0.139,0.023) & (0.186,0.012) \\
\hline
$f_1$  &  (0.112,0.002)  & (0.061,0.017)  & (0.050,0.020) \\
\hline 
$f_2$  & (0.112,0.002) & (0.061,0.017) & (0.050,0.020) \\
\hline
$f_3$ & (0.099,-0.003) & (0.061,0.011) & (0.056,0.015) \\ \hline
$f_4$   &  (0.106,-0.0004)  & (0.002,0.015)   &  (-0.026,0.014) \\ \hline
$f_5$  & (0.116,0.001) & (0.183,0.0004) & (-0.019,0.054) \\ 
\hline
\hline
\end{tabular}
}
\label{tabTs}
\end{table}

\begin{table*}[htb]
\caption{Fitting parameters of $T'_0$ and $T'_1$ of Eq.(\ref{Tm}) at various energies:}
\scalebox{1.3}[1.3]{
\begin{tabular}{c|c|c|c|c|c|c}
\hline
\hline
($T'_0, T'_1$) & 62.4 GeV & 200 GeV & 900 GeV & 2.76 TeV & 5.02 TeV & 7 TeV \\
\hline
$f_{BG}$ & (0.238,0.008) & (0.255,-0.007) & (0.266,0.018) & (0.495,-0.087) & (0.593,-0.083) & (0.624,-0.063) \\
\hline
$f_1$  &  (0.183,0.027)  & (0.155,0.019) &  (0.181,0.027) & (0.228,0.049) & (0.237,0.053) & (0.243,0.055) \\
\hline 
$f_2$  &   (0.183,0.027)  & (0.155,0.019) & (0.181,0.027) & (0.228,0.049) & (0.237,0.053) & (0.243,0.055) \\
\hline
$f_3$  & (0.148,0.028) & (0.128,0.023) & (0.148,0.035) & (0.179,0.046) & (0.186,0.050) & (0.192,0.053) \\ \hline
$f_4$   &  (0.118,0.0003)  & (0.048,-0.027)   &  (0.059,-0.025) & (0.130,0.010) & (0.133,0.011) & (0.133,0.010) \\ \hline
$f_5$  & (0.258,0.067) & (0.205,0.048) & (0.352,0.112) & (0.239,0.059) & (0.253,0.065) & (0.264,0.070) \\ 
\hline
\hline
\end{tabular}
}
\label{tabTm}
\end{table*}

Moving towards physical interpretation issues, we now investigate the dependence of $T$ on colliding energies $\sqrt{s}$ and the particle mass $m$. Fig.\ref{fig3} shows results both for pions and protons. 
The inverse slope parameters, $T$, look almost independent of the beam-energy apart from the BG fittings, which gives the worst result. 
This can be also inspected in Fig.\ref{figspe1} and Fig.\ref{figspe2} in Appendix \ref{sec:secB}: the spectral shapes of the same hadron stay similar while the colliding energies change.
Motivated by pQCD, we propose that the fitting temperature $T$ follows a DGLAP-like evolution~\cite{Gabor-2018, Gabor-2018-1}:
\begin{eqnarray}
T=T_0+T_1\cdot \ln \sqrt{s} ,
\label{Ts}
\end{eqnarray}
which goes through all the points. 
As also shown in Fig.\ref{fig3}, the $\sqrt{s}$-dependence is log-like and gets stronger with higher hadron mass.
In formulas under the logarithms we use GeV units.

The parameter, $T$, on the other hand, exhibits a mass-dependence, cf. Fig.\ref{fig3-1}.
The larger uncertainties of $f_{BG}$ in $pp$ at 7 TeV (the black stars which is too large to be shown in the plot) also prove that the BG distribution fails in describing the transverse momentum spectra in $pp$ collisions at larger beam energies. 
Hereby we also apply a similar guess:
\begin{eqnarray}
T=T'_0+T'_1\cdot \ln m~,
\label{Tm}
\end{eqnarray}
to check the dependence of this parameter on the hadron mass $m$ within the same $pp$ collision.
Characteristically lower $T$ values for the lighter particles might occur due to the fact that the low-$p_T$ pions mostly stem from resonance decays; another possible explanation can be a collective hydrodynamical behavior in the small $pp$ system, cf. Ref.~\cite{all-16}.

In order to quantify the statements given above, we list all fitted parameters from Eq.(\ref{Ts}) to Eq.(\ref{Tm}) in Tables \ref{tabTs} and \ref{tabTm}.
We could see that all the parameters obtained by functions $f_1$ and $f_2$ are exactly the same, meaning that the normalized constant does not affect the fitting parameters ($T$ or $q$) at all.
In addition, we can also tell the log-dependence of the center-of-mass energy $\sqrt{s}$ and hadron mass $m$ in the analysis of all fitting parameters $T$. 
For instance, the values of the slope parameter $T_1$ in Table \ref{tabTs} develop larger with $m$ increasing, indicating that the $\ln \sqrt{s}$-dependence is getting stronger with heavier particles observed in the same collisions, agreeing with the discussion above.
Apart from the worse fittings of $f_{BG}$ and $f_5$, it is also found that all the $T_0$ values go close for the same particle.

Combining Eq.(\ref{Ts}) and (\ref{Tm}) together we can re-write the fitting temperature as
\begin{eqnarray}
T=T_c+T_s\cdot \ln\frac{\sqrt{s}}{m}~.
\label{Tsm}
\end{eqnarray}
As shown in the upper panel of Fig.\ref{figTqsm}, all the fitting parameters are well described by this relationship except for $f_{BG}$.
The simple $m_T$ function $f_4$ seems to give the best fitting of Eq.(\ref{Tsm}) to the data set.

\subsection{The scaling of the Tsallis $q$}

\begin{figure}[!htb]
\includegraphics[width=0.95\linewidth]{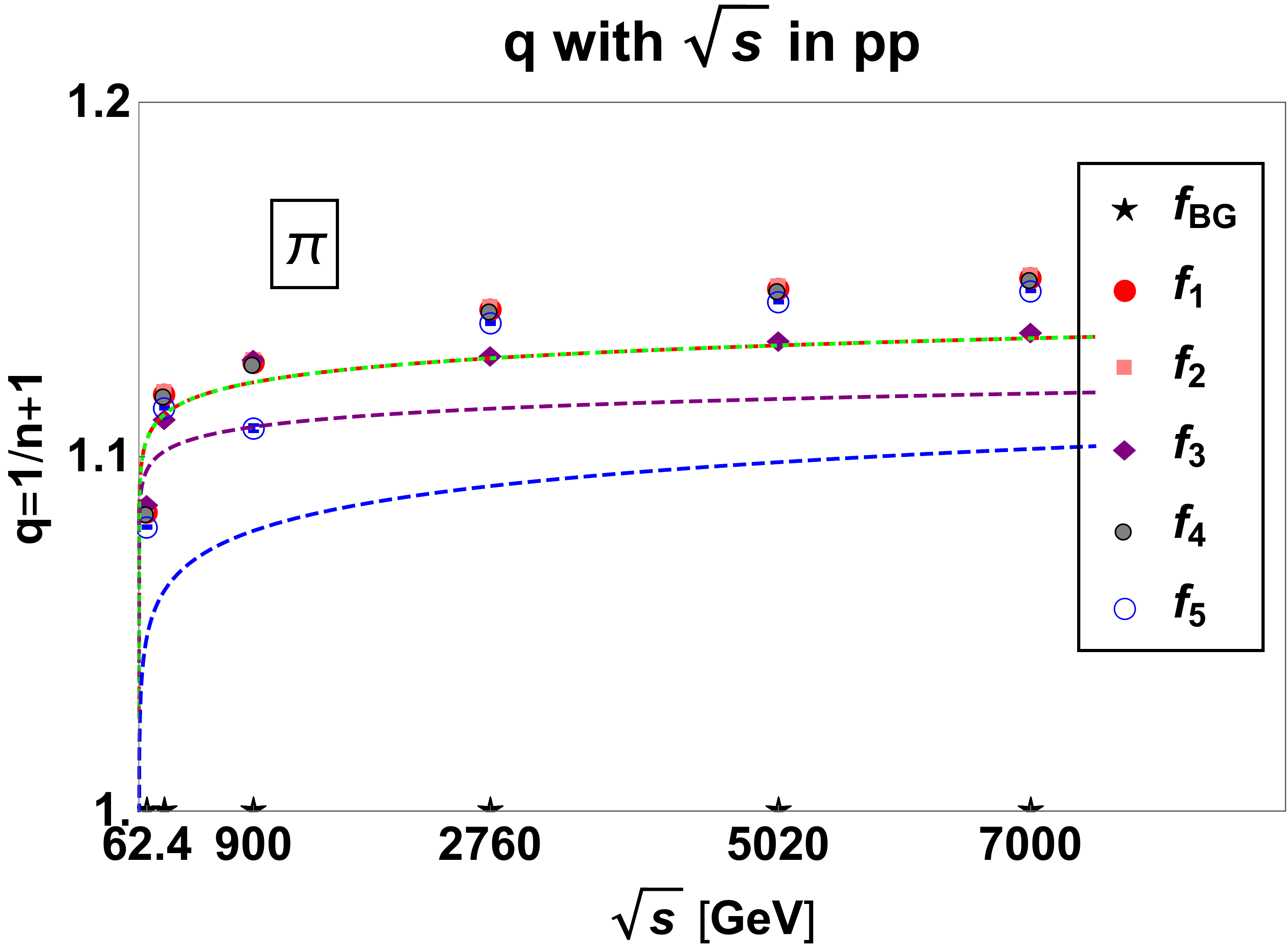}
\includegraphics[width=0.95\linewidth]{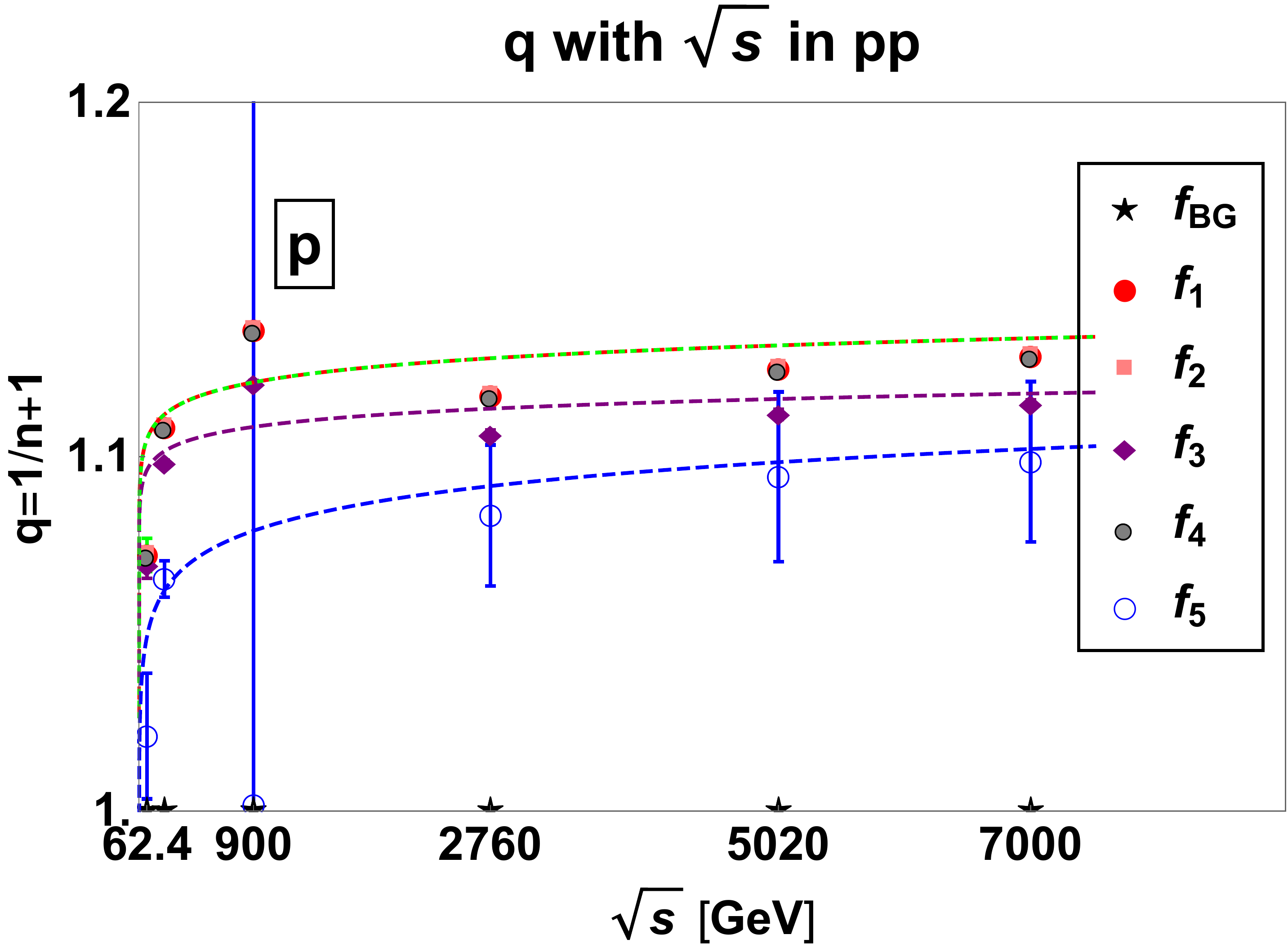}
\caption{The center-of-mass energy $\sqrt{s}$-dependence of the Tsallis parameter $q$ from $pp$. Results of $\pi$ spectra are shown in the upper panel and the lower is of protons as examples. All dashed lines are fitted by: $q=q_0+q_1\cdot\ln\sqrt{s}$.}
\label{fig33}
\end{figure}

\begin{figure}[!htb]
\includegraphics[width=0.95\linewidth]{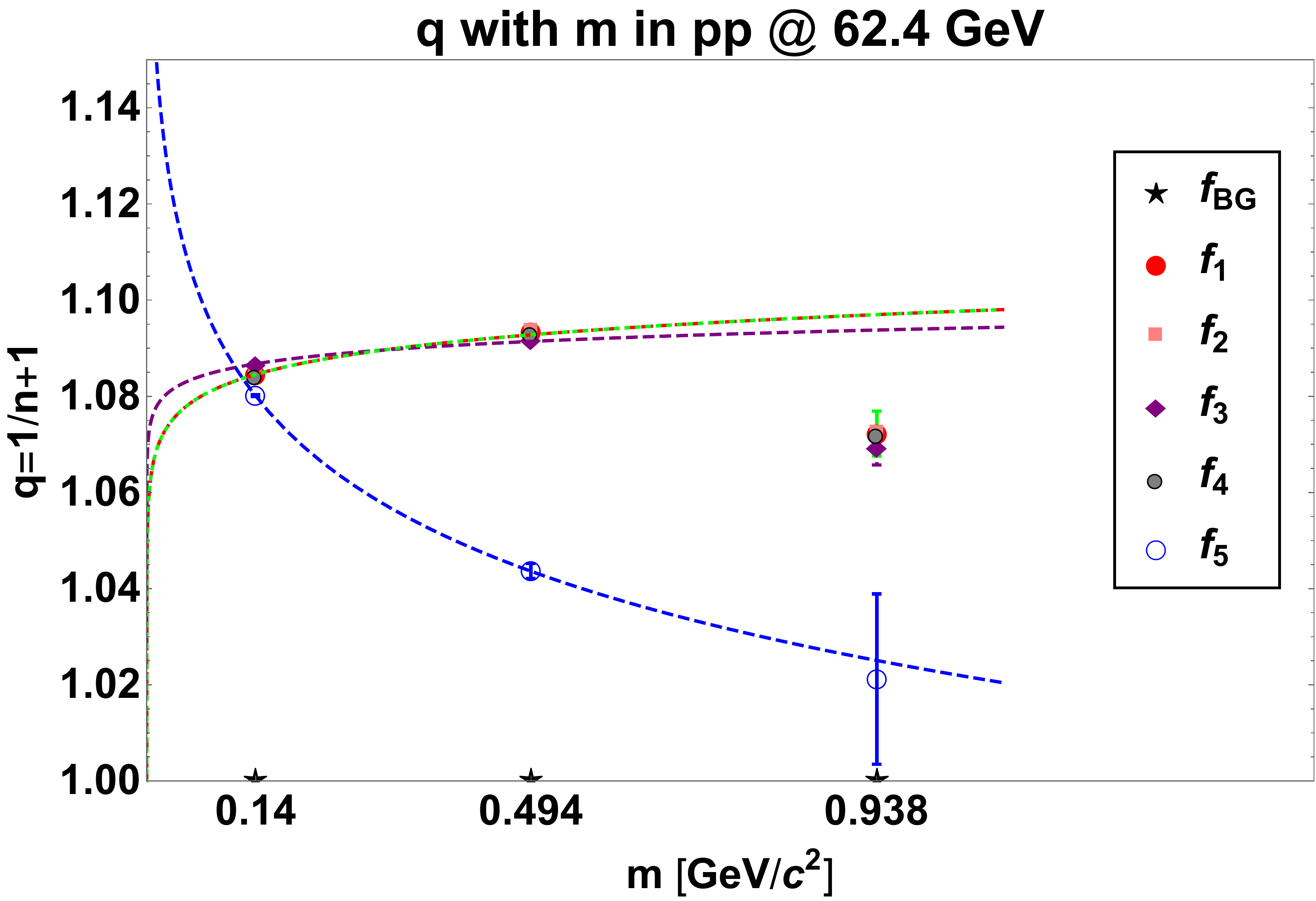}
\includegraphics[width=0.95\linewidth]{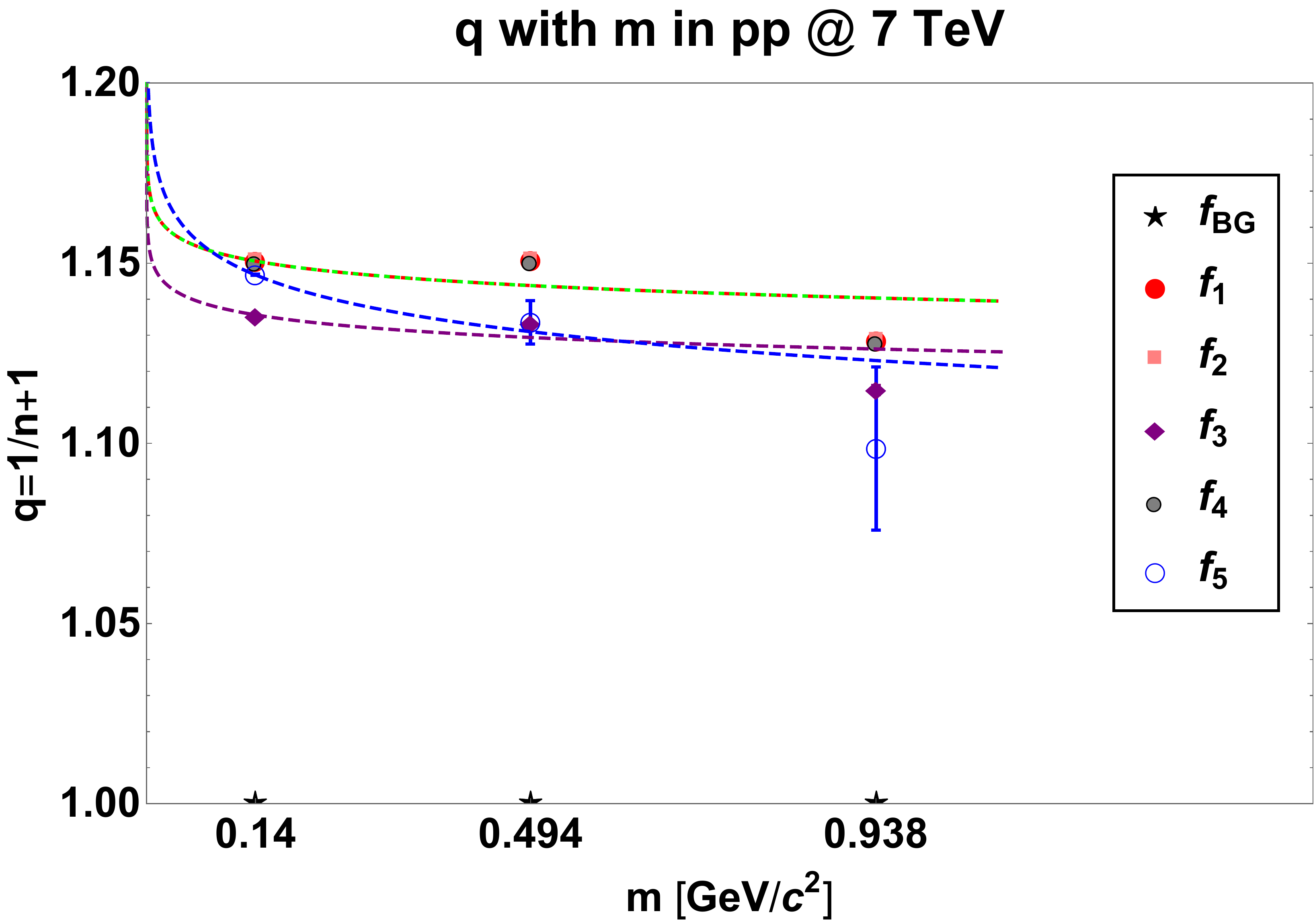}
\caption{The particle mass $m$-dependence analysis of the Tsallis parameter $q$ from $pp$. We list the analysis in $pp$ at 62.4 GeV and 7 TeV as examples. All the dashed lines are obtained by the relationship: $q=q'_0+q'_1\cdot\ln m$.}
\label{fig33-1}
\end{figure}

\begin{table}[htb]
\caption{Fitting parameters of $q_0$ and $q_1$ of Eq.(\ref{qs}) at different hadron masses:}
\scalebox{1.2}[1.2]{
\begin{tabular}{c|c|c|c}
\hline
\hline
($q_0,q_1$) & $m_\pi$ & $m_K$ & $m_p$ \\
\hline
$f_1$ & (1.042,0.012) & (1.125,0.002) & (1.080,0.006) \\
\hline
$f_2$  &   (1.042,0.012)  & (1.125,0.002) &  (1.080,0.006) \\
\hline 
$f_3$  & (1.066,0.009) & (1.112,0.002) & (1.077,0.005) \\
\hline
$f_4$  &  (1.042,0.012) & (1.125,0.002) & (1.080,0.006) \\ \hline
$f_5$   &  (1.034,0.013)  & (1.005,0.017)   &  (1.003,0.011) \\  
\hline
\hline
\end{tabular}
}
\label{tabqs}
\end{table}

\begin{table*}[htb]
\caption{Fitting parameters of $q'_0$ and $q'_1$ of Eq.(\ref{qm}) at various energies:}
\scalebox{1.3}[1.3]{
\begin{tabular}{c|c|c|c|c|c|c}
\hline
\hline
($q'_0, q'_1$) & 62.4 GeV & 200 GeV & 900 GeV & 2.76 TeV & 5.02 TeV & 7 TeV \\
\hline
$f_1$ & (1.097,0.007) & (1.163,0.023) & (1.162,0.018) & (1.140,-0.001) & (1.142,-0.003) & (1.140,-0.005) \\
\hline
$f_2$  &  (1.097,0.007)  &  (1.163,0.023) &  (1.162,0.018) & (1.140,-0.001) & (1.142,-0.003) & (1.140,-0.005) \\
\hline 
$f_3$  & (1.094,0.004) & (1.142,0.015) & (1.140,0.006)  & (1.123,-0.003) & (1.127,-0.003) & (1.126,-0.005) \\
\hline
$f_4$  & (1.097,0.007) & (1.163,0.023) & (1.162,0.018) & (1.140,-0.001) & (1.142,-0.003) & (1.140,-0.005)\\ \hline
$f_5$   &  (1.023,-0.029)  & (1.090,-0.012)   &  (1.069,-0.020) & (1.118,-0.010) & (1.122,-0.011) & (1.122,-0.012) \\ 
\hline
\hline
\end{tabular}
}
\label{tabqm}
\end{table*}

Similar behavior occurs on the Tsallis non-extensive parameter $q$,
\begin{eqnarray}
q=q_0+q_1\cdot \ln \sqrt{s}~,
\label{qs}
\end{eqnarray}
as shown in Fig.\ref{fig33} and 
\begin{eqnarray}
q=q'_0+q'_1\cdot \ln m~,
\label{qm}
\end{eqnarray}
in Fig.\ref{fig33-1}. 

In the plots in Fig.\ref{fig33-1} one inspects that the simple formula of $f_5(p_T)$, displays a typical relation between $q$ and $m$ especially at the low beam energies.
The results obtained by fitting the $f_{BG}$ distribution are removed because $q=1$ in this case.

Tables~\ref{tabqs} and \ref{tabqm} list all the parameters in Eqs.(\ref{qs}) and (\ref{qm}).
Similarly, we observe that fitting results by $f_1$ and $f_2$ are exactly the same. 
Furthermore, the $f_4(m_T)$ function gets the same non-extensive parameter $q$ as the $f_1(m_T-m)$ and $f_2(m_T-m)$.
As a consequence, the parameters ($q_0,q_1$) and ($q'_0,q'_1$) are also the same for the three.
Considering the $m$-dependence in Eq.(\ref{qm}), Table~\ref{tabqm} shows that all the $q'_0$ values are nearly the same for all kinds of beam energies.

\begin{figure}[!htb]
\includegraphics[width=0.95\linewidth]{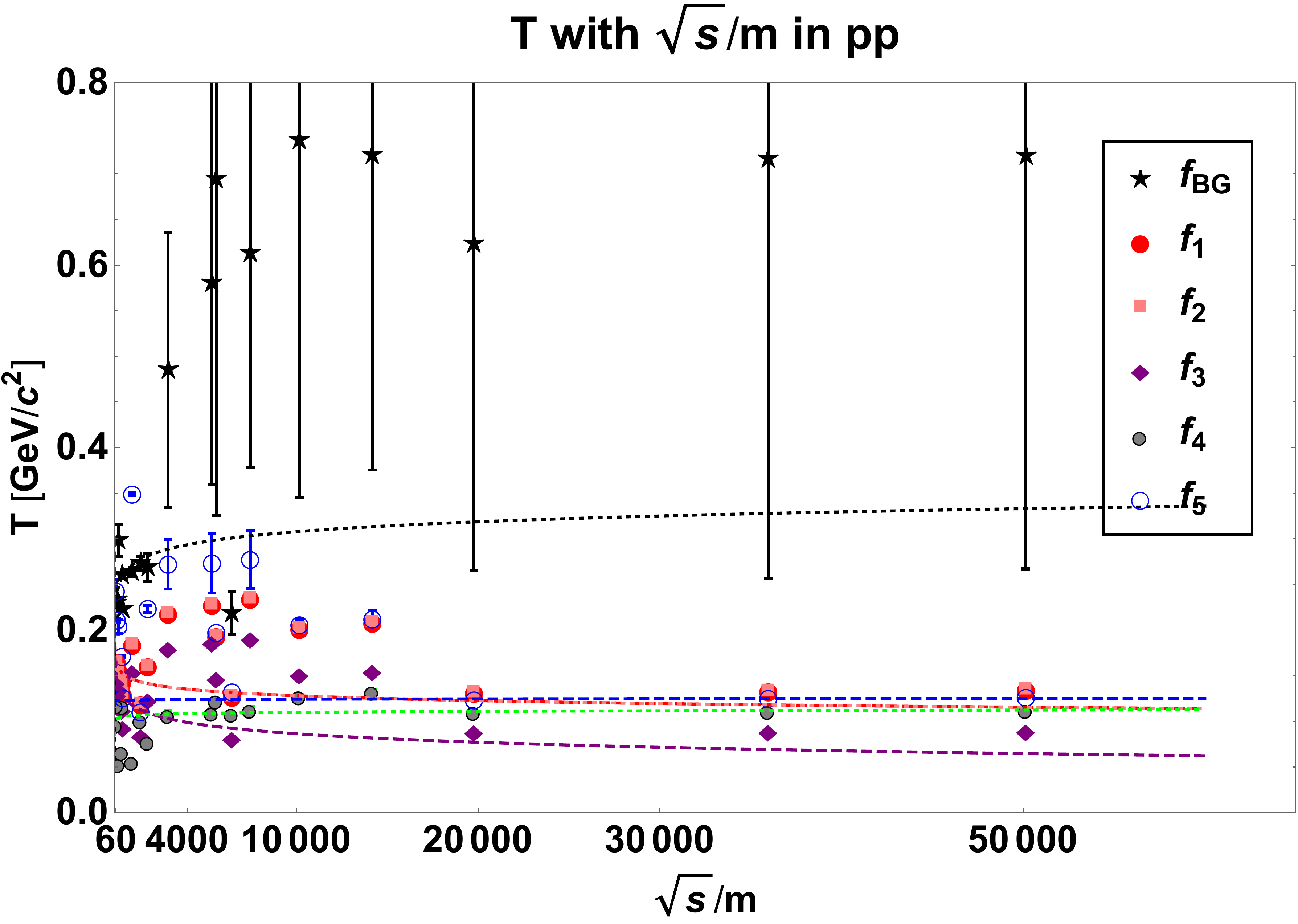}
\includegraphics[width=0.95\linewidth]{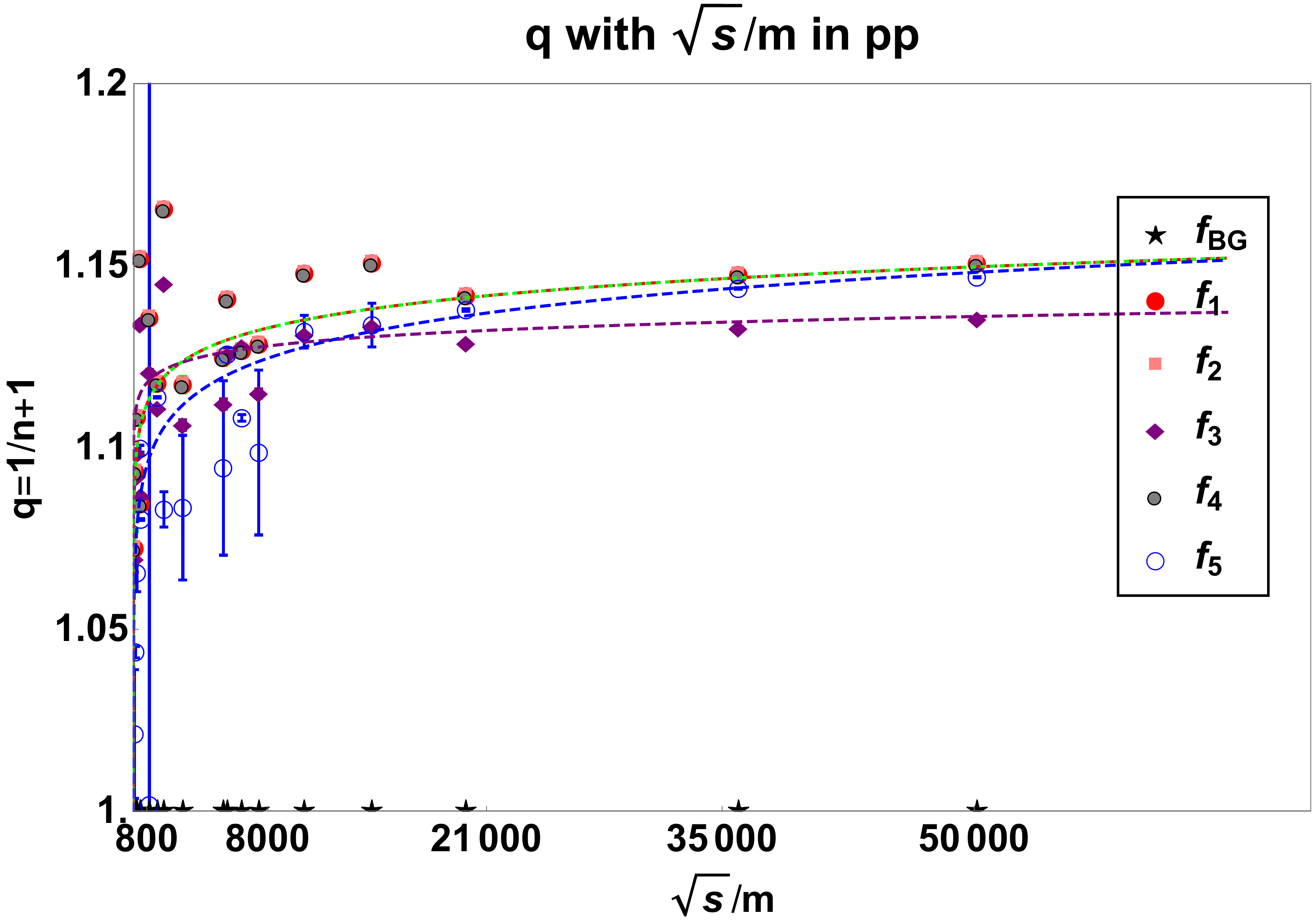}
\caption{Both the center-of-mass energy $\sqrt{s}$ and hadron mass $m$ distributions of the fitting temperature $T$ and the Tsallis parameter $q$. We analyze all the fitting results of various hadrons at all given colliding energies in $pp$ collisions. All the dashed lines are by the function: $T=T_c+T_s\cdot\ln(\sqrt{s}/m)$ (top) and $q=q_c+q_s\cdot\ln(\sqrt{s}/m)$ (bottom).}
\label{figTqsm}
\end{figure}

We also investigate the Tsallis parameter $q$ as a combined function of both:
\begin{eqnarray}
q=q_c+q_s\cdot\ln\frac{\sqrt{s}}{m}~.
\label{qsm}
\end{eqnarray}
The lower panel of Fig.\ref{figTqsm} tells us that this relation represents a good description.
The first two functions from Eqs.(\ref{functions}) give the common running on Eq.(\ref{qsm}) as well as the fourth one, due to their same fitting Tsallis parameter $q$.

\subsection{Mean transverse mass scaling}

In addition, the mean transverse mass $\langle m_T\rangle$ is determined by integrating the $p_T$ spectra for each particle species:
\begin{eqnarray}
\langle m_T\rangle =\frac{\int_m^\infty m_T\cdot \frac{1}{2\pi m_T}\frac{\dd ^2N}{\dd y\dd m_T}\cdot 2\pi m_T \dd m_T}{\int_m^\infty \frac{1}{2\pi m_T}\frac{\dd ^2N}{\dd y\dd m_T}\cdot 2\pi m_T \dd m_T}
\end{eqnarray}
where $\frac{1}{2\pi m_T}\frac{\dd ^2N}{\dd y\dd m_T}=\frac{1}{2\pi p_T}\frac{\dd ^2N}{\dd y\dd p_T}$ is the corresponding $p_T$ spectrum. 
The mean value of the transverse mass $m_T=\sqrt{m^2+p_T^2}$, based on the integration over all $p_T$ for the spectra, is given as:
\begin{align}
\langle m_T\rangle_0&=\langle m_T \rangle_{BG}=m+T+\frac{T^2}{m+T},  \nonumber \\
\langle m_T\rangle_1&=\frac{m(n-3)nT+2n^2T^2}{(n-3)[m(n-2)+nT]}+m,  \nonumber \\
\langle m_T\rangle_2&=\langle m_T\rangle_1,  \nonumber \\
\langle m_T\rangle_3&=m+ \nonumber \\
&\frac{m^2(n-4)(n-3)nT+4m(n-4)n^2T^2+6n^3T^3}{(n-4)[m^2(n-3)(n-2)+2m(n-3)nT+2n^2T^2]},  \nonumber \\
\langle m_T\rangle_4&=\frac{2nT}{n-3}+\frac{(n-2)(n-1)}{(n-3)[nT+m(n-1)]}m^2.
\label{mTmean}
\end{align}

\begin{table}[h]
\caption{Values of $\langle m_T\rangle$ in $pp$ at 62.4 GeV with different fittings:}
\scalebox{1.2}[1.25]{
\begin{tabular}{c|c|c|c}
\hline
\hline
$\langle m_T\rangle$ & $\pi$ [GeV/c$^2$]  & $K$ [GeV/c$^2$]  & $p$ [GeV/c$^2$]   \\
\hline
$\langle m_T \rangle_{BG}$ & 0.499$\pm$0.003 & 0.802$\pm$0.003 & 1.213$\pm$0.011  \\
\hline
$\langle m_T \rangle_1$  & 0.395$\pm$0.001  & 0.769$\pm$0.003 & 1.169$\pm$0.001   \\
\hline 
$\langle m_T \rangle_2$  & 0.395$\pm$0.001 & 0.769$\pm$0.003 & 1.169$\pm$0.001  \\
\hline
$\langle m_T \rangle_3$  & 0.415$\pm$0.004 & 0.772$\pm$0.002 & 1.170$\pm$0.001  \\ \hline
$\langle m_T \rangle_4$ & 0.395$\pm$0.001 & 0.769$\pm$0.003 & 1.169$\pm$0.001  \\
\hline
data\cite{all-16} & 0.395$\pm$0.015 & 0.740$\pm$0.013 & 1.176$\pm$0.032 \\
\hline
\hline
\end{tabular}
}
\label{tab3}
\end{table}

\begin{figure*}[htb]
\scalebox{1}[1]{
\includegraphics[width=0.49\linewidth]{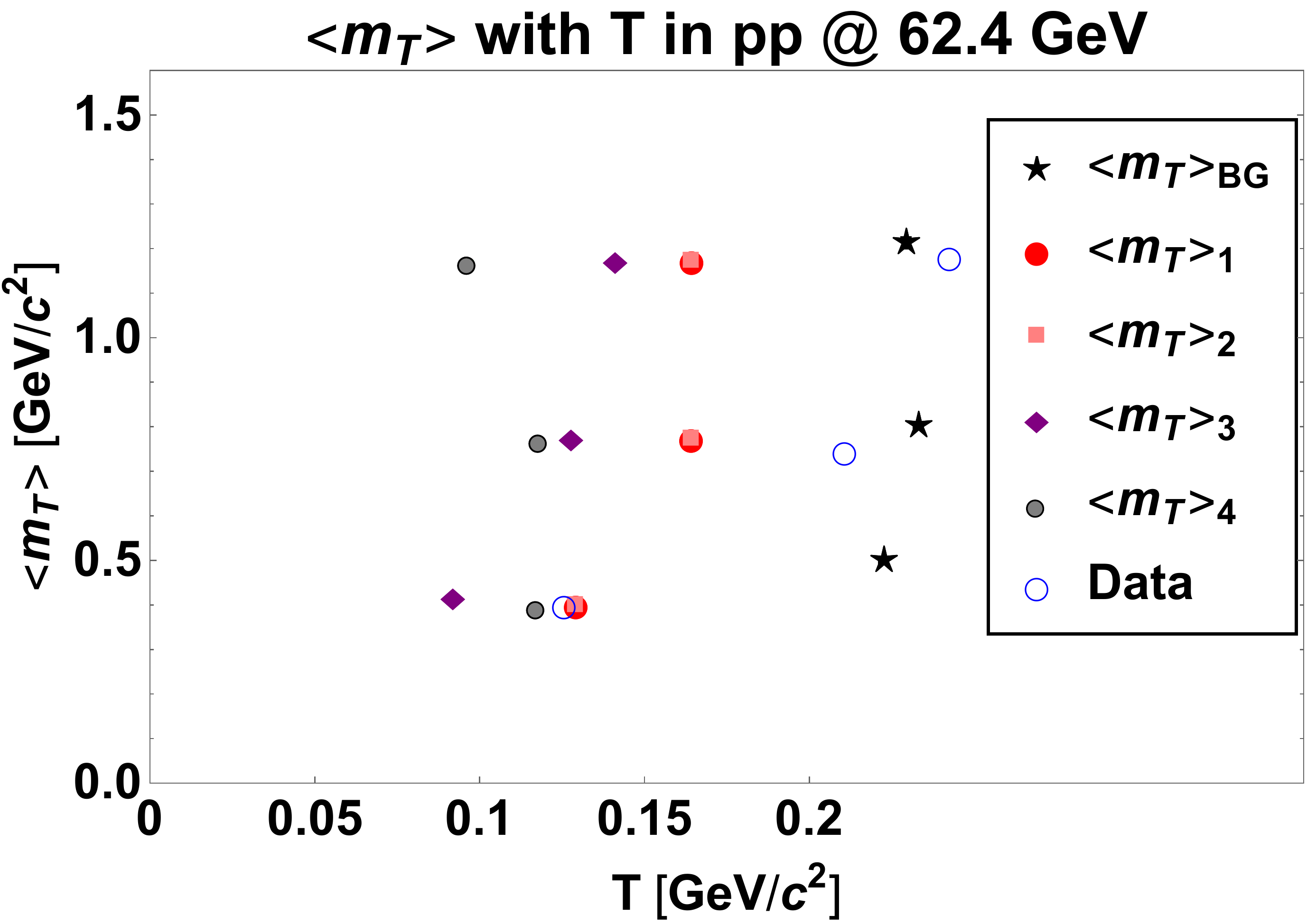}
\includegraphics[width=0.49\linewidth]{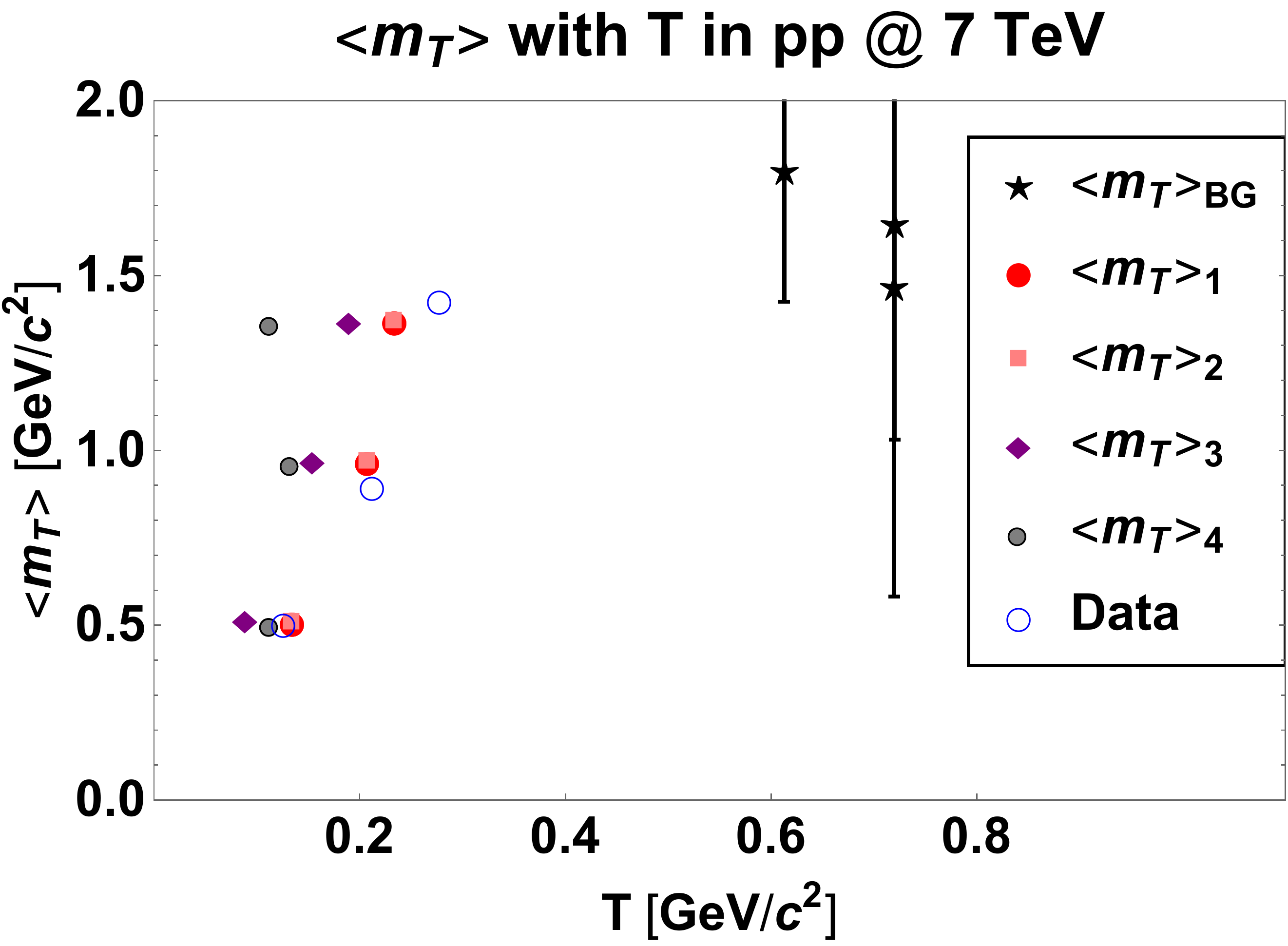}
}
\scalebox{1}[1]{
\includegraphics[width=0.49\linewidth]{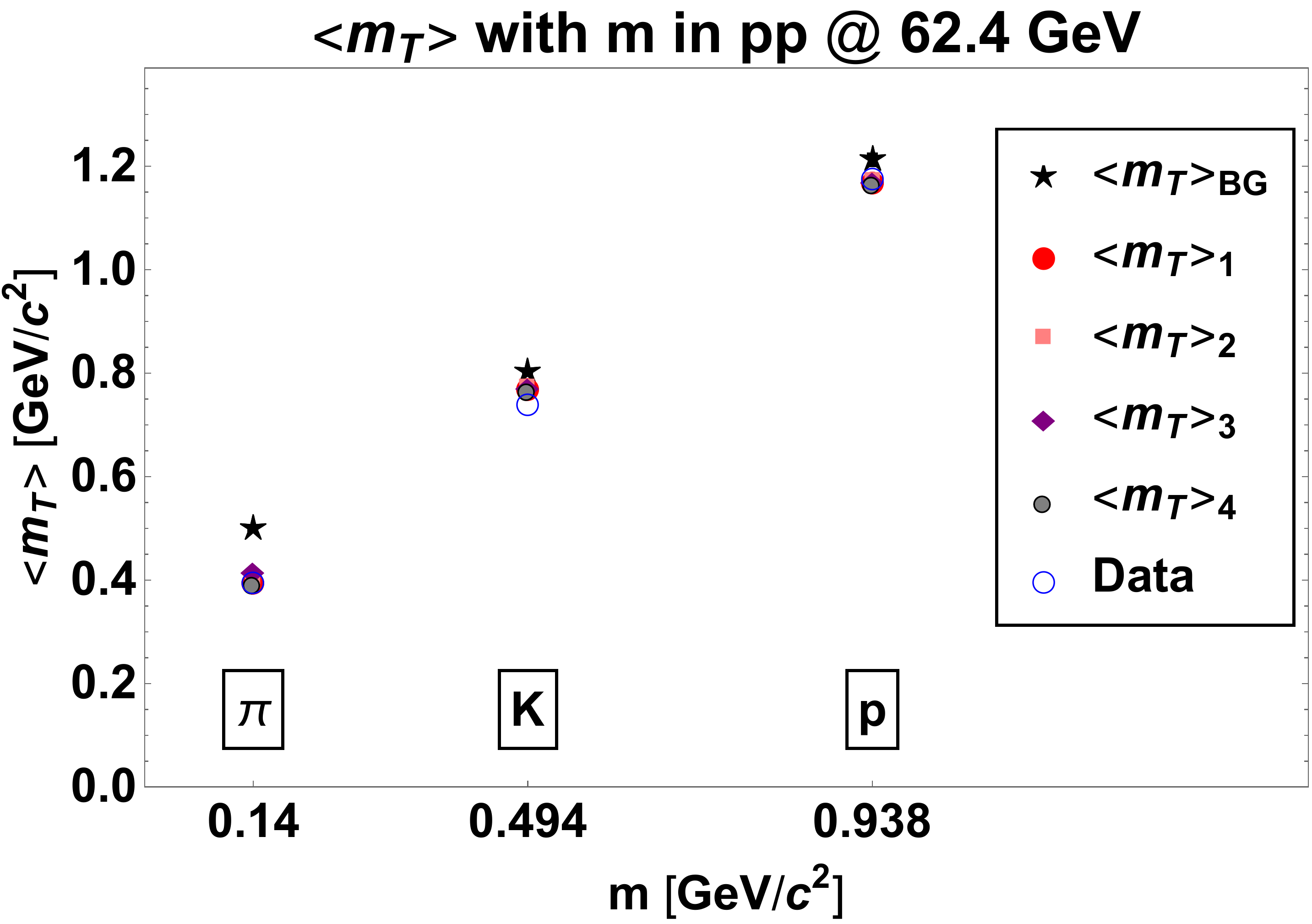}
\includegraphics[width=0.49\linewidth]{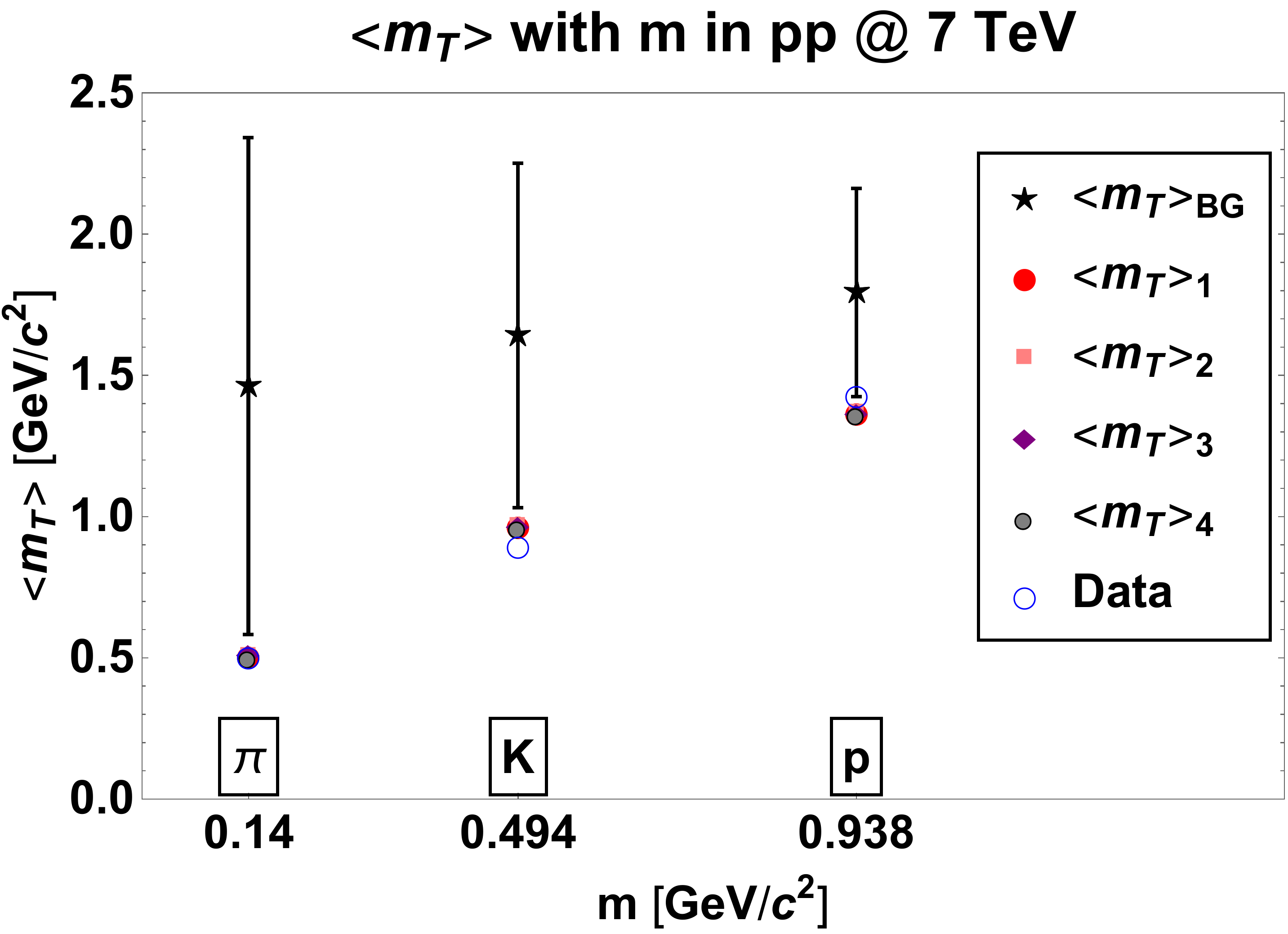}
}
\caption{Upper panel: Distributions of the fitting temperature $T$ with respect to the mean transverse mass $\langle m_T \rangle$ for $\pi$, $K$ and $p$. Lower panel:  Mean transverse mass $\langle m_T \rangle$ for $\pi$, $K$ and $p$ as a function of particle mass $m$. Here we list the results in $pp$ at 62.4 GeV and 7 TeV as examples. Data of the values of $\langle m_T \rangle$ (blue circles) are from Ref.~\cite{all-16} and \cite{pp-2} as comparisons.}
\label{fig4}
\end{figure*}

\begin{table}[h]
\caption{Values of $\langle m_T\rangle$ in $pp$ at 7 TeV with different fittings:}
\scalebox{1.2}[1.25]{
\begin{tabular}{c|c|c|c}
\hline
\hline
$\langle m_T\rangle$ & $\pi$ [GeV/c$^2$]  & $K$ [GeV/c$^2$]  & $p$ [GeV/c$^2$]   \\
\hline
$\langle m_T \rangle_{BG}$ & 1.462$\pm$0.880 & 1.641$\pm$0.610 & 1.793$\pm$0.368  \\
\hline
$\langle m_T \rangle_1$  & 0.503$\pm$0.001  & 0.963$\pm$0.0003 & 1.364$\pm$0.0005   \\
\hline 
$\langle m_T \rangle_2$  & 0.503$\pm$0.001 & 0.963$\pm$0.0003 & 1.364$\pm$0.0005  \\
\hline
$\langle m_T \rangle_3$  & 0.512$\pm$0.003 & 0.966$\pm$0.0002 & 1.365$\pm$0.0005  \\ \hline
$\langle m_T \rangle_4$ & 0.503$\pm$0.001 & 0.963$\pm$0.0003 & 1.364$\pm$0.0005  \\
\hline
data\cite{pp-2} & 0.500$\pm$0.007 & 0.892$\pm$0.0308 & 1.424$\pm$0.0515 \\
\hline
\hline
\end{tabular}
}
\label{tab3-7}
\end{table}

We compare our results with the one from the measured hadron spectra  in $pp$ at 62.4 GeV and 7 TeV for pions, kaons and protons in Table \ref{tab3} and \ref{tab3-7}. 
This study tells us that both the $m_T-m$ and $m_T$ distributions ($f_1$, $f_2$ and $f_4$ of Eqs. (\ref{functions})) share the same values for $\langle m_T\rangle$. 
This is also easy to check when we focus on the connections between parameters as discussed in the subsection above.
In Fig.\ref{fig4} the fitting temperature $T$ and $\langle m_T \rangle$ are plotted as a function of each other within different shape functions for various hadron kinds.
In order to clarify the relationship, we consider the $m$-dependence of the mean transverse mass in the same collision as well.
In the lower panel of Fig.\ref{fig4}, this quantity $\langle m_T\rangle$ indeed exhibits a linear connection to the hadron mass, cf. Eqs.(\ref{mTmean}).
Values obtained by the non-extensive approaches nicely coincide with the data set from Ref.~\cite{all-16} and \cite{pp-2}.
Results of the classical BG distribution $\langle m_T \rangle_{BG}$, the black stars, are somewhat far from the experimental observations. 
This again reflects the fact that the BG distribution cannot fit the transverse momentum spectra well.

~

In summary, we have systematically studied the fitting $p_T$ spectra of different hadrons in $pp$ collisions at several beam energies within various types of functions in Eqs.(\ref{functions}).
Our study indicates that the BG distribution, is no longer suitable for describing the hadron spectra over a wide range of $p_T$.
Comparisons of the corresponding fitting errors $\chi^2/ndf$ show that both $m_T-m$ and $m_T$ functions share the same goodness with or without the normalization.
Together with the thermodynamically derived formula, $f_3$, these non-extensive distributions ($f_1\sim f_4$) do follow the experimental data accurately. 
The simple $p_T$ function, $f_5$, on the other hand, gives the best fitting results on the pion spectra because of its small mass.

\section{Analysis of the $pPb$ and $PbPb$ results}\label{sec:sec4}

\begin{figure*}[!htb]
\scalebox{1}[1]{
\includegraphics[width=0.4\linewidth]{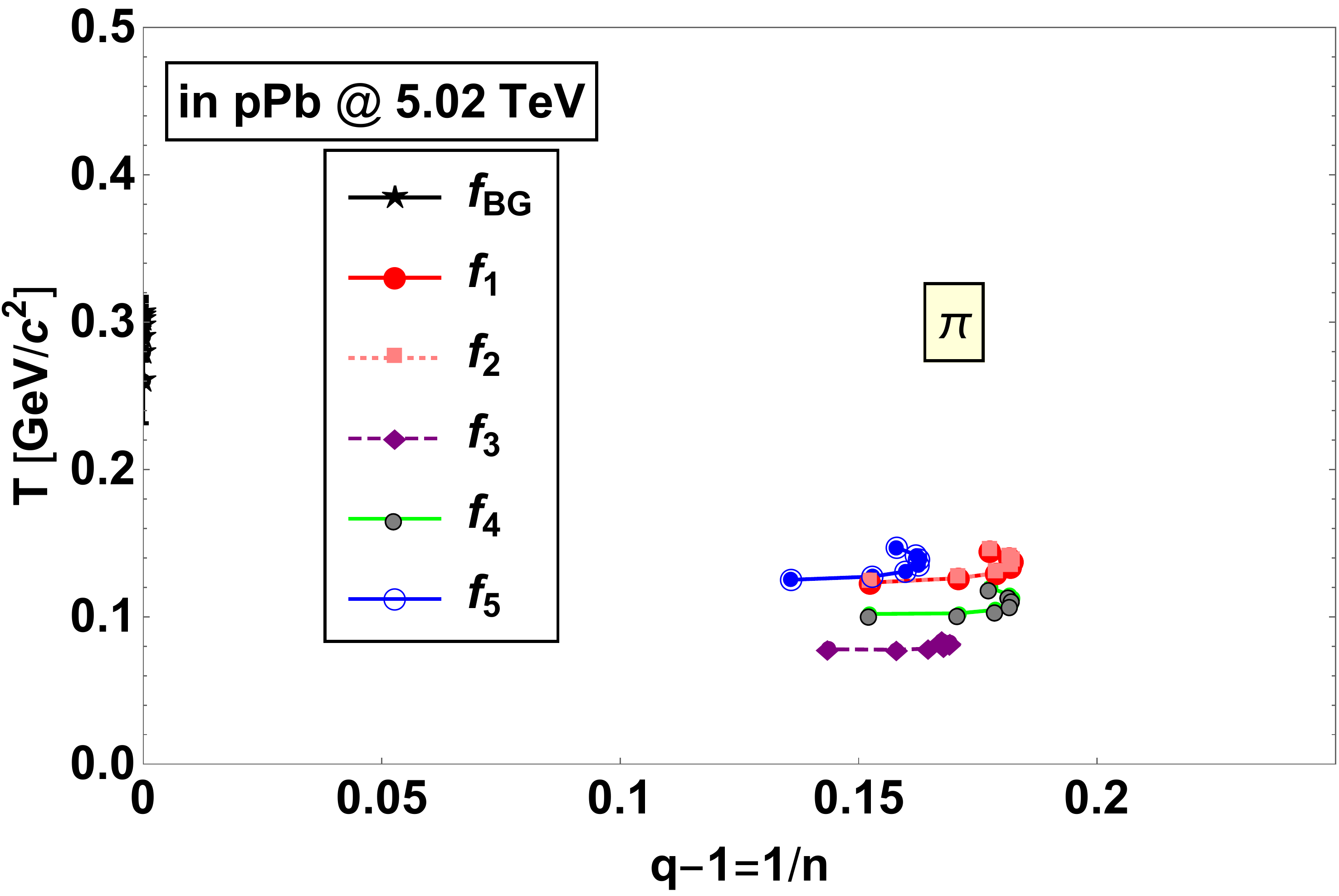}
\includegraphics[width=0.4\linewidth]{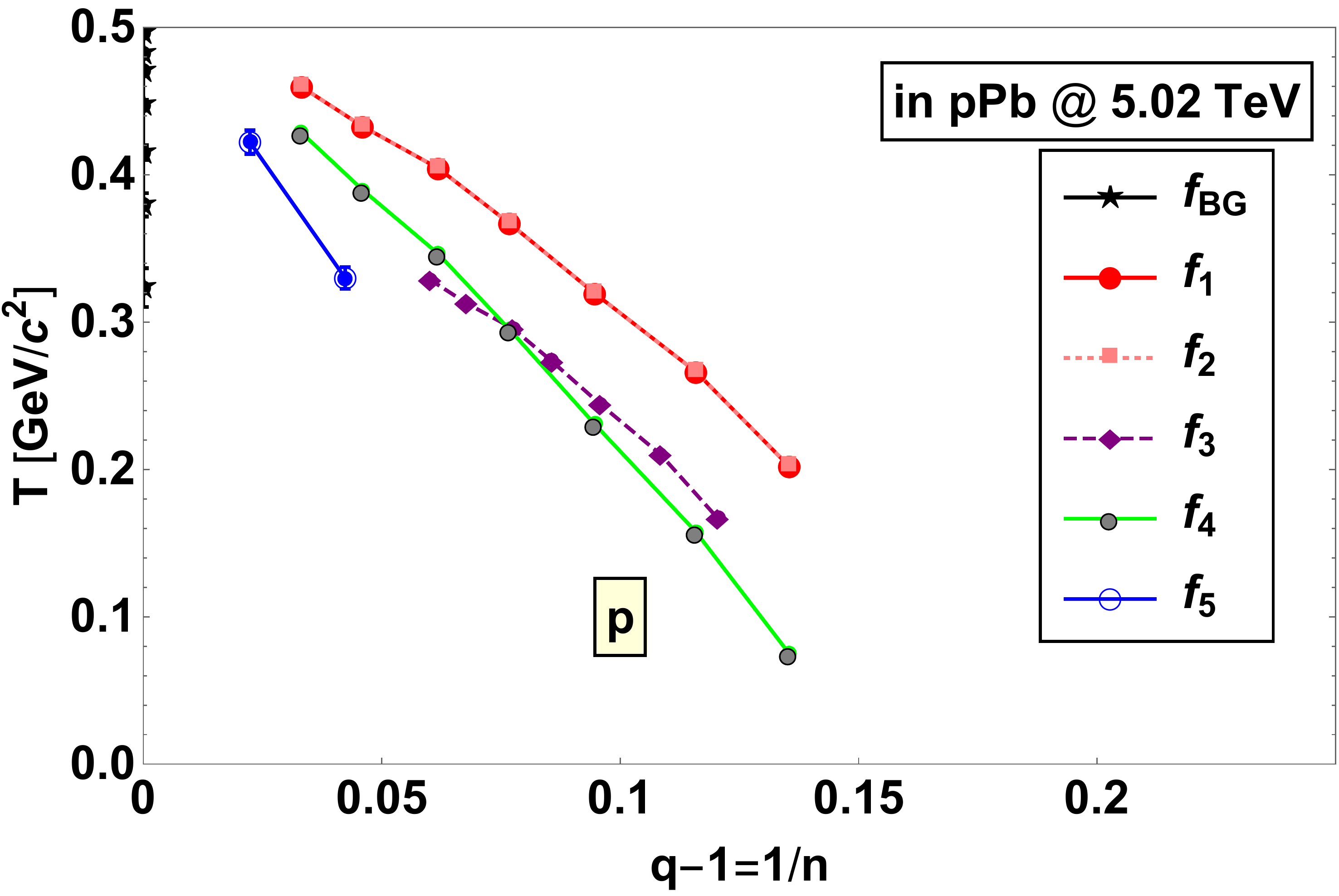}
}
\scalebox{1}[1]{
\includegraphics[width=0.4\linewidth]{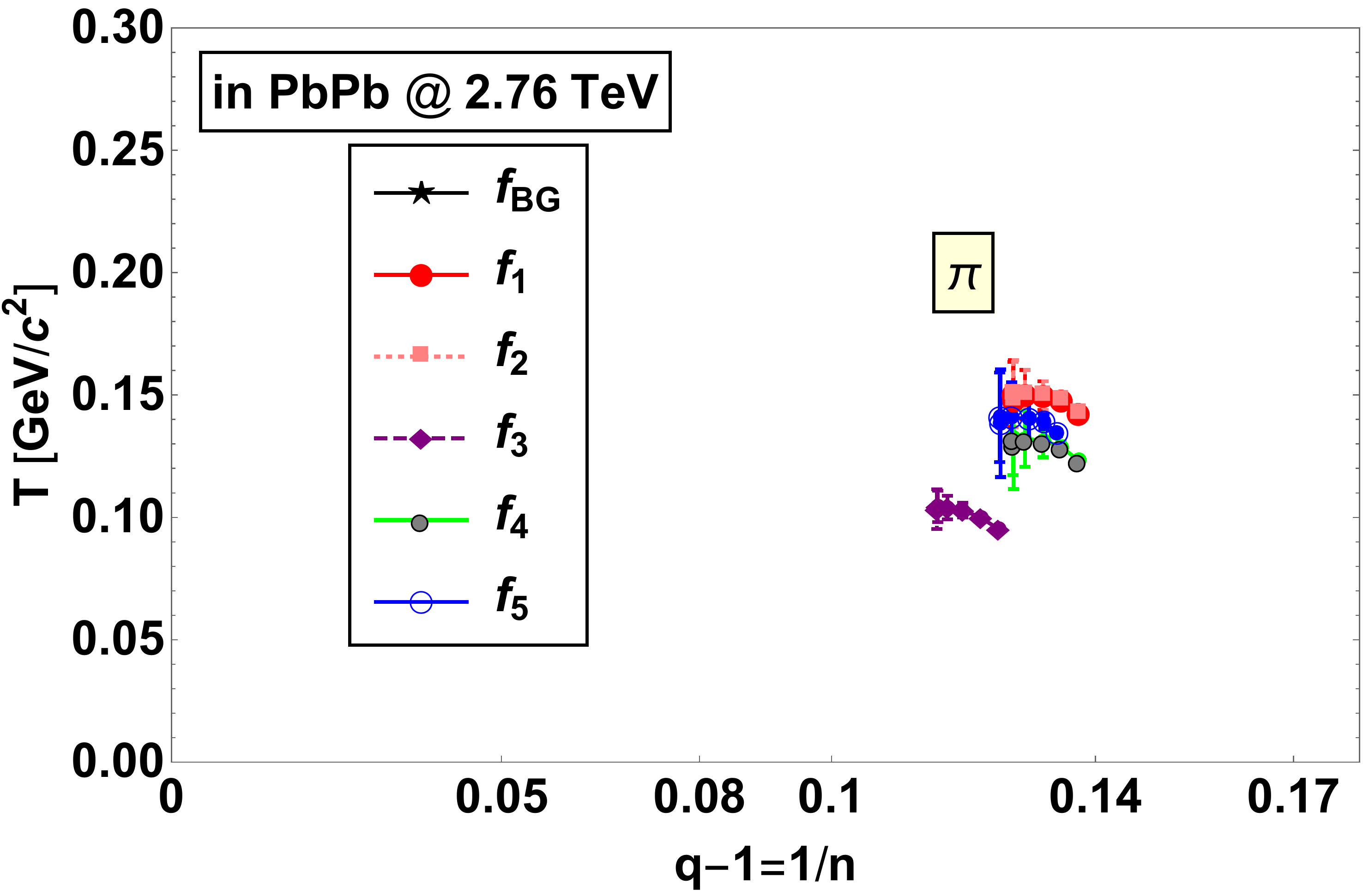}
\includegraphics[width=0.4\linewidth]{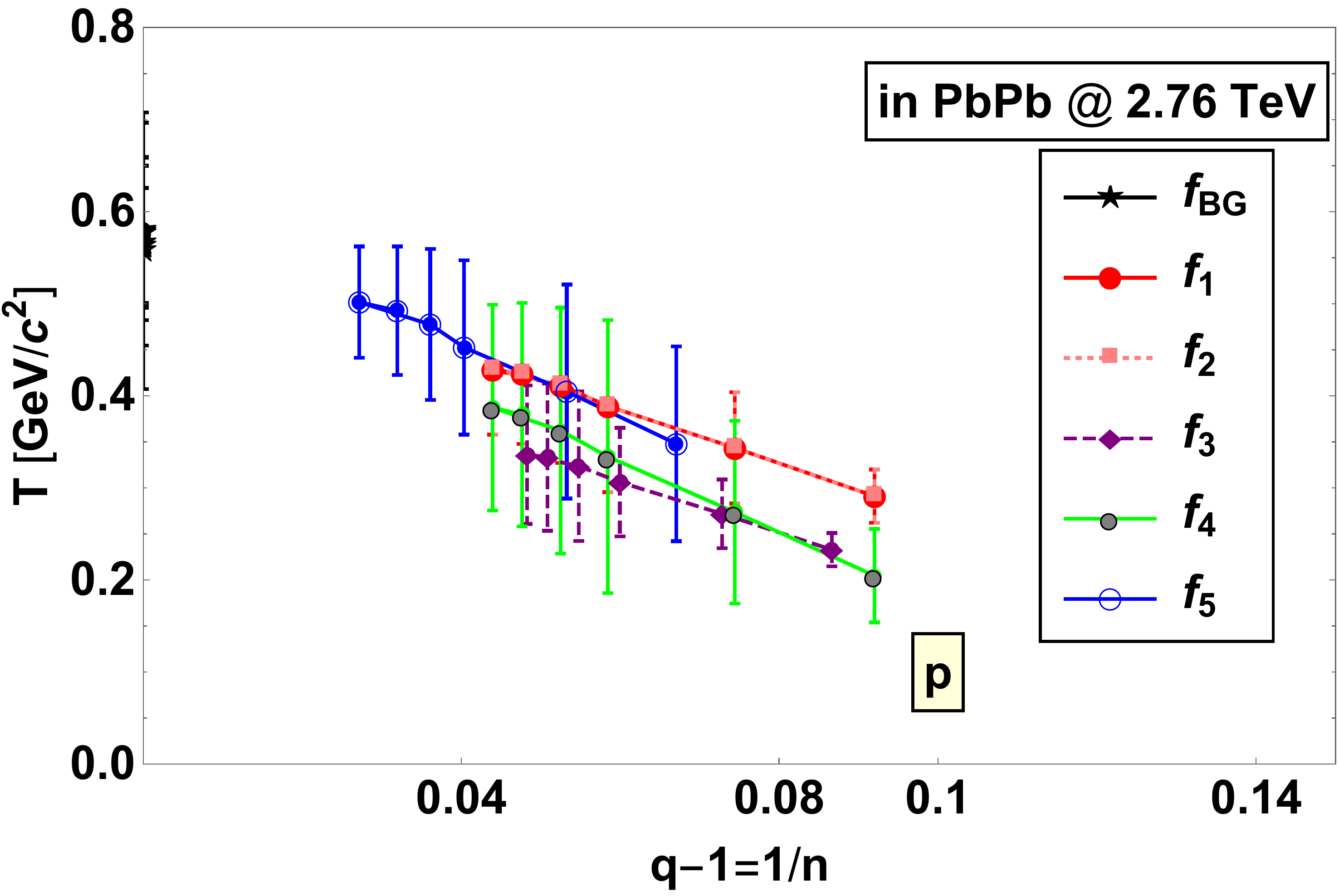}
}
\caption{Correlations between $T$ and $q-1=1/n$ for spectra of pions and protons in $pPb$ (top) and $PbPb$ (bottom) collisions.
Data sets are from different centrality bins for each particle spectra.}
\label{figpApar}
\end{figure*}

\begin{figure}[htb]
\includegraphics[width=0.9\linewidth]{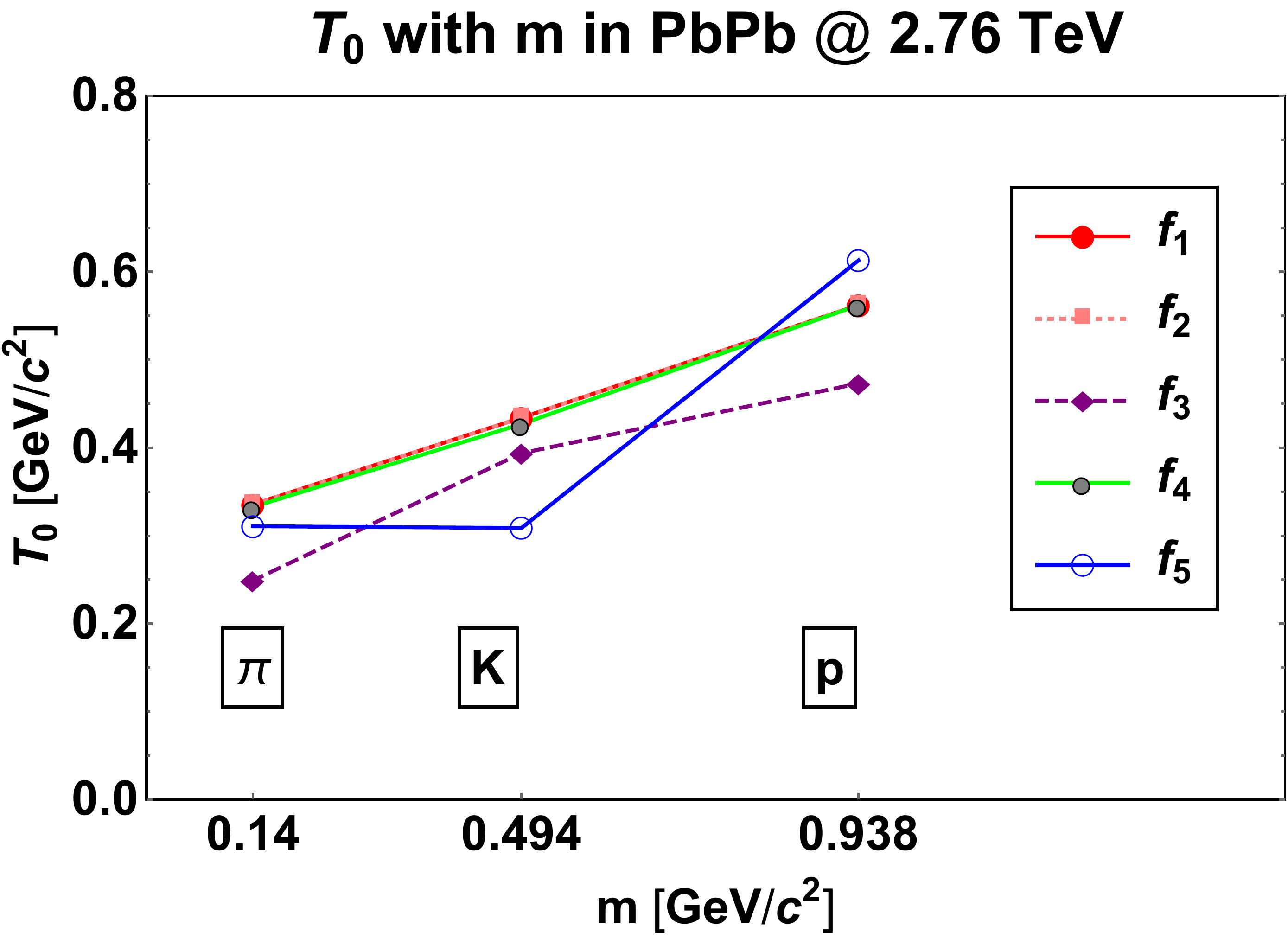}
\includegraphics[width=0.9\linewidth]{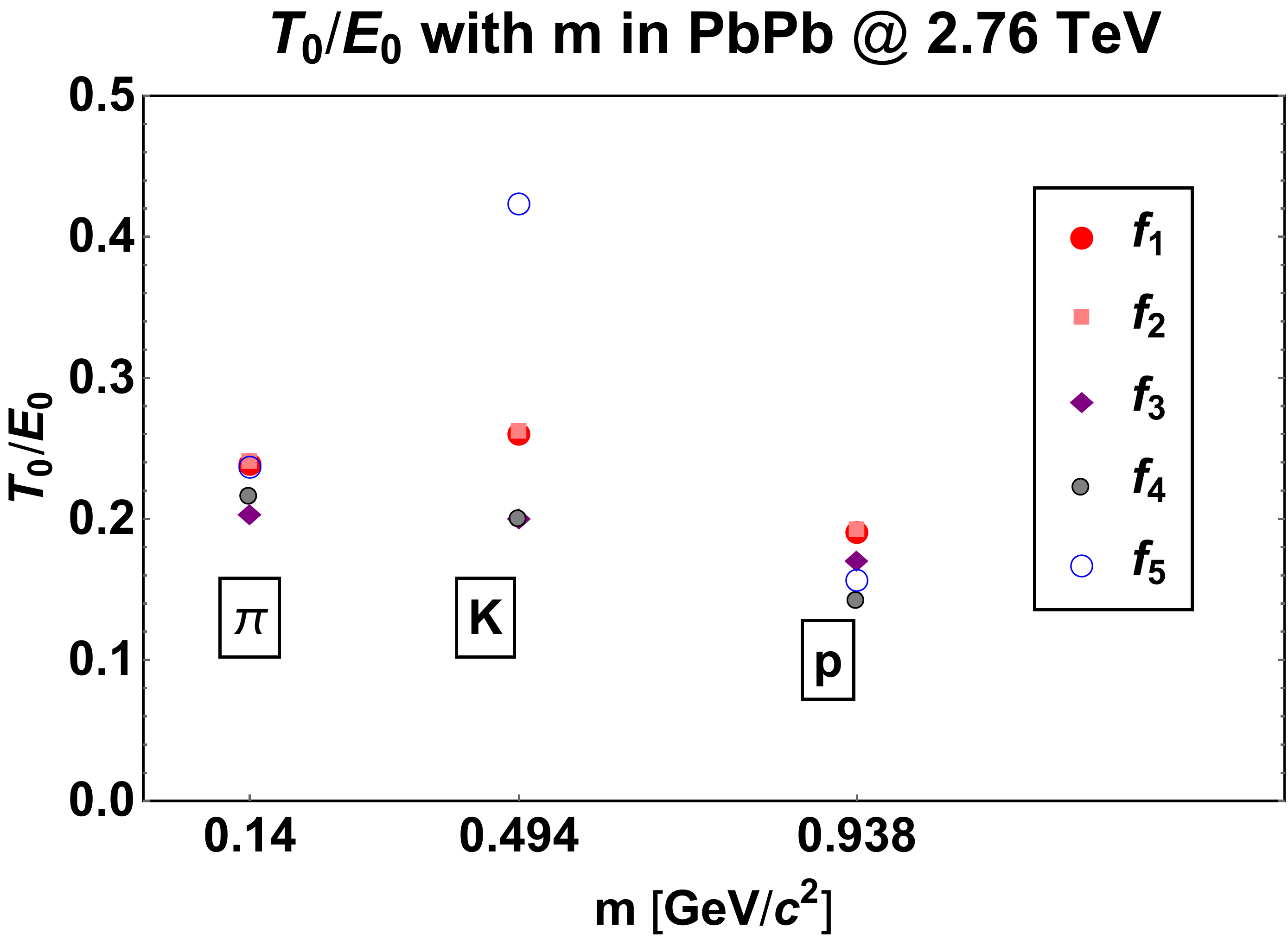}
\caption{Upper panel: Distributions of the fitting temperature $T_0$ with respect to the hadron mass $m$  for $\pi$, $K$ and $p$ in $PbPb$ at 2.76 TeV. Here we list the results for all fitting functions of Eqs. (\ref{functions}). Lower panel: The corresponding $T_0/E_0$ ratio is plotted as a function of $m$.}
\label{fig8}
\end{figure} 

In order to further check how well the Tsallis distribution performs, we have used Eqs.(\ref{functions}) to fit $p_T$ spectra in $pPb$~\cite{all-30} collisions at 5.02 TeV and in $PbPb$~\cite{AA-1,AA-2,AA-3,AA-4} collisions at 2.76 TeV. 
Data were taken by the ALICE Collaboration within wide $p_T$ ranges and for more identified charged particles. 
In Fig.\ref{figAA} and \ref{figA} in Appendix~\ref{sec:secB} we show the $p_T$ distributions of the most central collision ($0-5 \%$), the peripheral ones ($60-80 \%$ and $80-100 \%$) in $pPb$ at 5.02 TeV and in $PbPb$ collisions at 2.76 TeV, as well as the corresponding fits within all six functions defined in Eqs.(\ref{functions}). 
The corresponding $p_T$ range is listed in Table \ref{tabB1}.
One observes that in the low $p_T$ region ($0<p_T<3$ GeV/c), the fitting formula, $f_{BG}$, still performs well in the fitting, much like the non-extensive approaches.
On the other hand, for the high $p_T$ part, this classical distribution no longer describes the hadron spectra well.
The non-extensive distributions from Eqs.\ref{functions}, on the other hand, present nice results over the whole fitting range of $p_T$.

In order to better clarify the differences and similarities, values of $\chi^2/ndf$ are compared in Fig.\ref{figB}. 
More central collisions lead to larger values of $\chi^2/ndf$ of all non-extensive fitting formulas used in this work. 
This is probably due to the fact that there are flow effects in heavy-ion collisions which were ignored in our fittings; while the most peripheral collisions are similar to the elementary ones.
All fits of the spectra by $f_5(p_T)$, perform better on pions than protons for the same reasons as in the $pp$ case. 
The formulas ($f_1$ and $f_4$) share the same values of $\chi^2/ndf$ with the ``normalized" one, $f_2$, as in the analysis of $pp$ collisions. 
Note that in heavy-ion collisions, the fitting function derived thermodynamically, $f_3$, results in the best fitting results for all centralities.

\begin{table}[h]
\caption{Fitting $p_T$ range [GeV/c] of different hadron spectra in heavy-ion collisions in this work~\cite{all-30,AA-1,AA-2,AA-3,AA-4}:}
\scalebox{1.2}[1.2]{
\begin{tabular}{c|c|c|c}
\hline
\hline
particles &  mass [GeV/$c^2$] & $pPb$ & $PbPb$ \\
\hline
$\pi$ & 0.140 & 0.11-2.85 & 0.11-19 \\
\hline
$K$ & 0.494 &  0.225-2.45 & 0.225-19 \\
\hline 
$K_S^0$ & 0.498 & 0.05-7 &  0.45-11 \\
\hline
$K^*$ &  0.896 &    &  0.55-4.5 \\
\hline
$p$ & 0.938 &  0.325-3.9  & 0.325-17.5 \\
\hline
$\phi$ &  1.019 &    &  0.65-4.5 \\
\hline
$\Lambda$ & 1.116 &  0.65-7 & 0.65-11 \\
\hline
$\Xi$ &  1.321 &    &  0.7-7.5 \\
\hline
$\Omega$ &  1.672 &    &  1.3-7.5 \\
\hline
\hline
\end{tabular}
}
\label{tabB1}
\end{table}

It is instructive to plot the relationship between the fitting temperature $T$ and the Tsallis non-extensive parameter $q$ (or $q-1=1/n$). 
Fig.\ref{figpApar} tells us that for the same produced particle yield in all centrality collisions, a strong correlation appears. 
In fact whatever kind of particle we take, all these fitting formulas result in a similar relationship between $T$ and $q-1=1/n$, approximating $T\approx T_0-(q-1)E_0$. 
Here $T_0$ stands for the limiting value of temperature in the BG limit when $q=1$. 
Hereby we list our results of $T_0$ values for $\pi$ and $p$ in $PbPb$ at 2.76 TeV. 
We find that $T_0$ also increases linearly with the particle mass, shown in Fig.\ref{fig8}, whereas the $T_0/E_0$ ratio displays an $m$-independence except for the worst fitting by $f_5$.

\section{Discussions}\label{sec:sec5}

We have considered various approximations to the energy in the Lorentz-invariant particle spectrum in order to account for their sensitivity to different fitting variables, in use within non-extensive approaches. 
The hadronization, as well as the $p_T$ distributions in high energy collisions, are being studied by the use of one or the other of these functions. 

When describing the $p_T$ spectra of the particles produced in high-energy collisions, there are many kind of assumptions based on the non-extensive expression, $\ee_q(x)$.
Since there are still fierce arguments on their theoretical backgrounds, we suggest to compare with different experimental data sets to shed light on the best fitting scenario with the strongest physical content.
Considering the values of $\chi^2/ndf$, we review the goodness of fits and connections of fitting functions.
Analysis on the fitting parameters, on the other hand, could also provide a tool for understanding these statistical and thermodynamical models. 

In order to illuminate the connections and discrepancies of different fitting functions applied,  we have investigated the most common ones, listed in Eqs. (\ref{functions}), for several kinds of identified charged particles, stemming both from $pp$ and $NN$ collisions. 

Our results reveal that the normalizing parameter has no major effect on the shape of these fittings,  unlike the mass of the particles, which is a decisive factor.
In other words, the fitting formulas of either $m_T-m$ or $m_T$ lead to the same goodness of fit as well as their normalized cases.
Comparing all these fitting functions, for all kinds of identified charged particles, we conclude that the simple $m_T-m$ or $m_T$ function presents the best fitting results in the elementary collisions while the one obtained by the thermodynamic calculation explores the best $\chi^2/ndf$ in high energy collisions.
This trend breaks when $\pi$ spectra are investigated, for which there is no large impact on the fitting factors by the various formulas.
It can be due to the small mass of pions and high multiplicities in collisions.
For heavier particles, the $m_T$ dependence becomes increasingly more explicit. 

\section{Summary}\label{sec:sec6}

Summarizing, we have investigated five types of non-extensive functions in Eqs.(\ref{functions}), $f_1 - f_5$, based on the $q$-exponential as well as the Boltzmann\,--\,Gibbs distribution, $f_{BG}$.
As discussed above, results show that for the same definitions $m_T-m$ functions behave as the $m_T$ ones, but with different normalization constants.
This leads us to a free choice of the fitting functions of either $m_T-m$ or $m_T$ when focusing on $p_T$ spectra.
Nevertheless, as shown in Section II and others' work, $E-\mu\approx m_T-m$ scaling applies better  for low $p_T$ experimental data.
On the other hand, since the volume $V$ is not experimentally available, the normalized function, $f_2$, is preferred by fitting the hadron spectra and analyzing the particle yield per unit rapidity, $\dd N/\dd y$.
There is no big difference in the spectral shape for they share the same fitting goodness whether the fitting formula is normalized or not. 
Note that the simple $p_T$ function is more proper to study the $\pi$ spectra in $pp$ or $ep$ collisions at low energy because of its small mass and the low multiplicities in small systems.
The usual Boltzmann\,--\,Gibbs distribution fails in describing the hadronization in all cases but in a quite narrow $p_T$ range.

Finally, we paid attention to the relationship between the fitting parameters, $T$ and $q$. All approaches seem to  reveal a linear dependence, which agrees well with our results in previous work\cite{doppler-p2} and that of others'\cite{Wilk-2015}. 
The mean transverse mass $\langle m_T\rangle$ was also analyzed and led to a linear dependence on the mass $m$, while almost no dependence on the temperature parameter $T$ in the investigated data range for pions, kaons and protons.

\vspace{0.3cm}

\noindent {\bf Acknowledgments}

\vspace{3mm}

This work has been supported by the Hungarian National Research, Development and Innovation Office (NKFIH) under the contract numbers K120660 and K123815 and THOR COST CA 15213.
The author Shen Keming also wants to thank the funding for the Doctoral Research of ECUT.

\appendix
\section{Coordinate system and spectra}\label{sec:secA}

In high energy physics, 
a commonly 
used, Lorentz invariant 
variable 
to describe the kinematic condition of a particle is the rapidity variable $y$. It is defined in terms of its energy and 
longitudinal 
momentum components, $E$ and $p_z$ respectively ($z$ is the beam axis). The formula
\begin{eqnarray}
y\equiv \frac{1}{2}\ln \frac{E+p_z}{E-p_z}=\frac{1}{2}\ln \frac{1+v_z}{1-v_z}~,
\end{eqnarray}
implies $v_z=\tanh y$ and $\gamma_z=\frac{1}{\sqrt{1-v_z^2}}=\frac{E}{\sqrt{E^2-p_z^2}}=\frac{E}{m_T}=\cosh y$. Thus,
$E=\gamma_zm_T=m_T\cosh y$ and $p_z=\gamma_zm_Tv_z=m_T\sinh y$.

In order to characterize the rapidity of a particle, it is then necessary to measure two quantities, such as energy and longitudinal momentum \cite{rap-p1}. 
In many experiments it is only possible to measure the angle of the detected particle 
relative to the beam axis. In this case it turns out to be more convenient to utilize 
the pseudo-rapidity $\eta$, 
given by 
\begin{eqnarray}
\eta=\frac{1}{2}\ln\frac{p+p_z}{p-p_z}
\approx
\ln\sqrt{\frac{1+\cos \theta_z}{1-\cos\theta_z}}=\ln [\cot\frac{\theta_z}{2}]~.
\end{eqnarray}
One easily 
retains 
their connection as being
\begin{eqnarray}
\frac{\dd y}{\dd\eta}=\frac{1}{2}\frac{\dd}{\dd\eta}\ln\frac{E+p_z}{E-p_z}
=\frac{p}{E}
\end{eqnarray}
leading to particle yields in rapidity and pseudorapidity units,
\begin{eqnarray}
\frac{\dd N}{\dd y}=\frac{E}{p}\frac{\dd N}{\dd \eta} .
\end{eqnarray}

In high energy physics, one investigates the Lorentz-invariant particle spectrum $E\frac{\dd N}{\dd^3p}$. Fix $p_T$ (or $m_T$), $\dd p_z=m_T\cosh y\dd y=E\dd y$, therefore, $\dd p_z/E=\dd y$,
result in
\begin{eqnarray}
E\frac{\dd N}{\dd ^3p}=\frac{\dd N}{\dd y\dd ^2p_T}=\frac{\dd N}{\dd yp_T\dd p_T\dd \phi_p}=\frac{\dd N}{\dd ym_T\dd m_T\dd \phi_p} ,
\label{mTpT}
\end{eqnarray}
and
\begin{eqnarray}
\frac{\dd N}{\dd \eta \dd^2p_T}\:=\: \sqrt{1-\frac{m^2}{m^2_T\cosh^2y}}\,\,\frac{\dd N}{\dd y\dd ^2p_T},
\end{eqnarray}
At mid-rapidity we obtain 
\begin{eqnarray}
\frac{\dd N}{\dd \eta \dd^2p_T}\bigg|_{\eta =0}=\frac{p}{E}\frac{\dd N}{\dd y\dd ^2p_T}\bigg|_{y=0}=\frac{p_T}{m_T}\frac{\dd N}{\dd y\dd ^2p_T}\bigg|_{y=0}~.
\end{eqnarray} 

\section{Figures and Results}\label{sec:secB}

For clarity and simplicity we put our figures and results of all hadron spectra fittings at the end of this paper.

~

\begin{figure*}[htb]
\scalebox{1}[1]{
\includegraphics[width=0.45\linewidth]{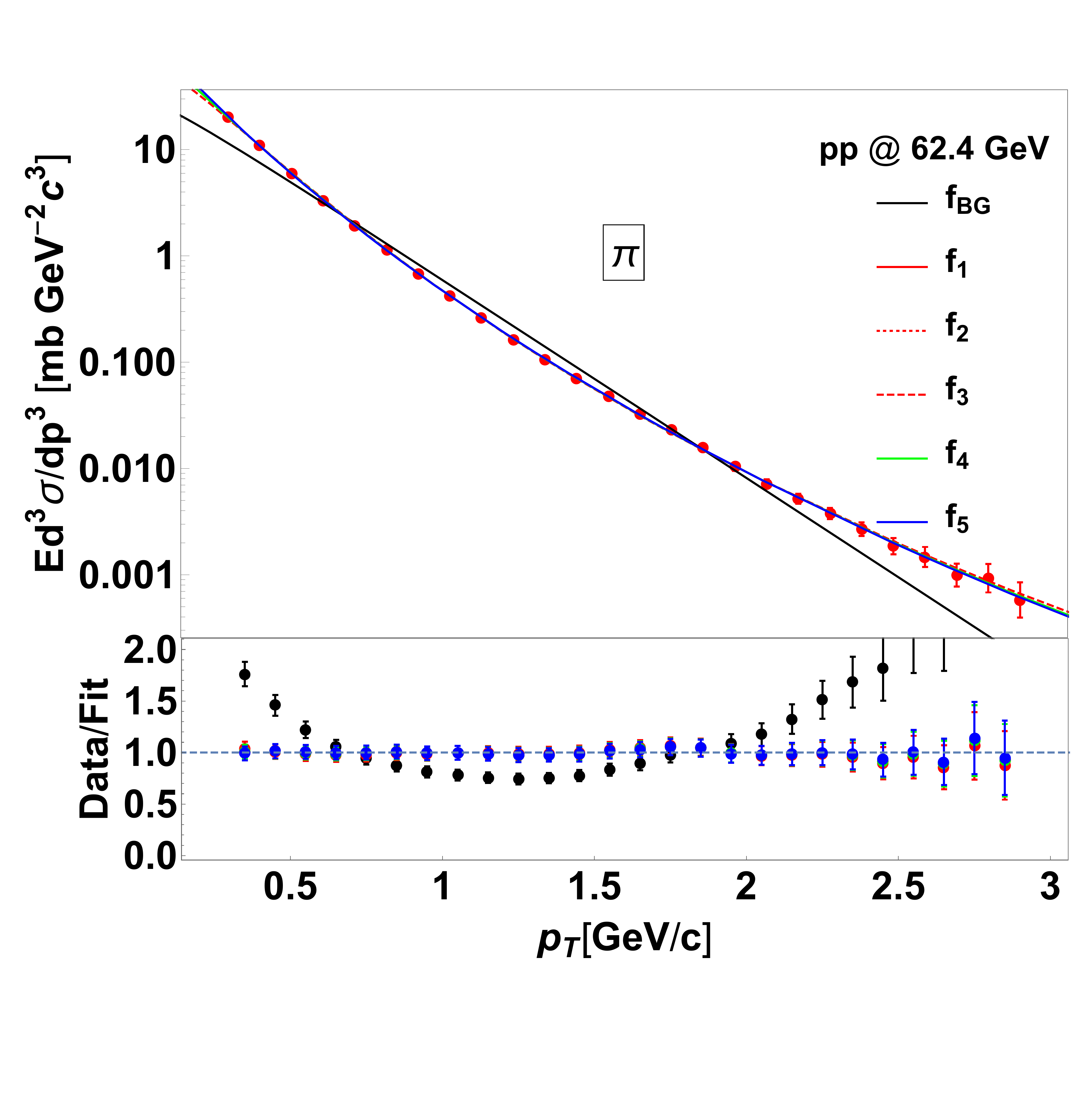}
\includegraphics[width=0.45\linewidth]{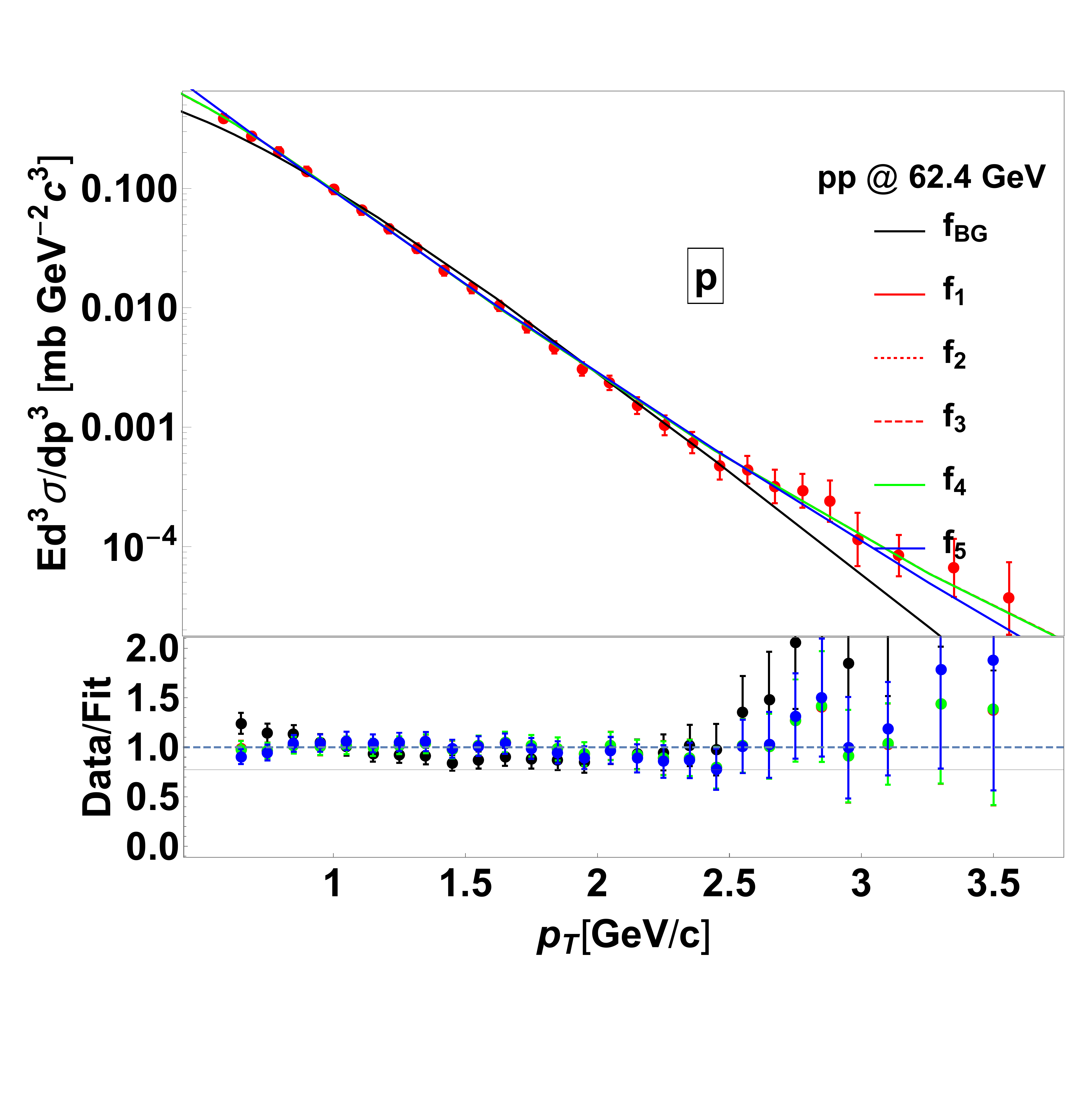}
}
\scalebox{1}[1]{
\includegraphics[width=0.45\linewidth]{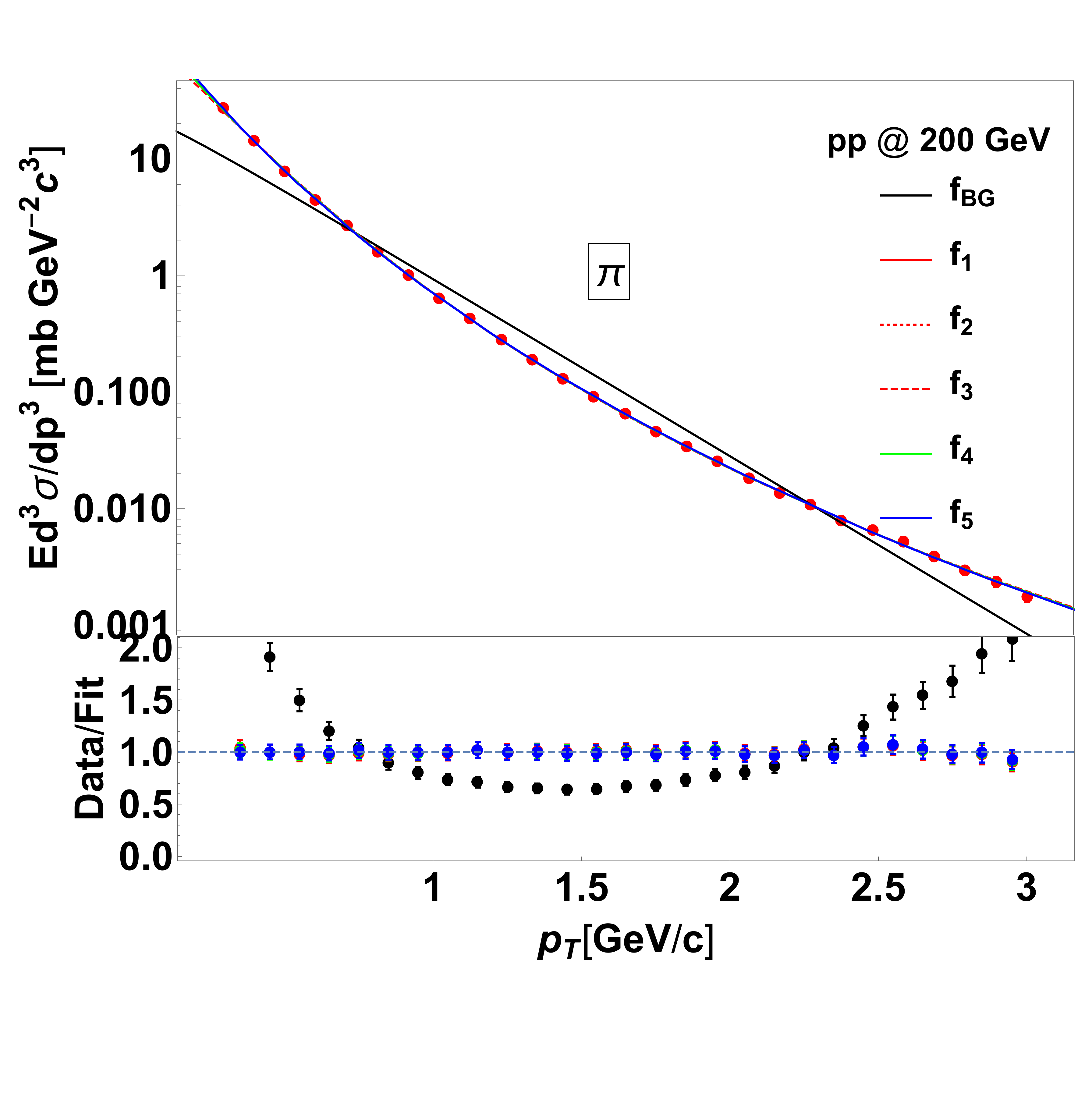}
\includegraphics[width=0.45\linewidth]{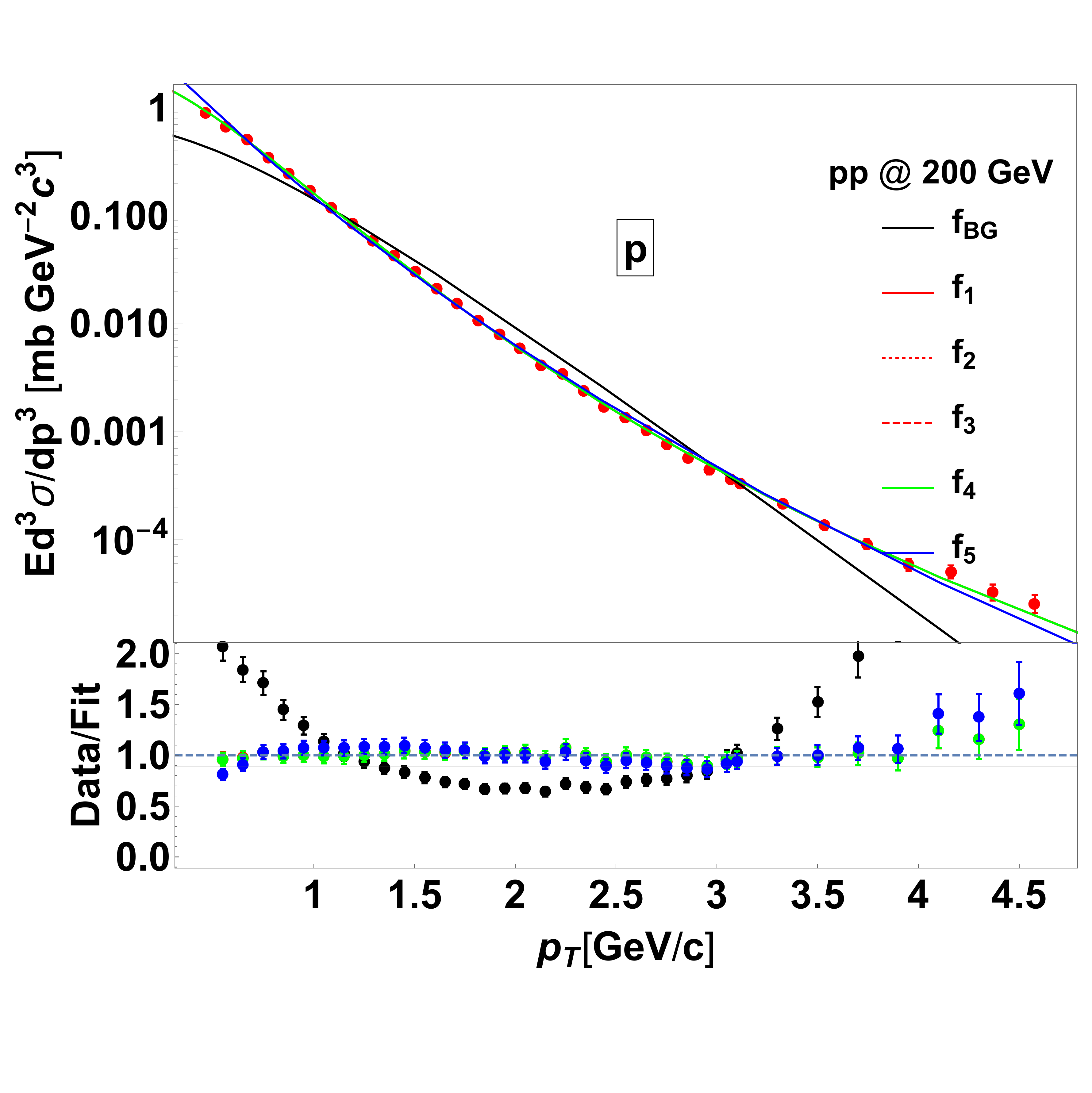}
}
\scalebox{1}[1]{
\includegraphics[width=0.45\linewidth]{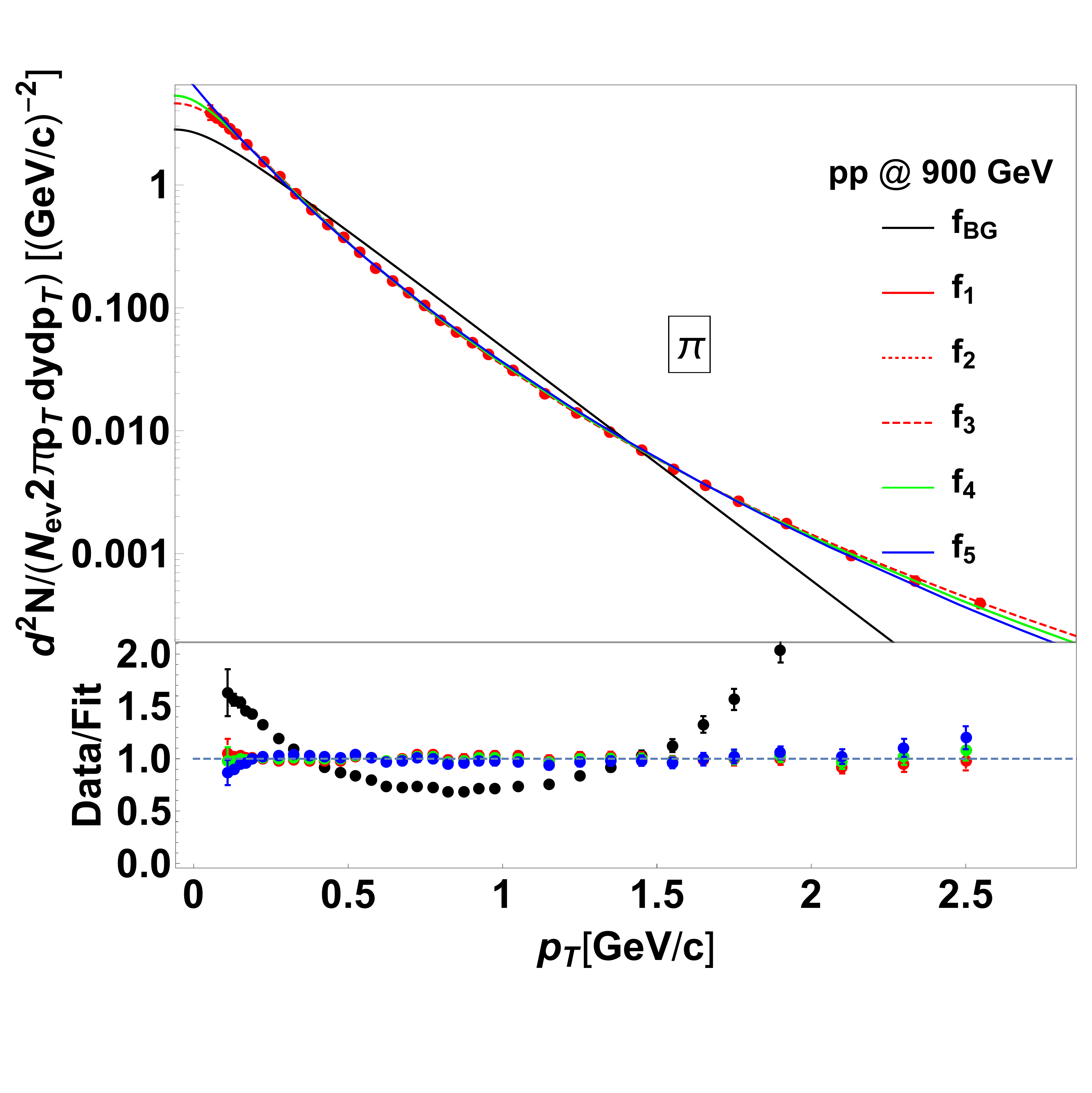}
\includegraphics[width=0.45\linewidth]{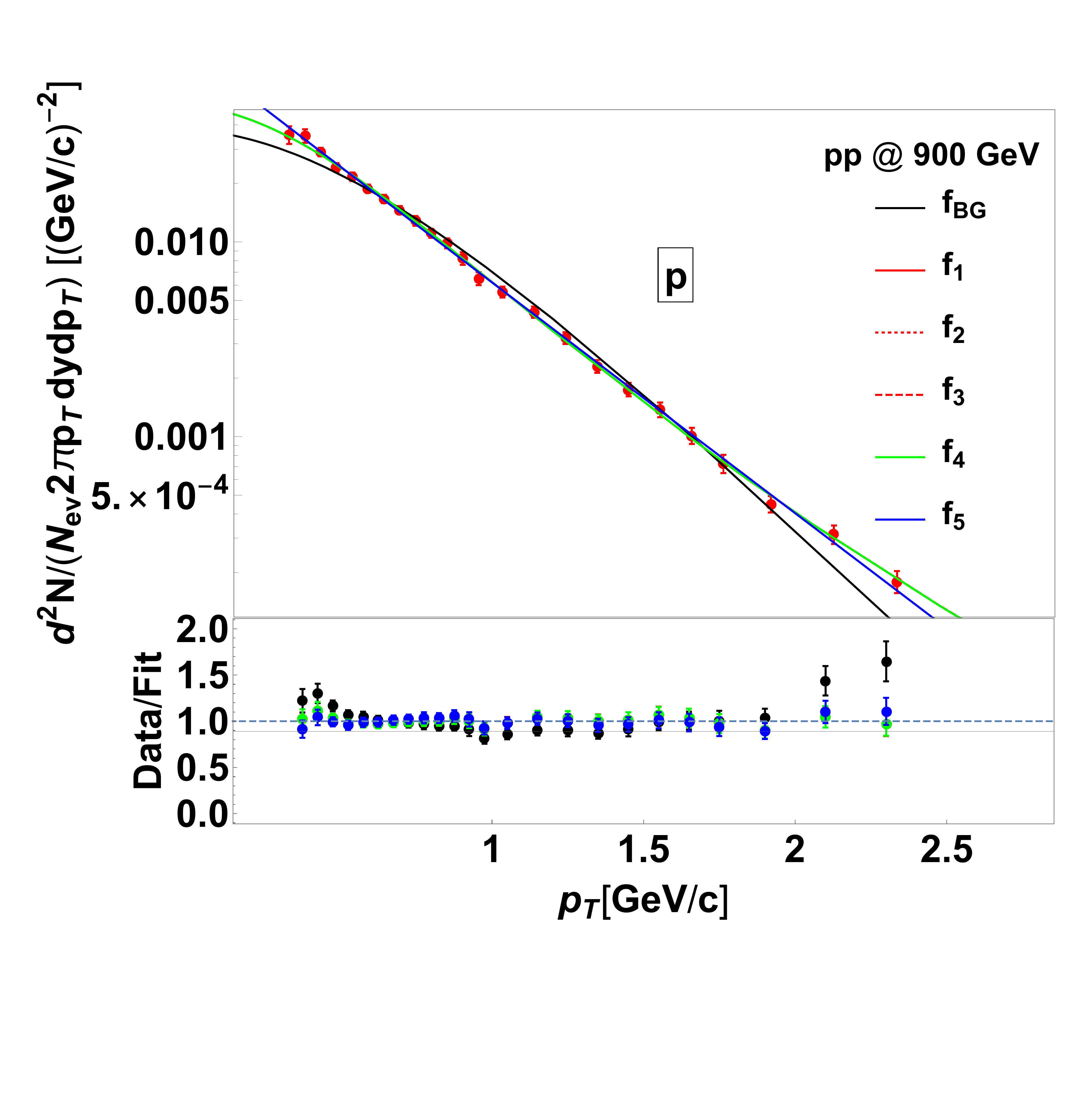}
}
\caption{The $p_T$ spectra for pions and protons in $pp$ collisions at $\sqrt{s}=$ 62.4, 200 and 900 GeV at midrapidity. Data are taken from \textcolor{blue}{Ref.~\cite{all-16}}. Each spectrum is fitted with all the six functions of Eqs.(\ref{functions}) in the $p_T$ range as shown in Table \ref{tab2}. Ratios of the net fits to data are also shown in each lower panel.}
\label{figspe1}
\end{figure*}

\begin{figure*}[htb]
\scalebox{1}[1]{
\includegraphics[width=0.45\linewidth]{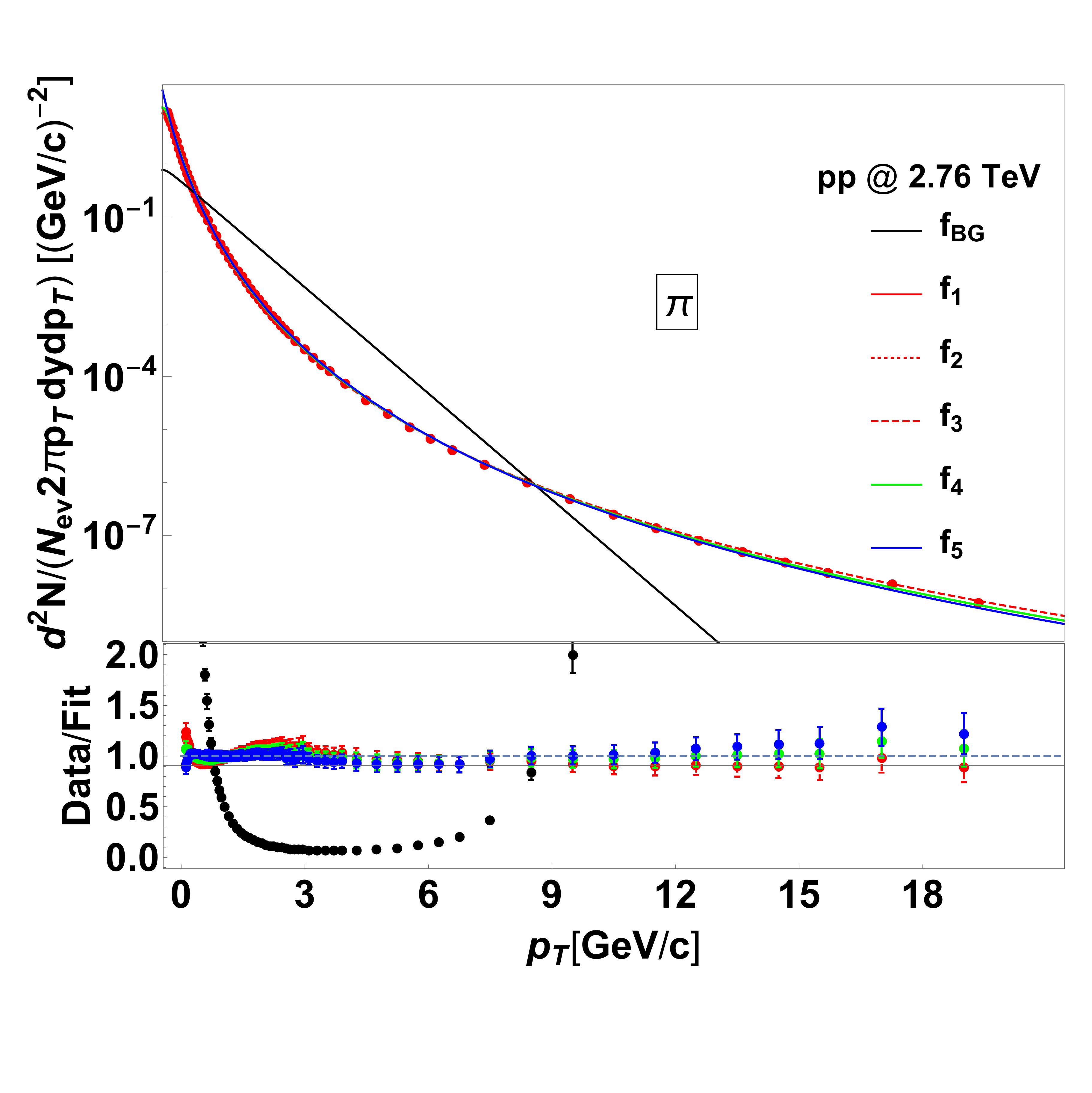}
\includegraphics[width=0.45\linewidth]{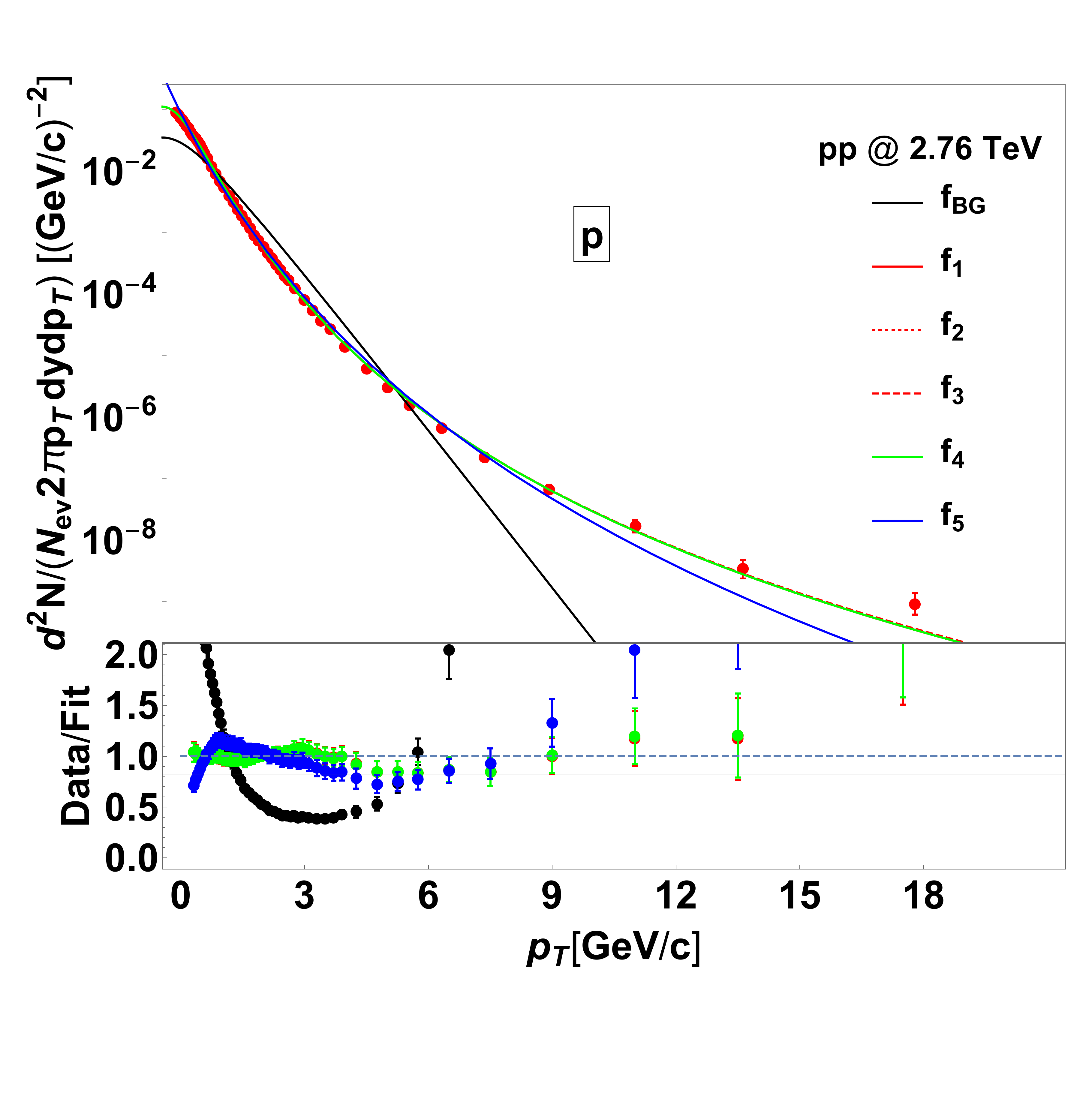}
}
\scalebox{1}[1]{
\includegraphics[width=0.45\linewidth]{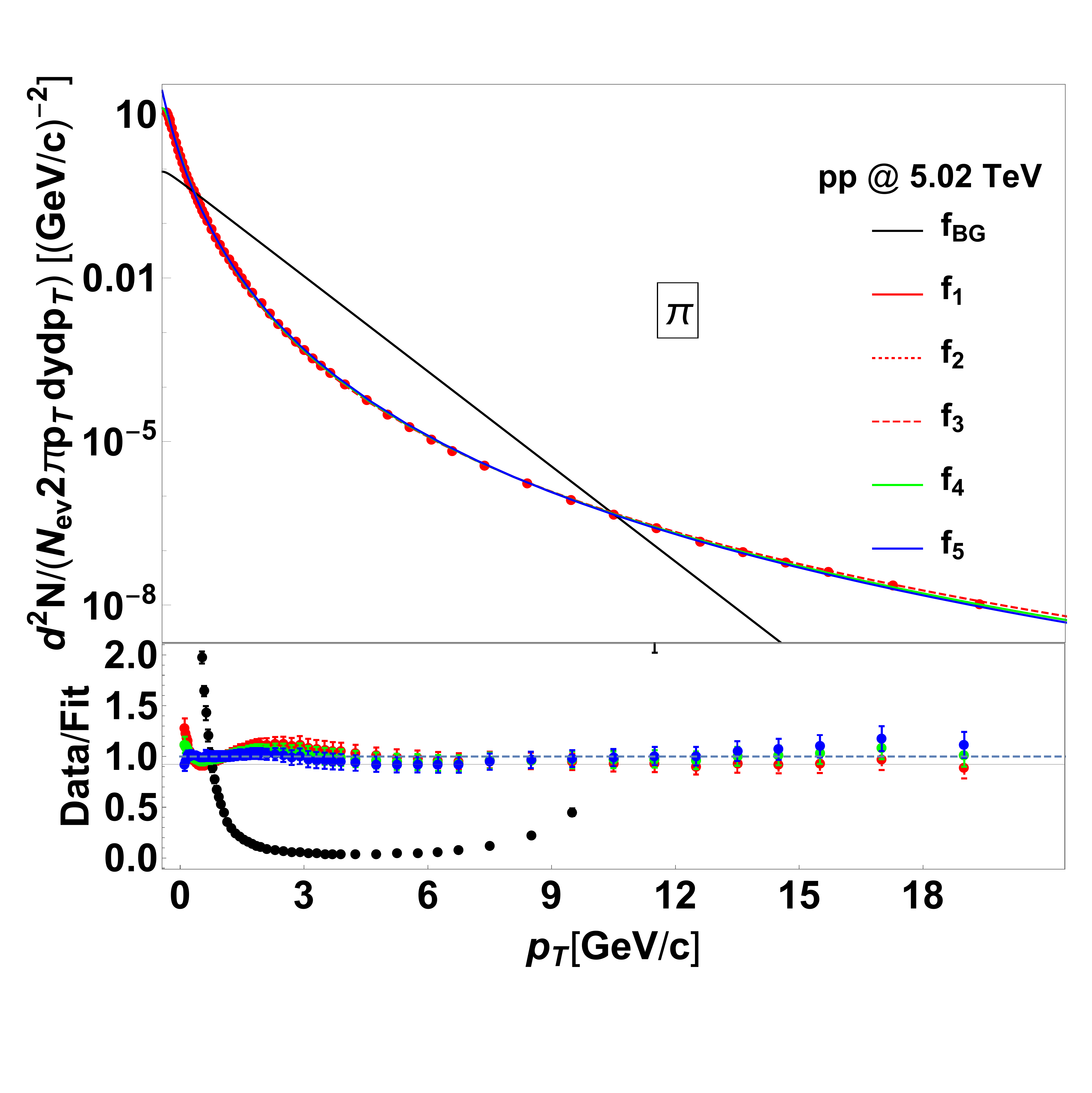}
\includegraphics[width=0.45\linewidth]{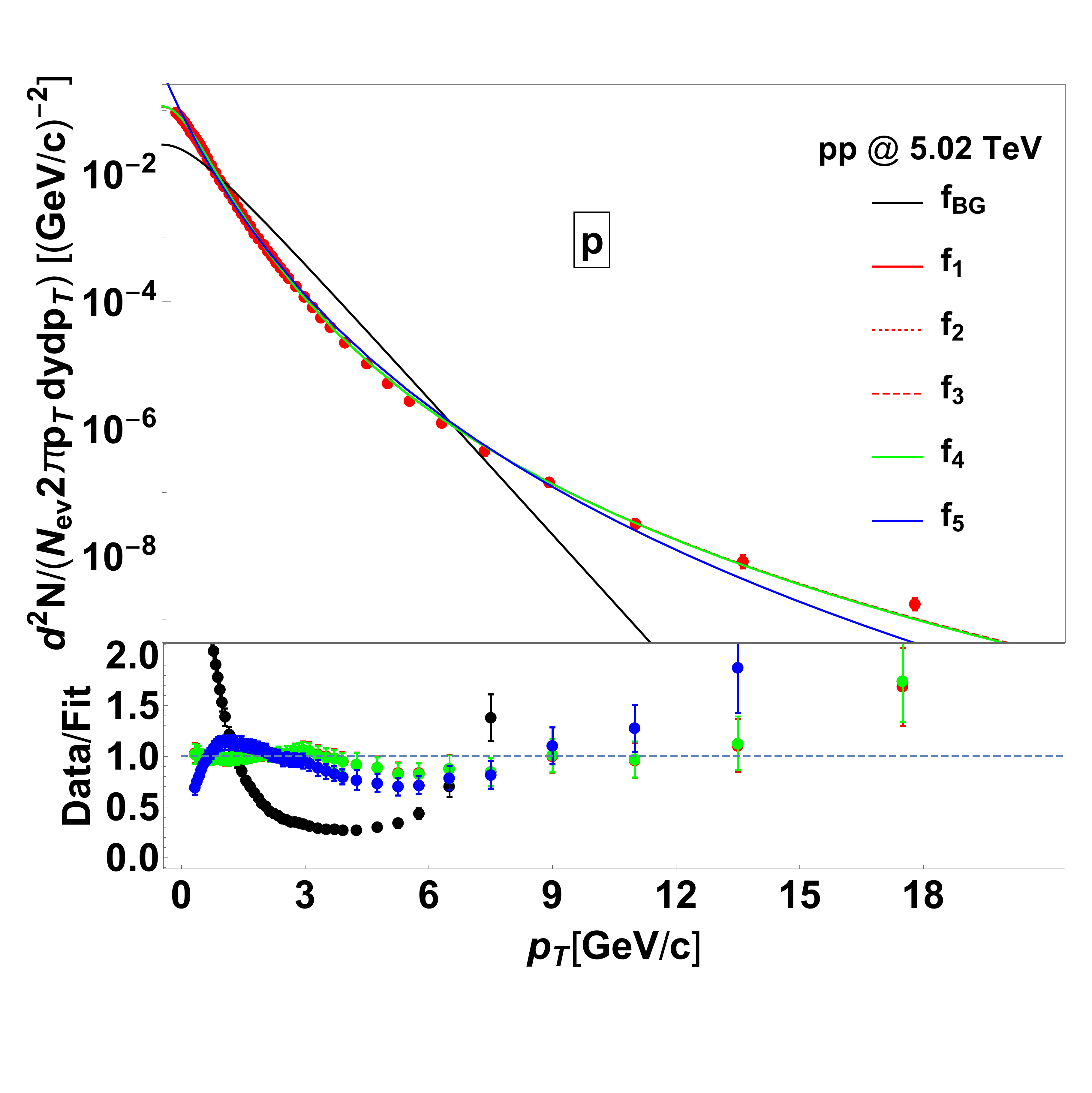}
}
\scalebox{1}[1]{
\includegraphics[width=0.45\linewidth]{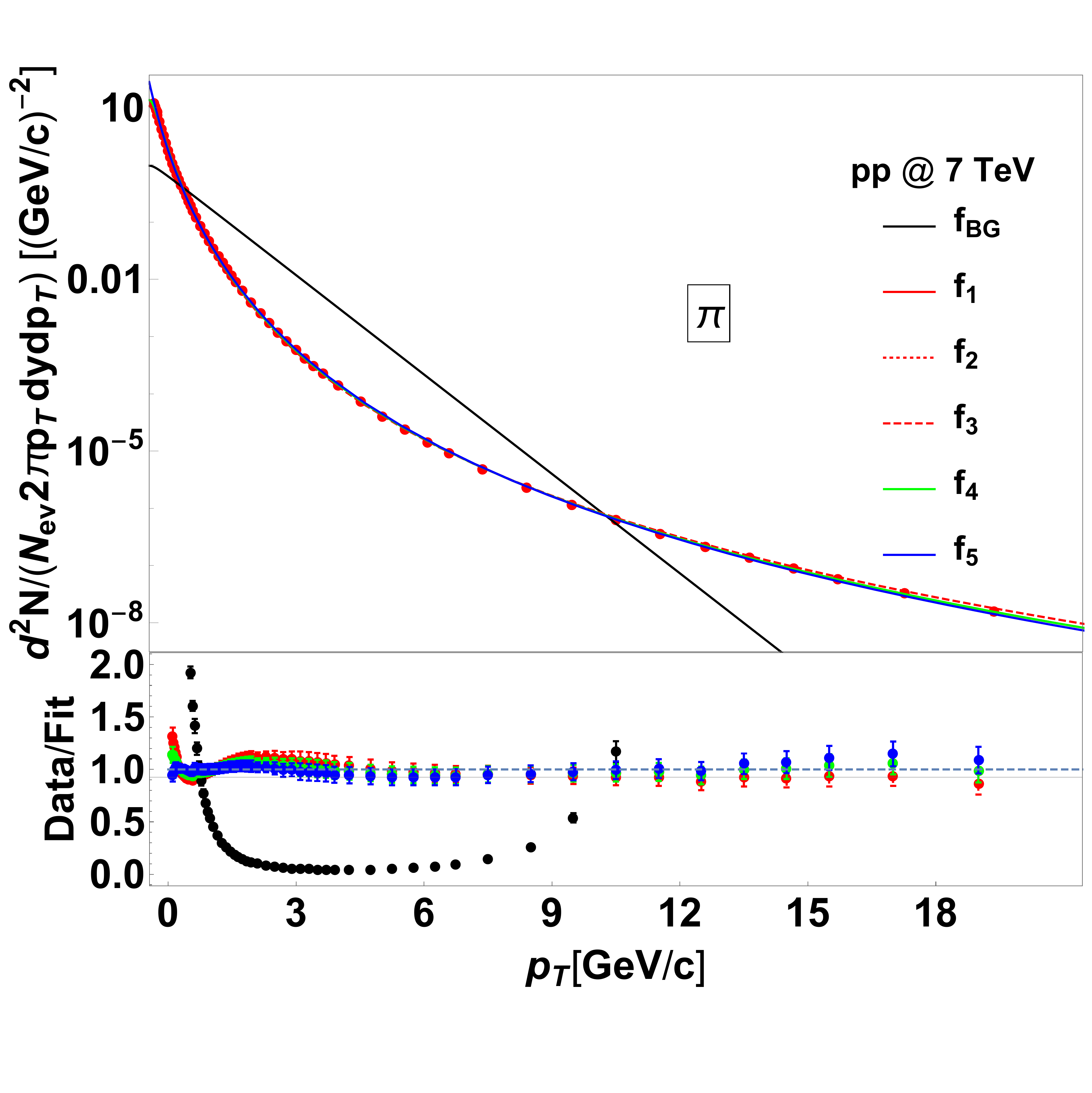}
\includegraphics[width=0.45\linewidth]{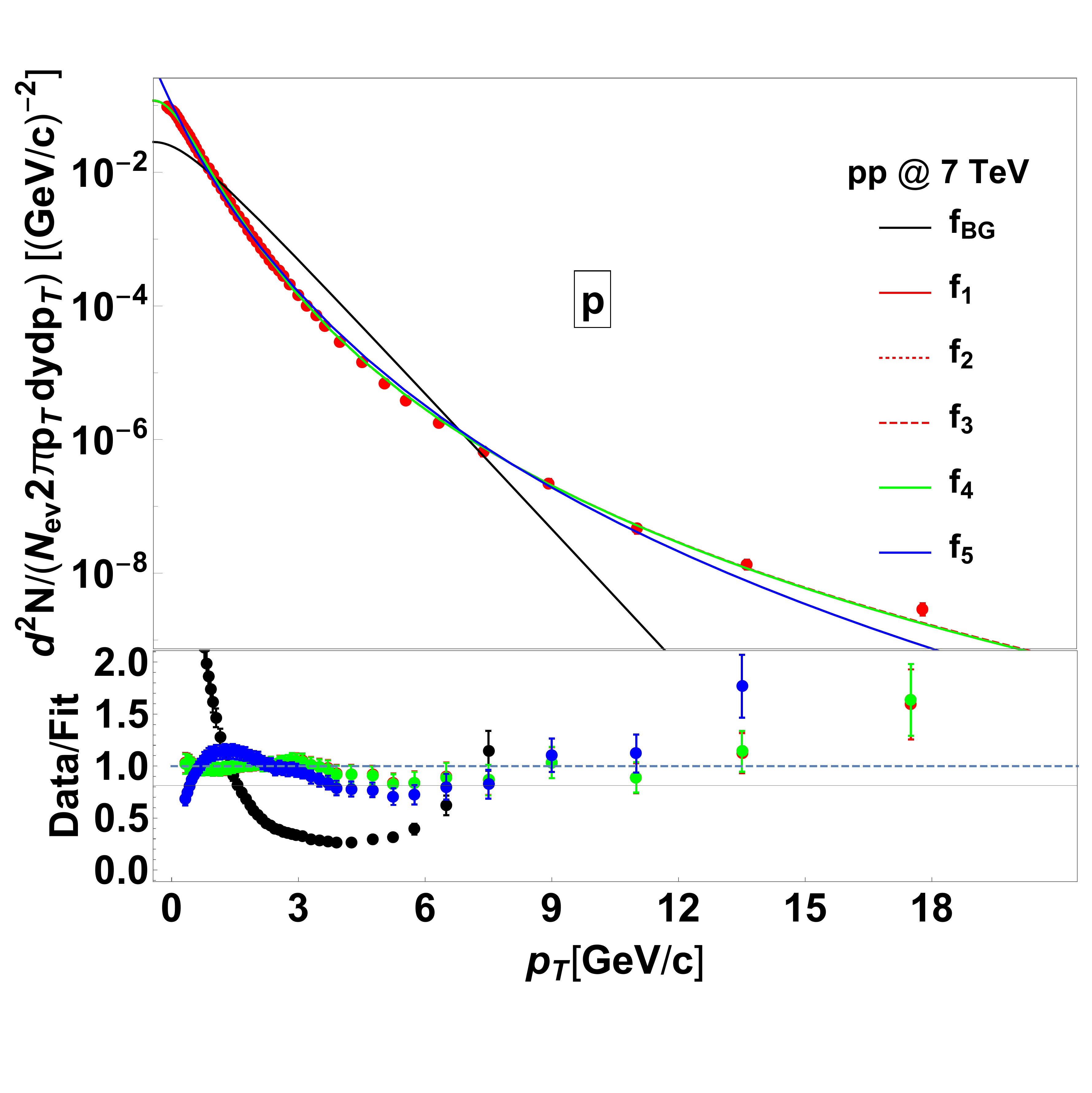}
}
\caption{The $p_T$ spectra for pions and protons in $pp$ collisions at $\sqrt{s}=$ 2.76, 5.02 and 7 TeV at midrapidity. Data are taken from Refs.~\cite{all-10-1,all-32}. Each spectrum is fitted with all the six functions of Eqs.(\ref{functions}) in the $p_T$ range as shown in Table \ref{tab2}. Ratios of the net fits to data are also shown in each lower panel. 
\textcolor{blue}{Note that as mentioned in Ref.~\cite{all-10-1}, the data for $pp$ collisions at 5.02 TeV are generated by interpolating the data measured at 2.76 TeV and 7 TeV.}}
\label{figspe2}
\end{figure*}

\begin{figure*}[htb]
\scalebox{1}[1]{
\includegraphics[width=0.45\linewidth]{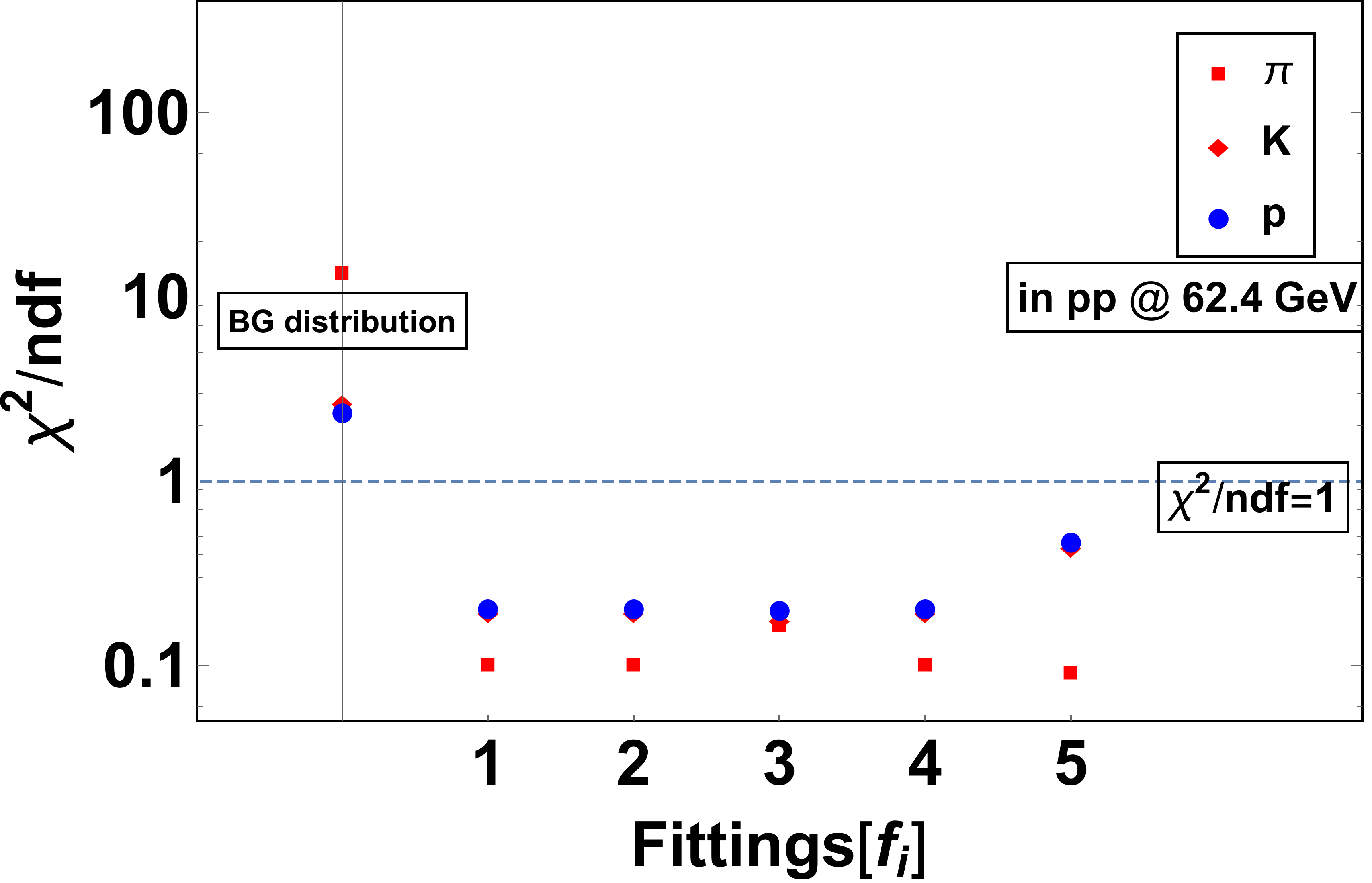}
\includegraphics[width=0.45\linewidth]{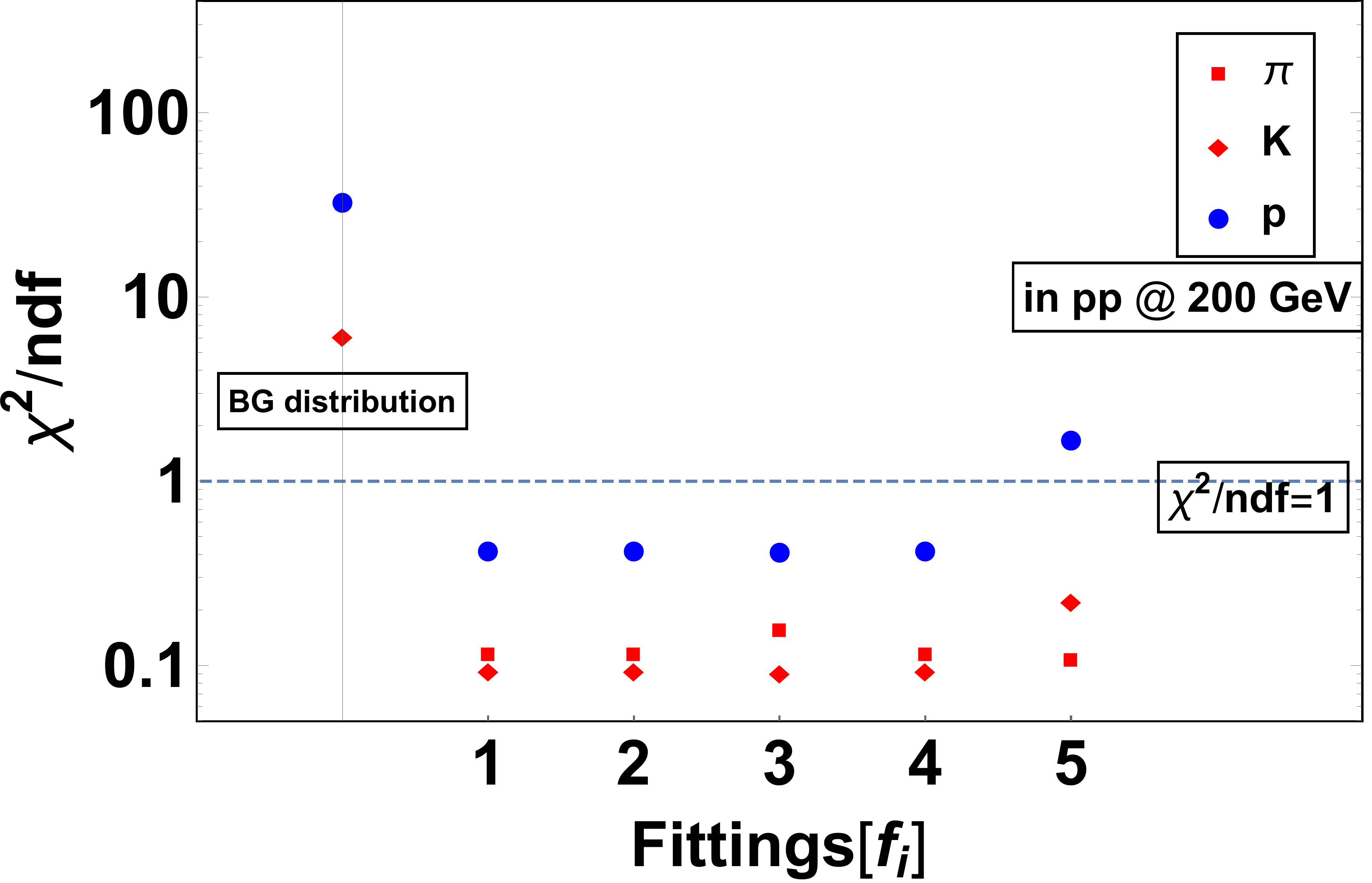}
}
\scalebox{1}[1]{
\includegraphics[width=0.45\linewidth]{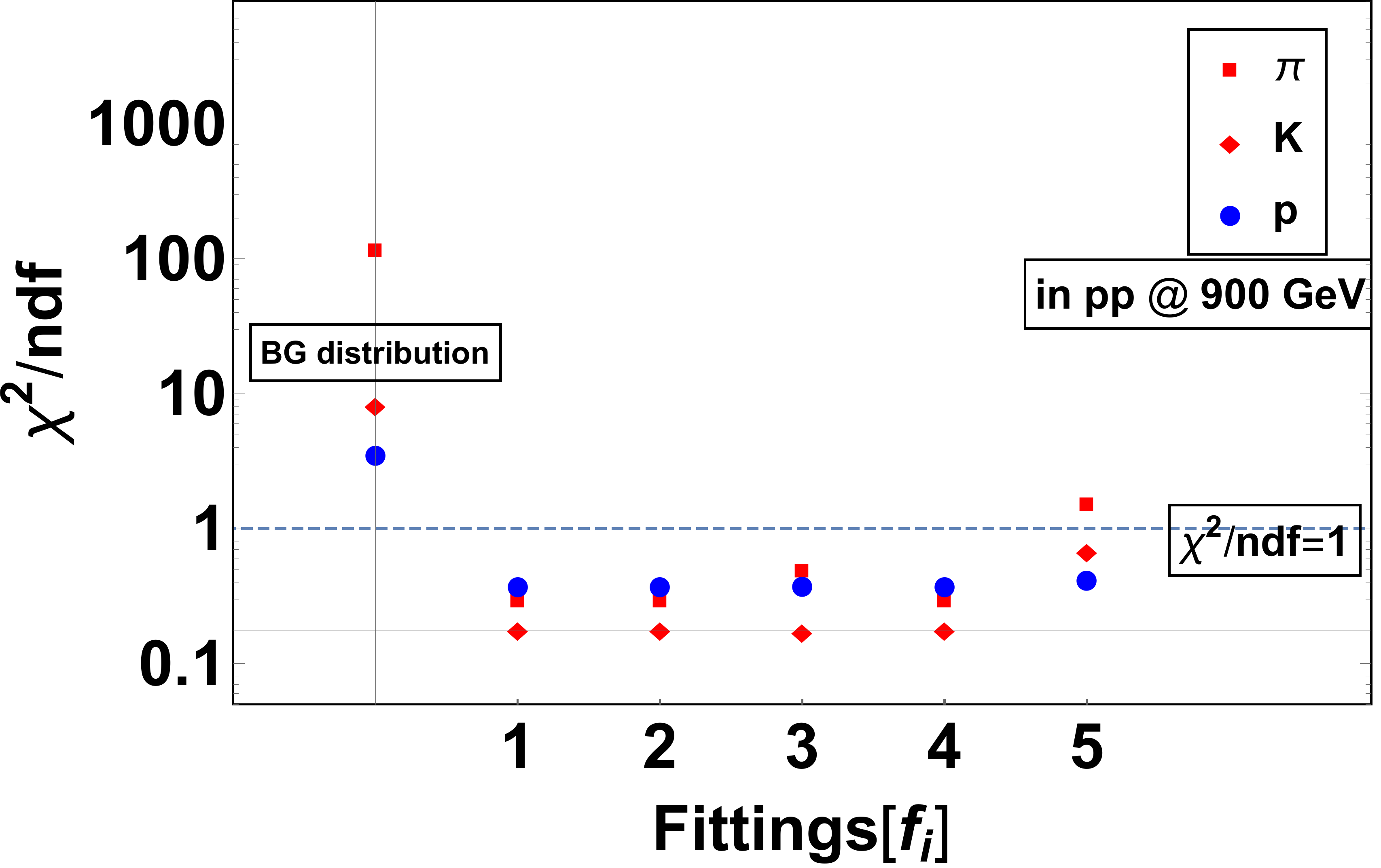}
\includegraphics[width=0.45\linewidth]{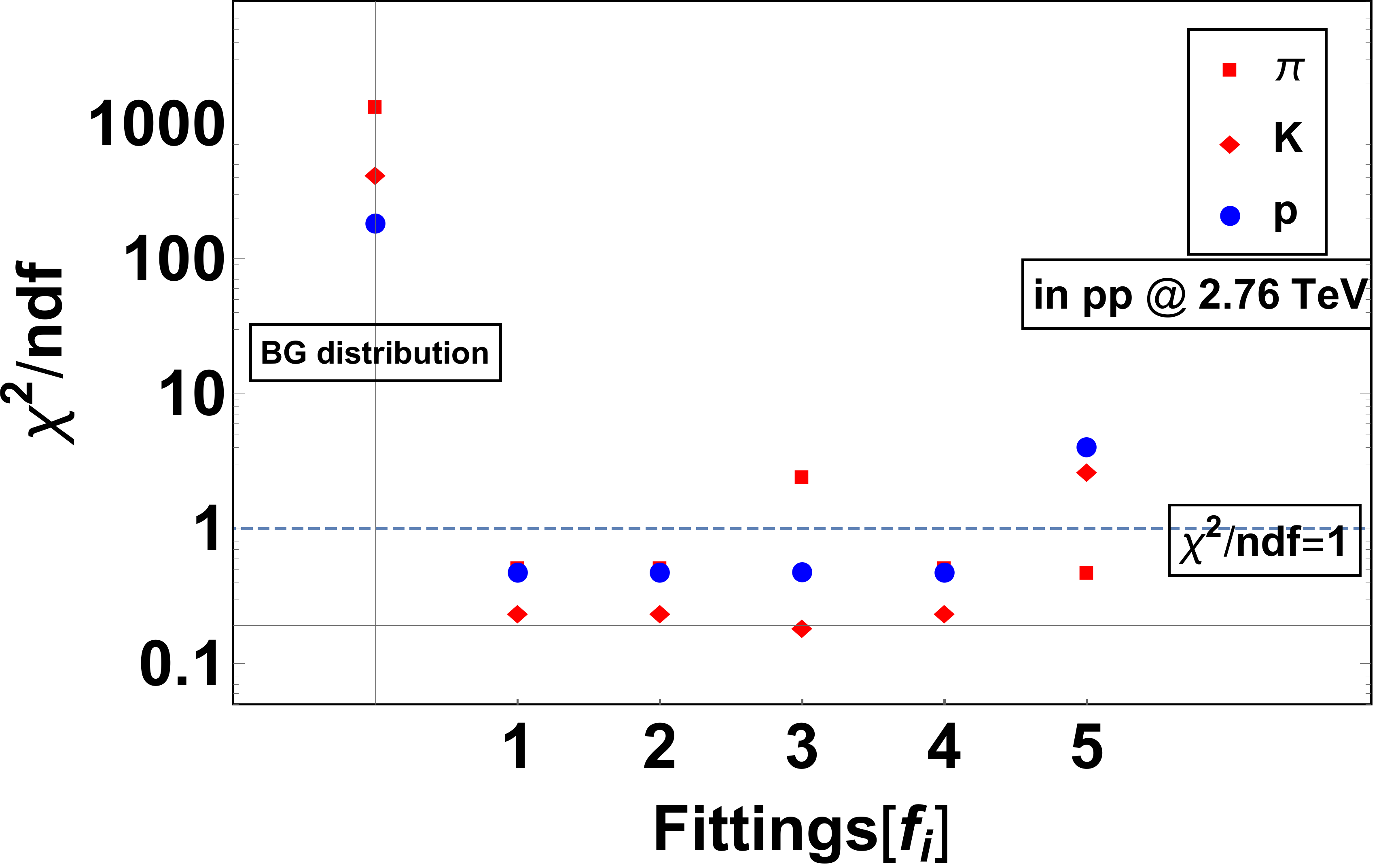}
}
\scalebox{1}[1]{
\includegraphics[width=0.45\linewidth]{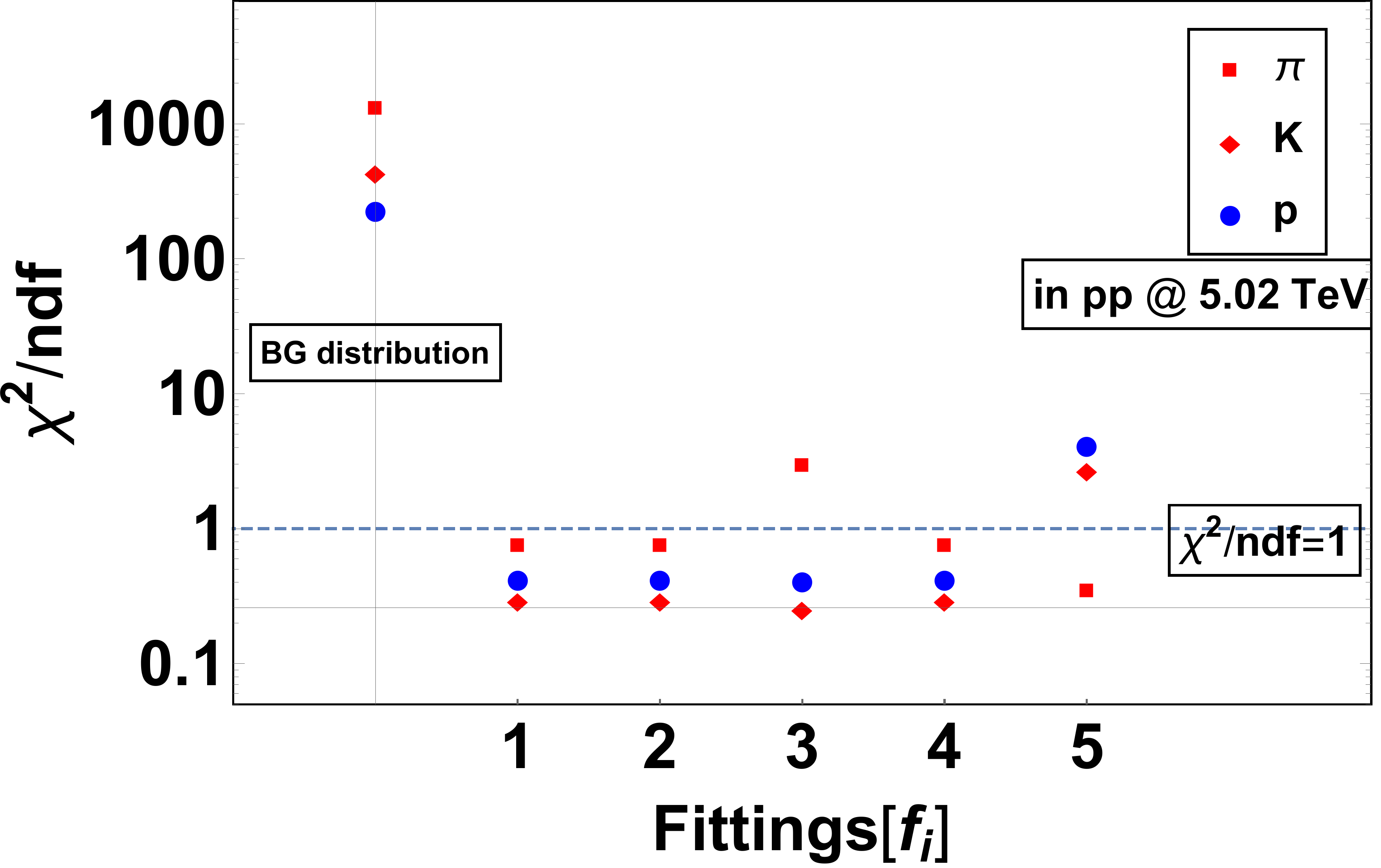}
\includegraphics[width=0.45\linewidth]{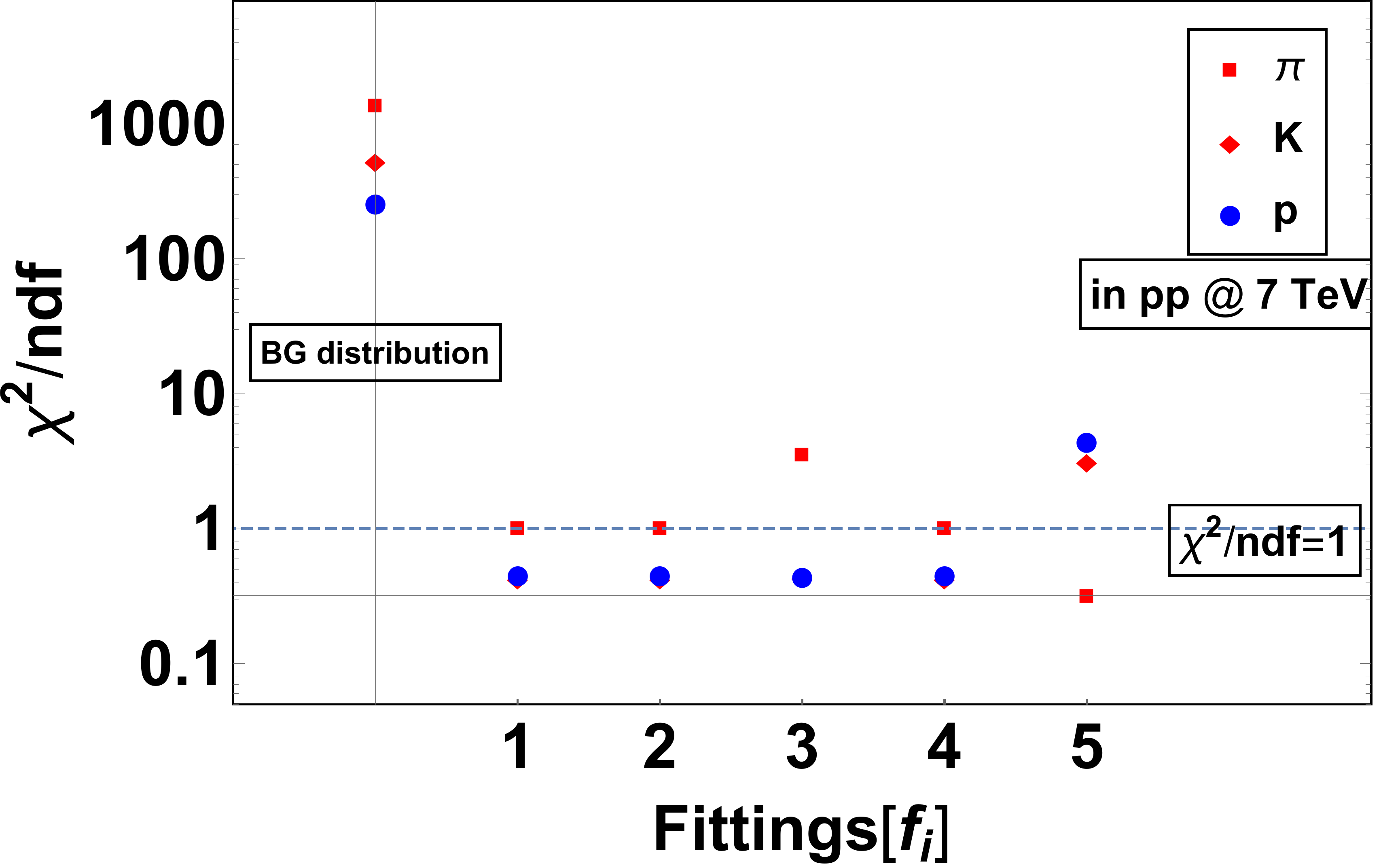}
}
\caption{Plot of $\chi^2/ndf$ for $p_T$ spectra fits by different Tsallis formulas as well as the usual BG one from Eqs.(\ref{functions}) in $pp$ collisions at given energies on the panels for identified particles ($\pi$, $K$ and $p$). The fitting $p_T$ range in each spectrum is listed in Table \ref{tab2}. The $\chi^2/ndf =1$ is also plotted as a reference.}
\label{figchi}
\end{figure*}

\begin{figure*}[htb]
\scalebox{1}[1]{
\includegraphics[width=0.44\linewidth]{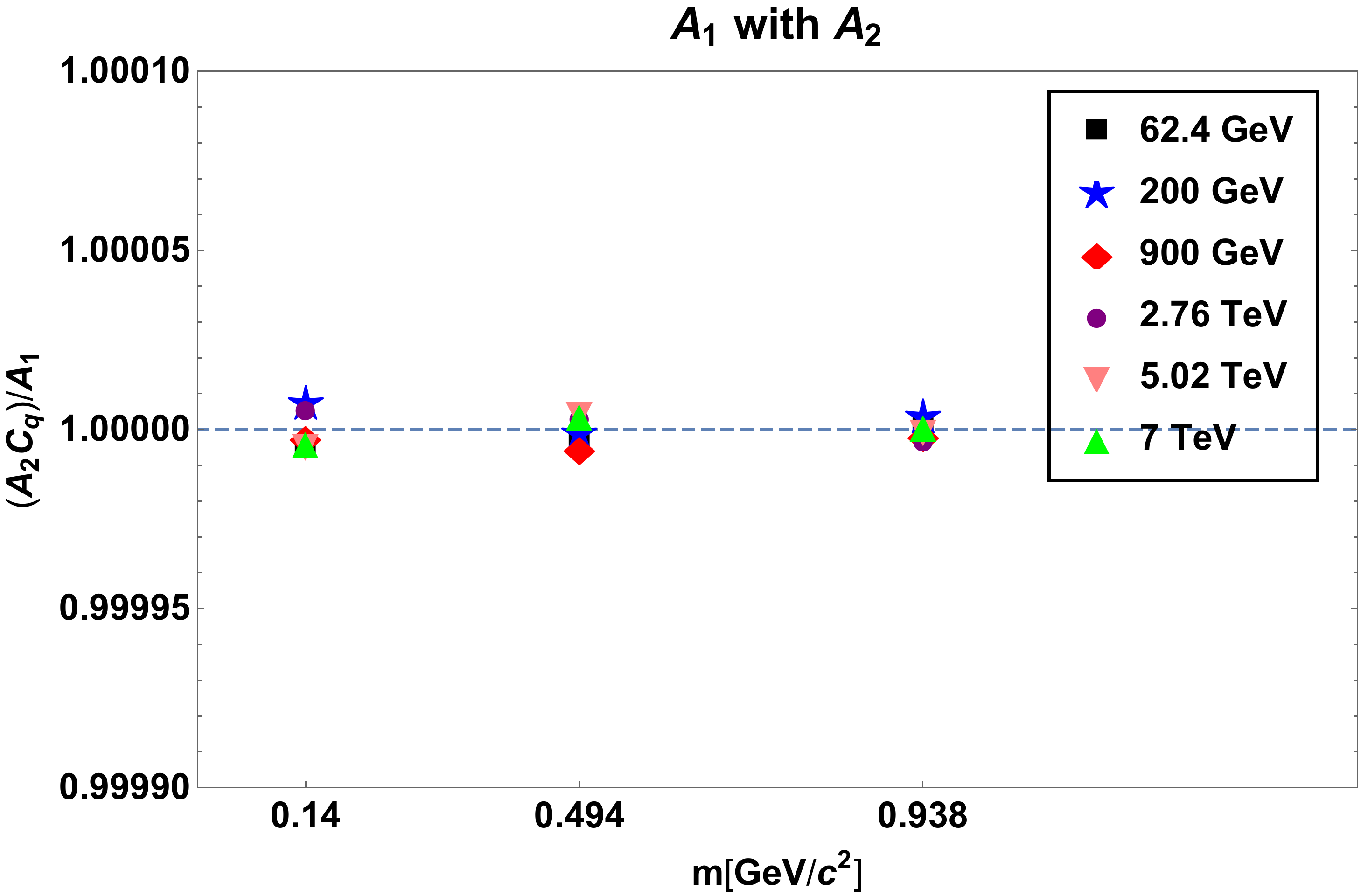}
\includegraphics[width=0.45\linewidth]{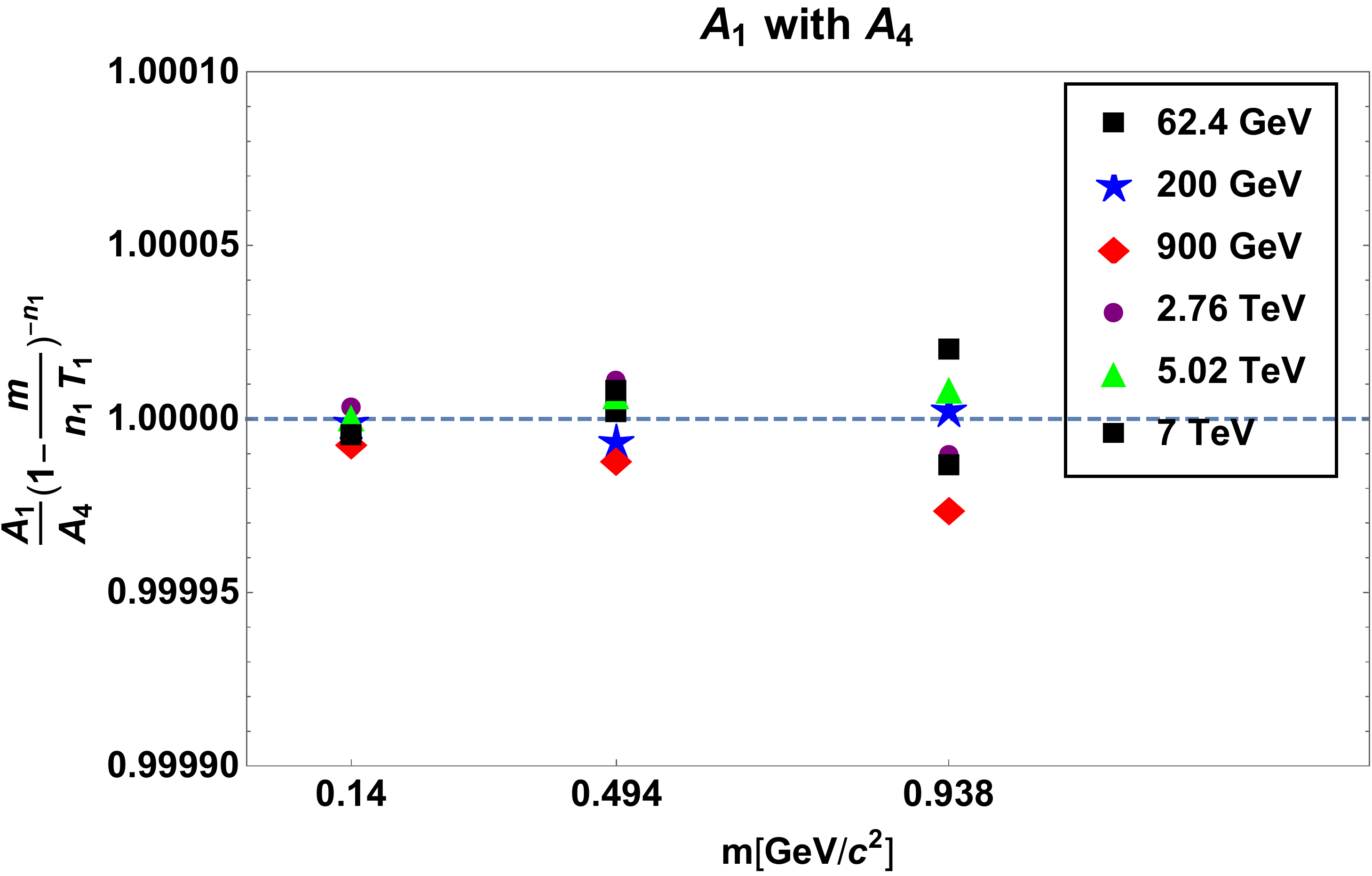}
}
\caption{Analysis of fitting parameters for different particles in $pp$ for various kinds of collisions.}
\label{figAT}
\end{figure*}


\begin{figure*}[!htb]
\scalebox{1}[1]{
\includegraphics[width=0.45\linewidth]{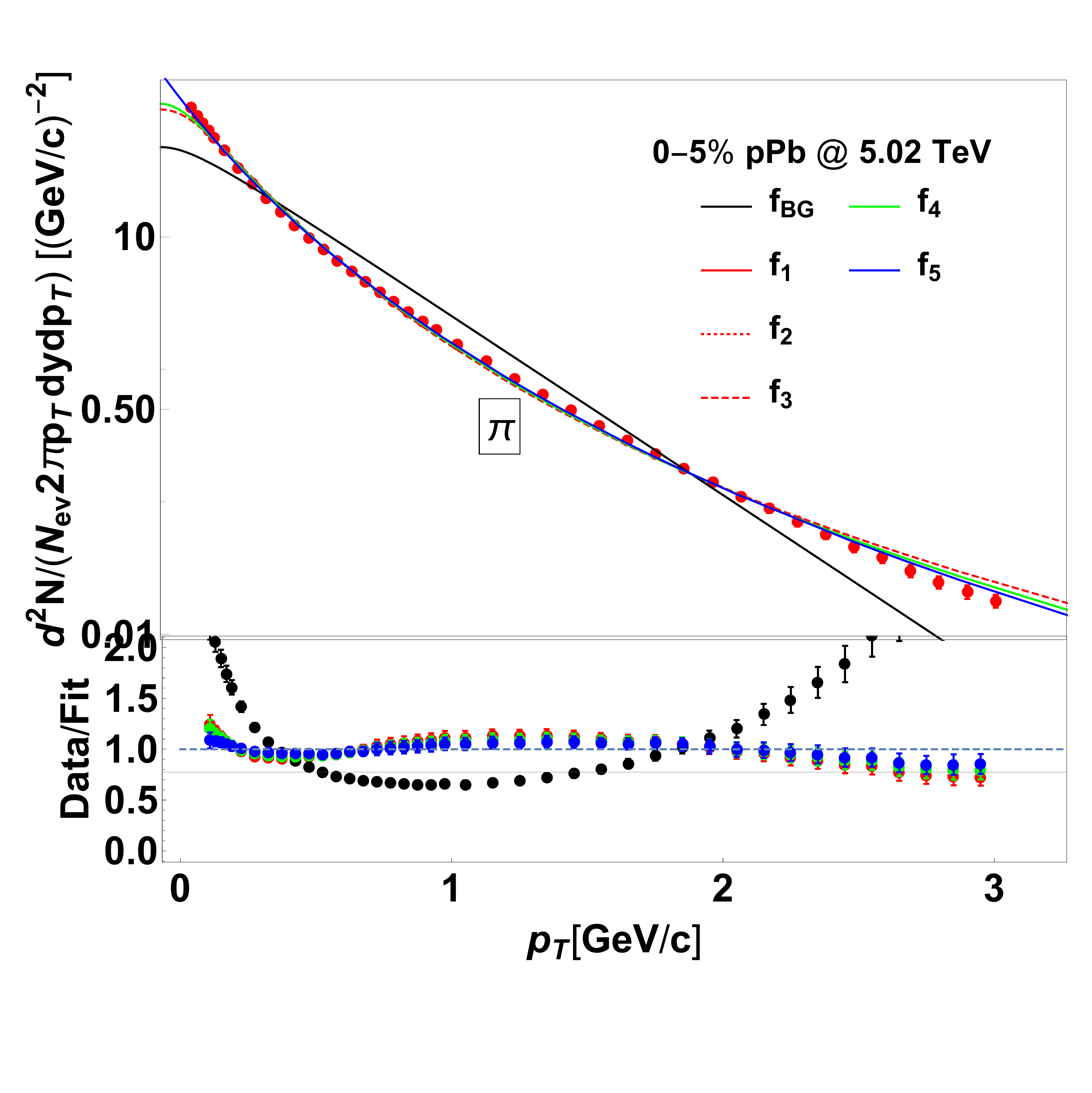}
\includegraphics[width=0.45\linewidth]{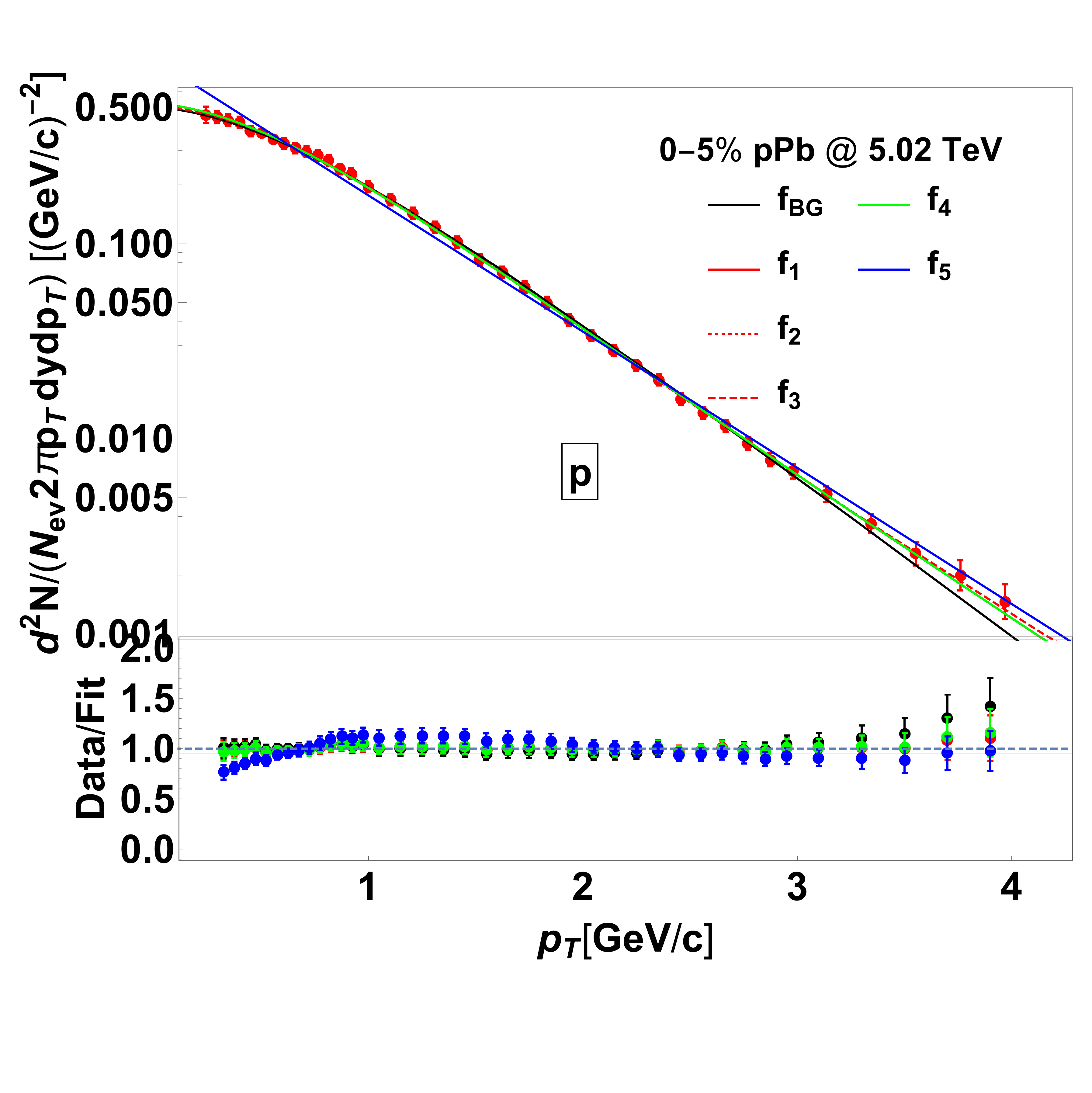}
}
\scalebox{1}[1]{
\includegraphics[width=0.45\linewidth]{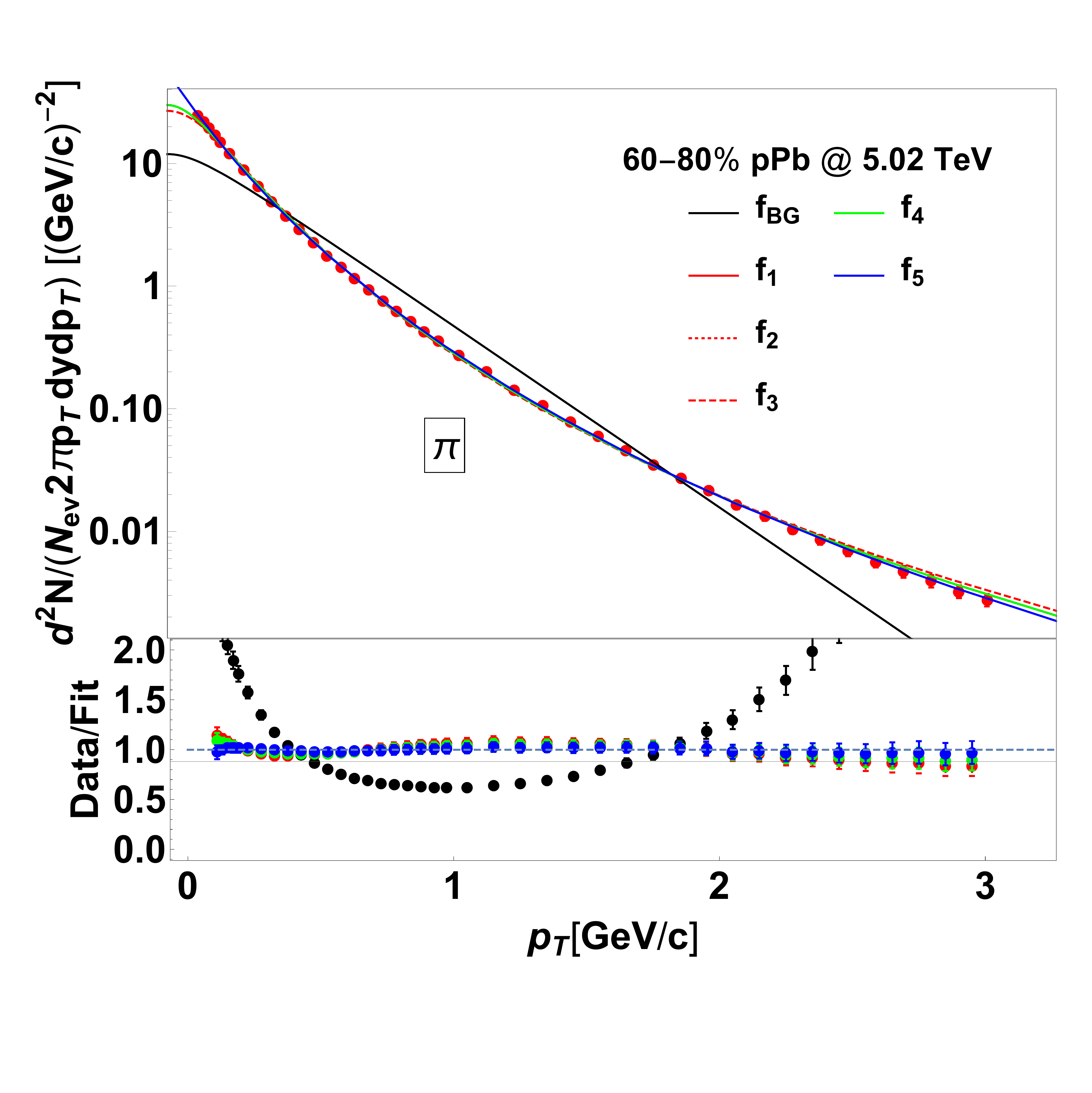}
\includegraphics[width=0.45\linewidth]{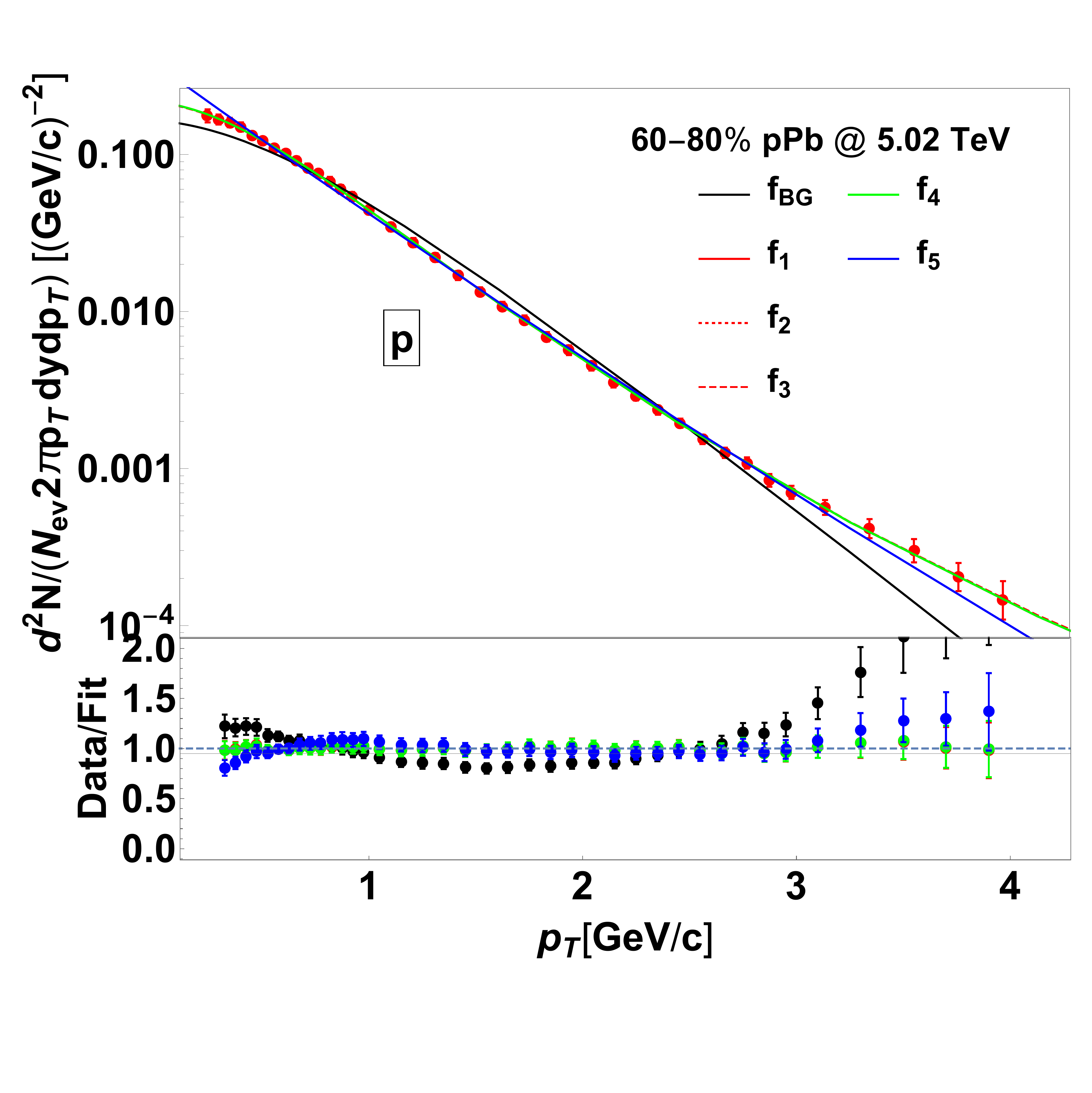}
}
\scalebox{1}[1]{
\includegraphics[width=0.45\linewidth]{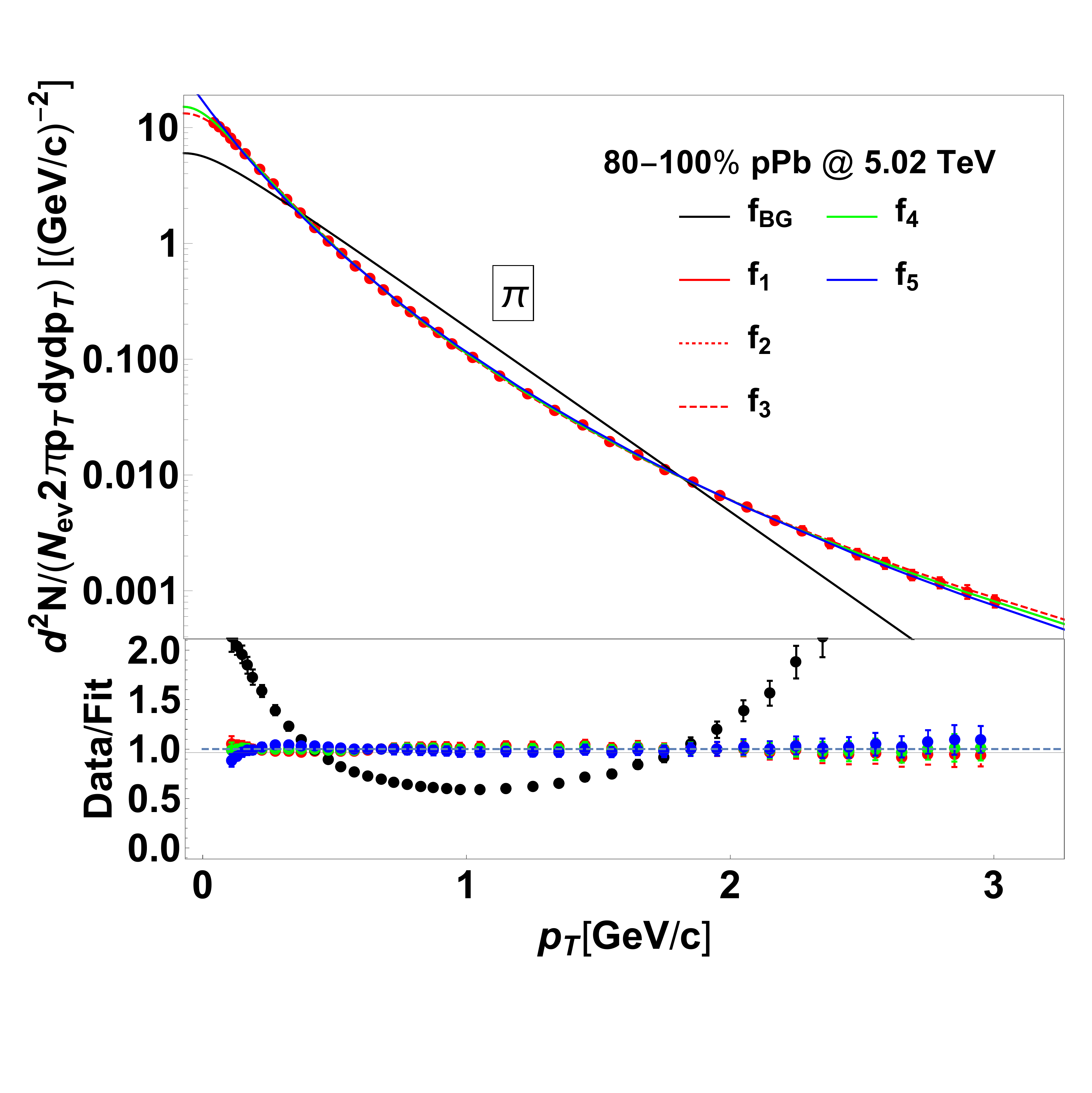}
\includegraphics[width=0.45\linewidth]{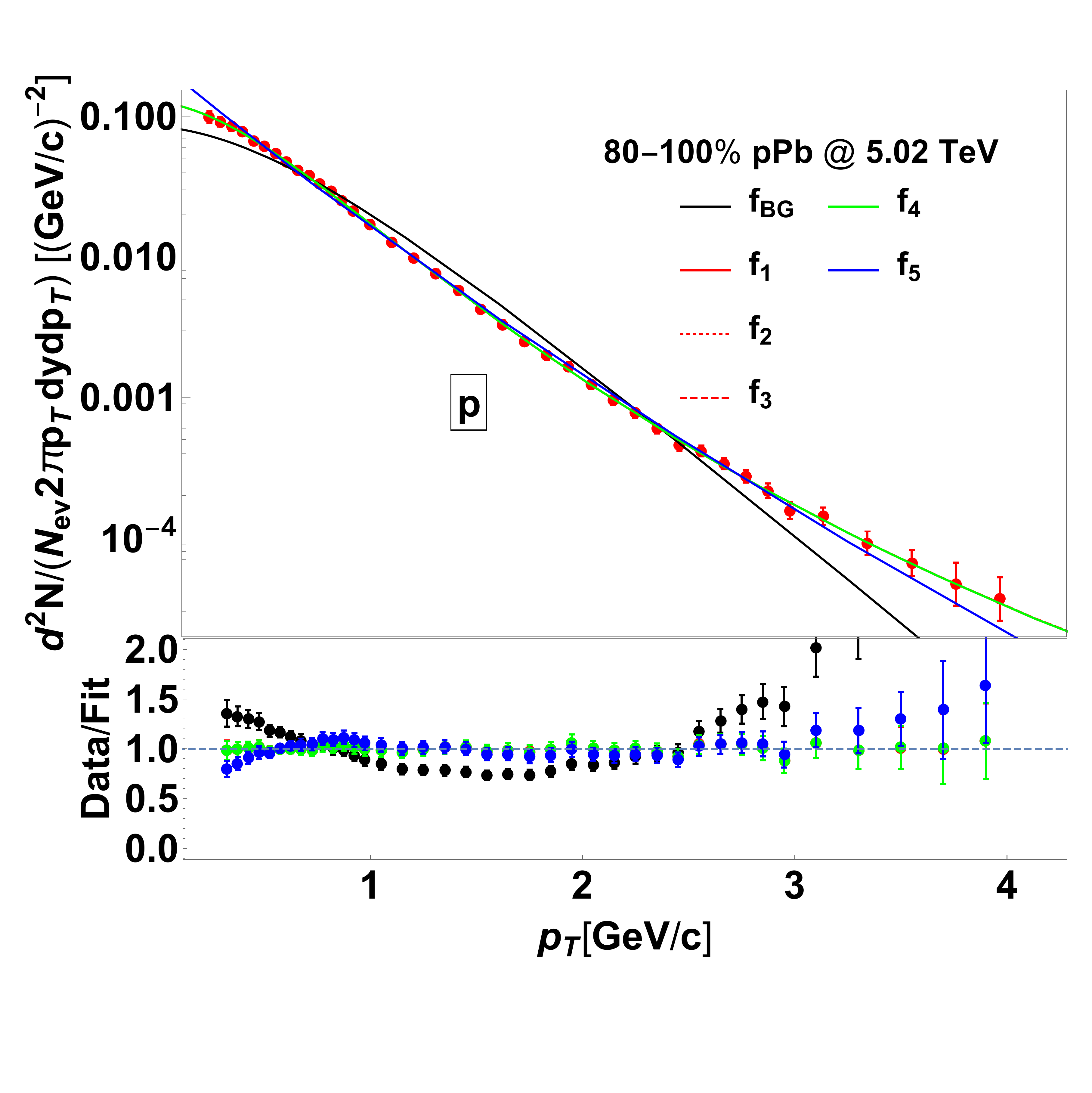}
}
\caption{The $p_T$ spectra for pions (left) and protons (right) in $pPb$ at 5.02 TeV within different centrality bins ($0-5 \%$ and $60-80 \%$, $80-100 \%$). Data are taken from Ref.~\cite{all-30}.
The fitting $p_T$ range in each spectrum is listed in Table \ref{tabB1}. Ratios of the six fittings to data are shown in the lower panels as well.}
\label{figAA}
\end{figure*}

\begin{figure*}[!htb]
\scalebox{1}[1]{
\includegraphics[width=0.5\linewidth]{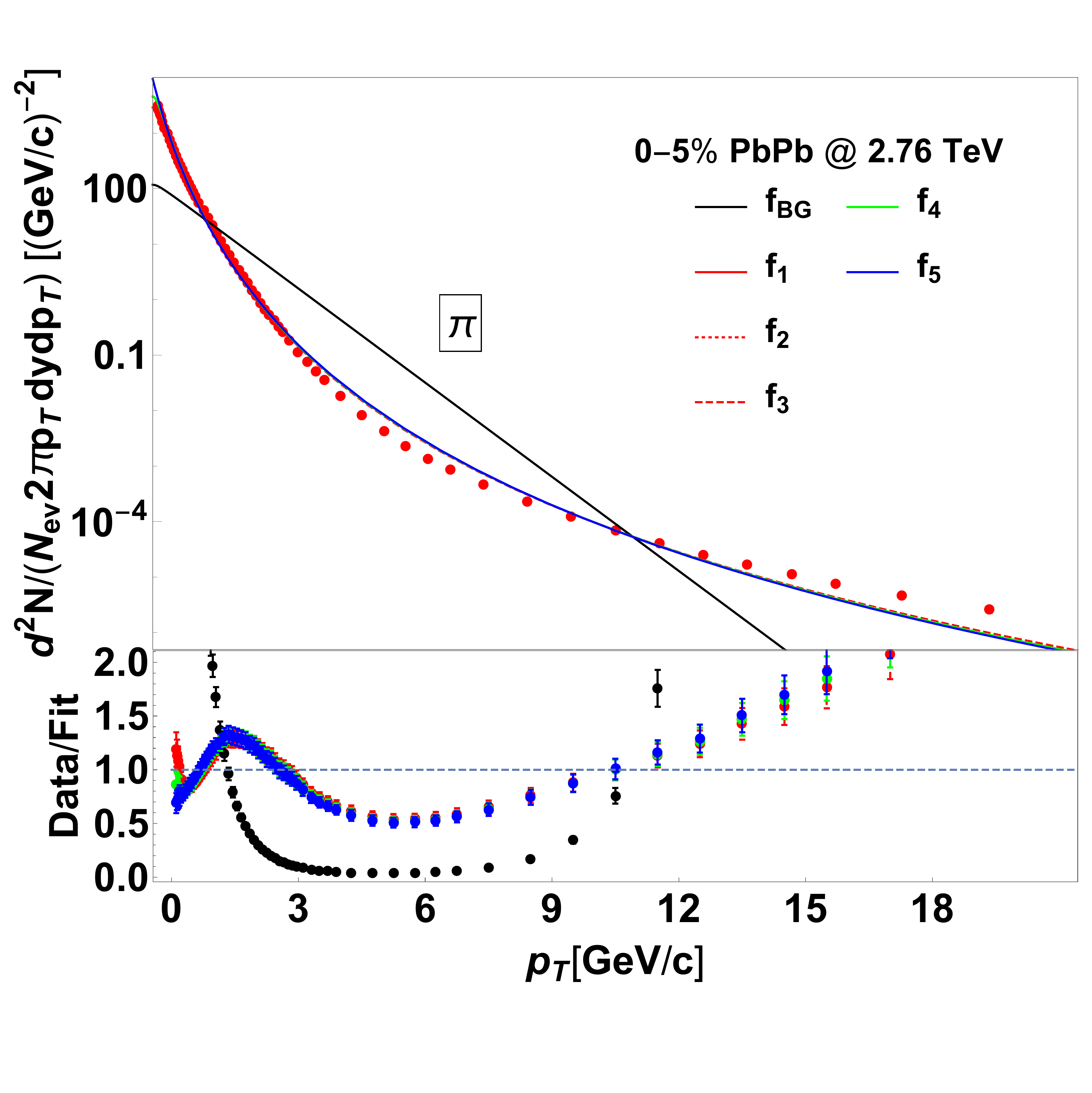}
\includegraphics[width=0.5\linewidth]{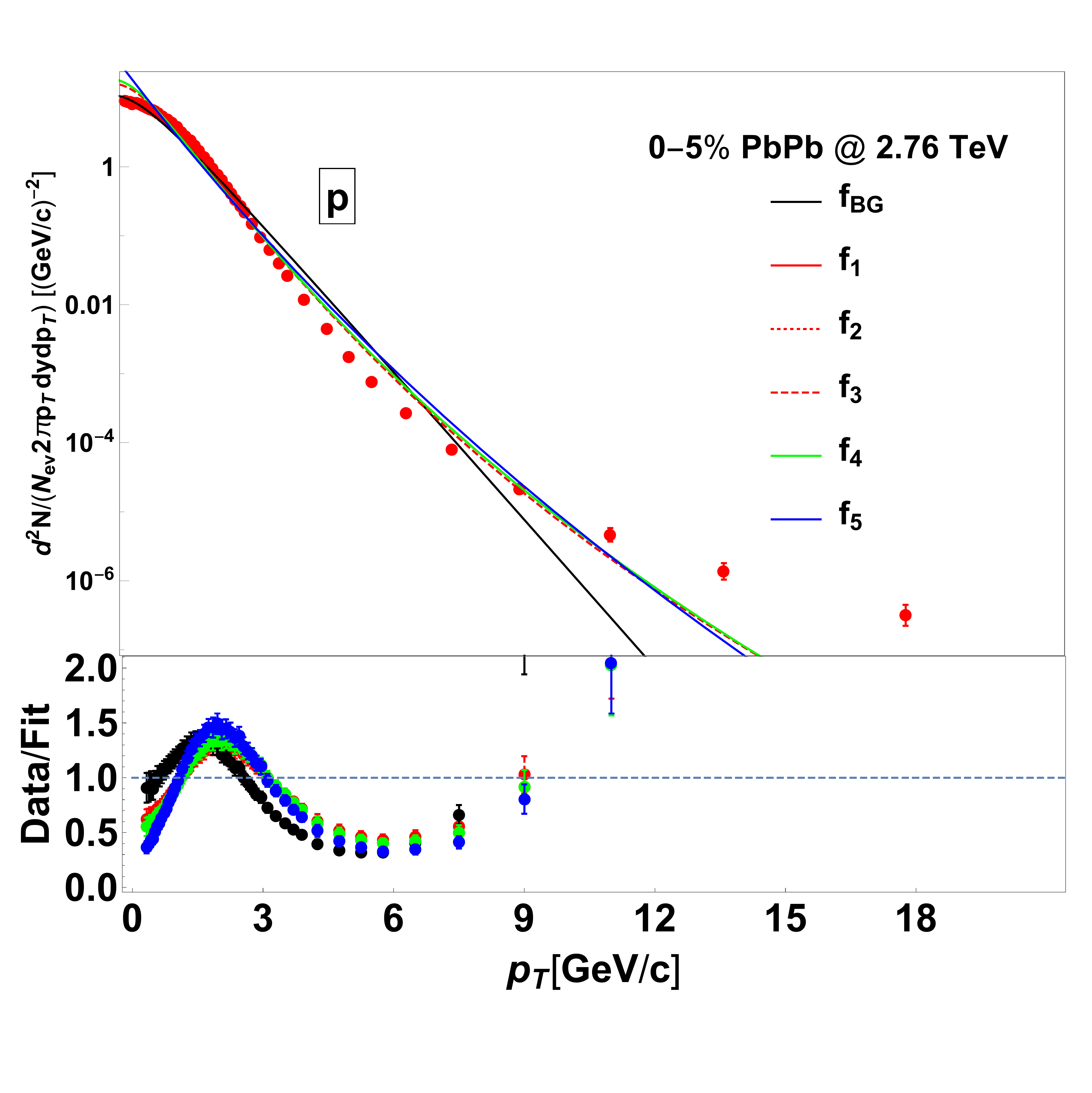}
}
\scalebox{1}[1]{
\includegraphics[width=0.5\linewidth]{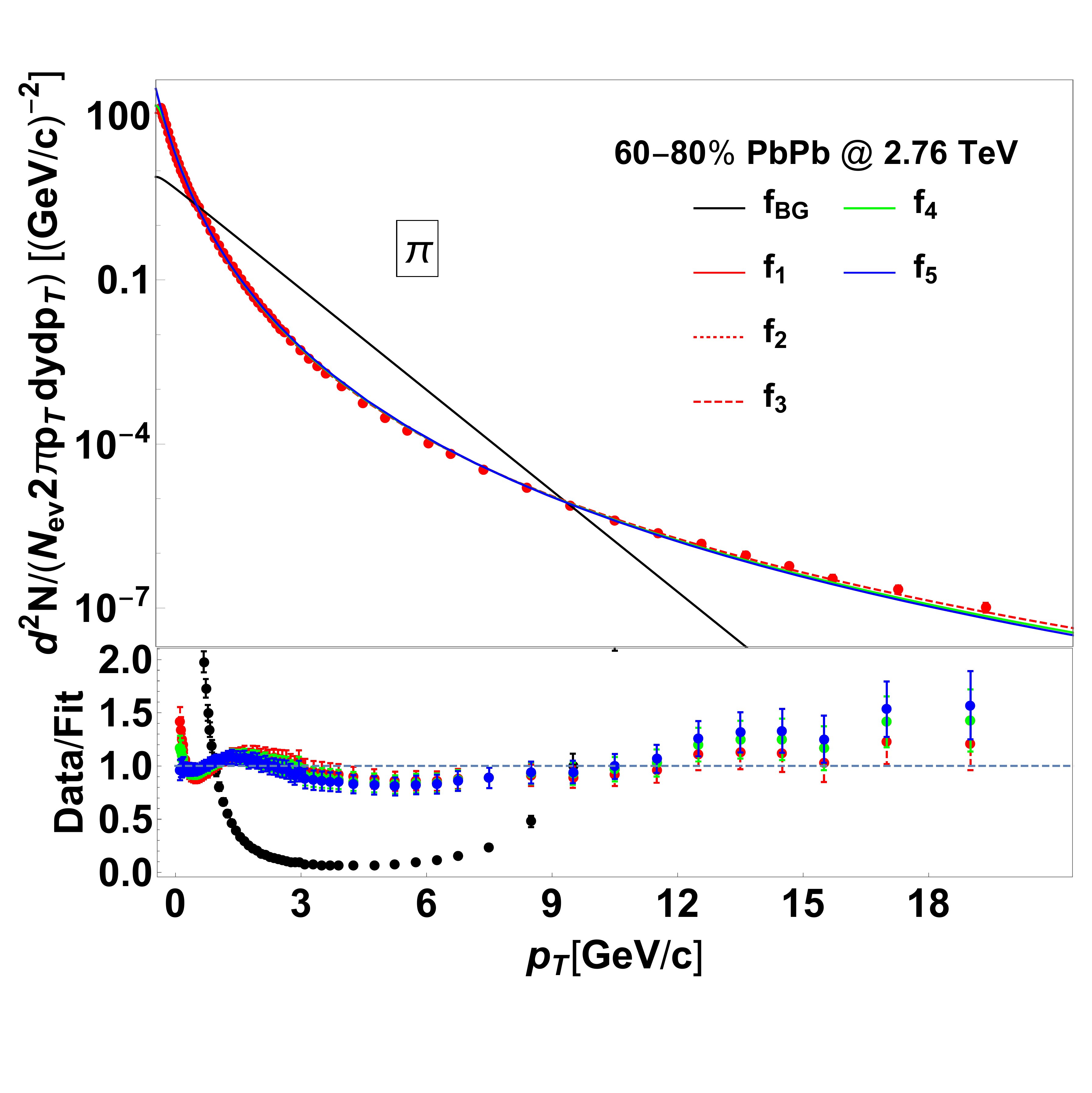}
\includegraphics[width=0.5\linewidth]{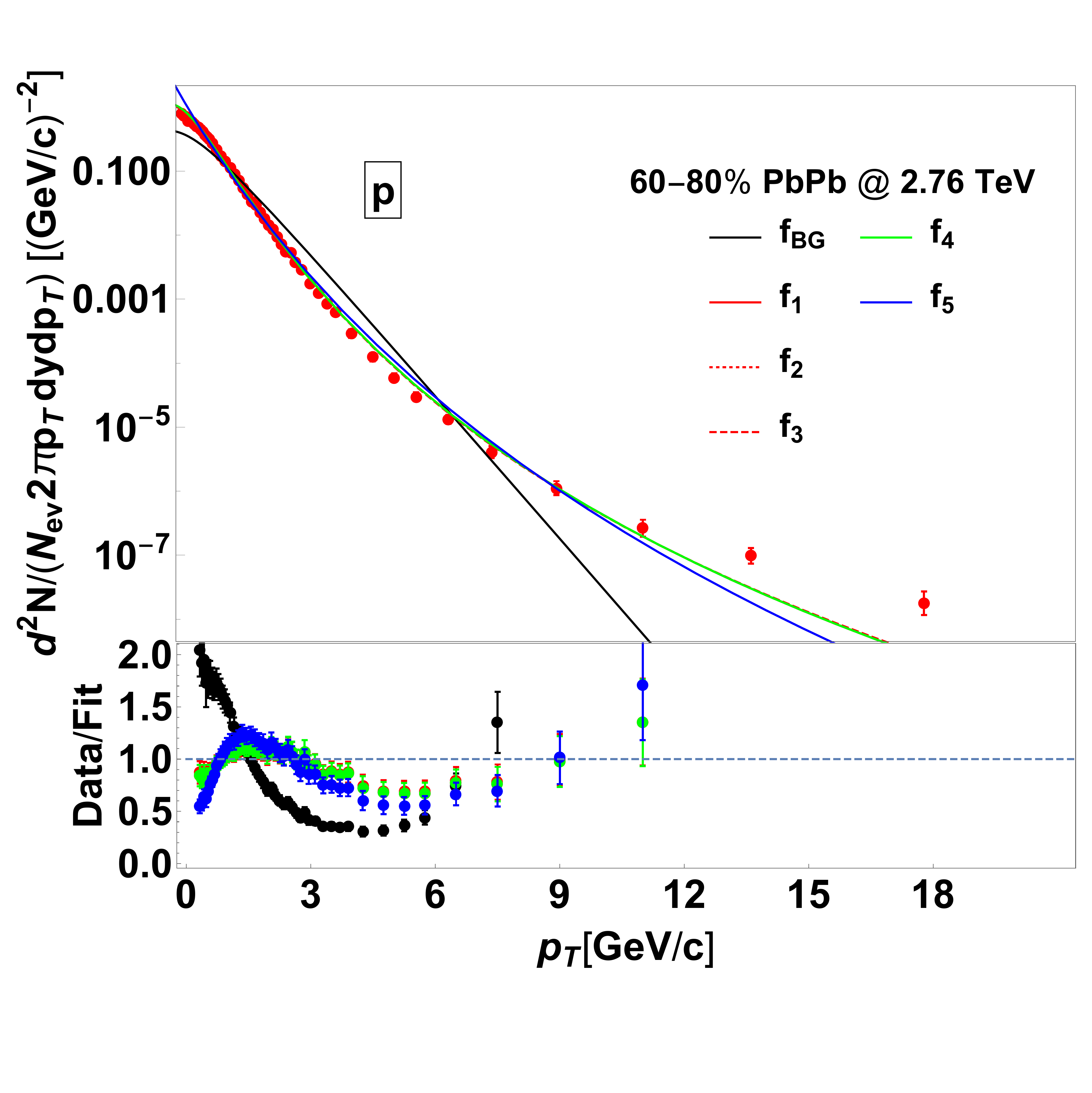}
}
\caption{The $p_T$ spectra for pions (left) and protons (right) in $PbPb$ at 2.76 TeV within different centrality bins ($0-5 \%$ and $60-80 \%$).
Data are taken from Refs.~\cite{AA-1,AA-2,AA-3,AA-4}.
The fitting $p_T$ range in each spectrum is listed in Table \ref{tabB1}. Ratios of the six fittings to data are shown in the lower panels as well.}
\label{figA}
\end{figure*}

\begin{figure*}[!htb]
\scalebox{1}[1]{
\includegraphics[width=0.4\linewidth]{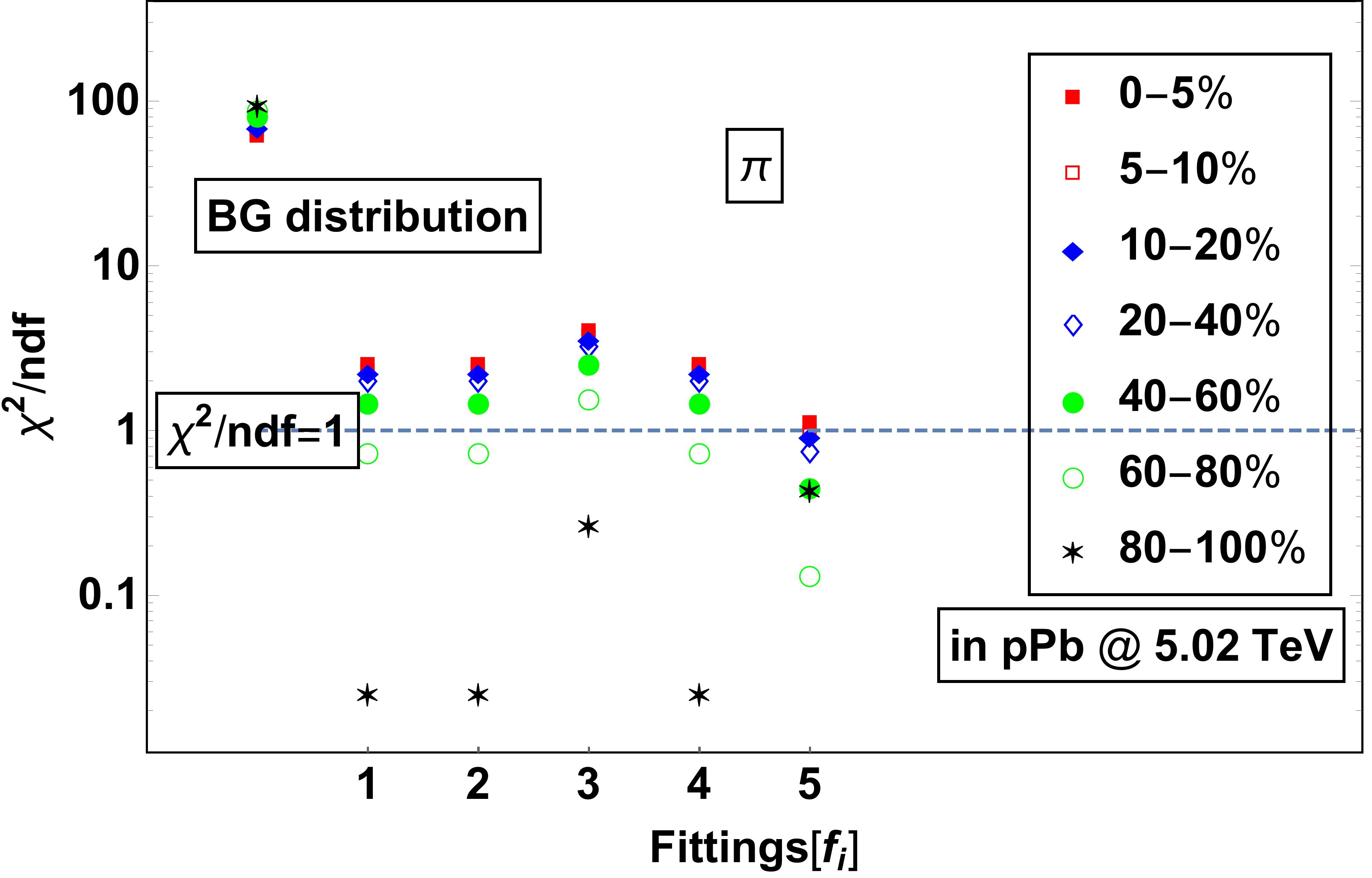}
\includegraphics[width=0.4\linewidth]{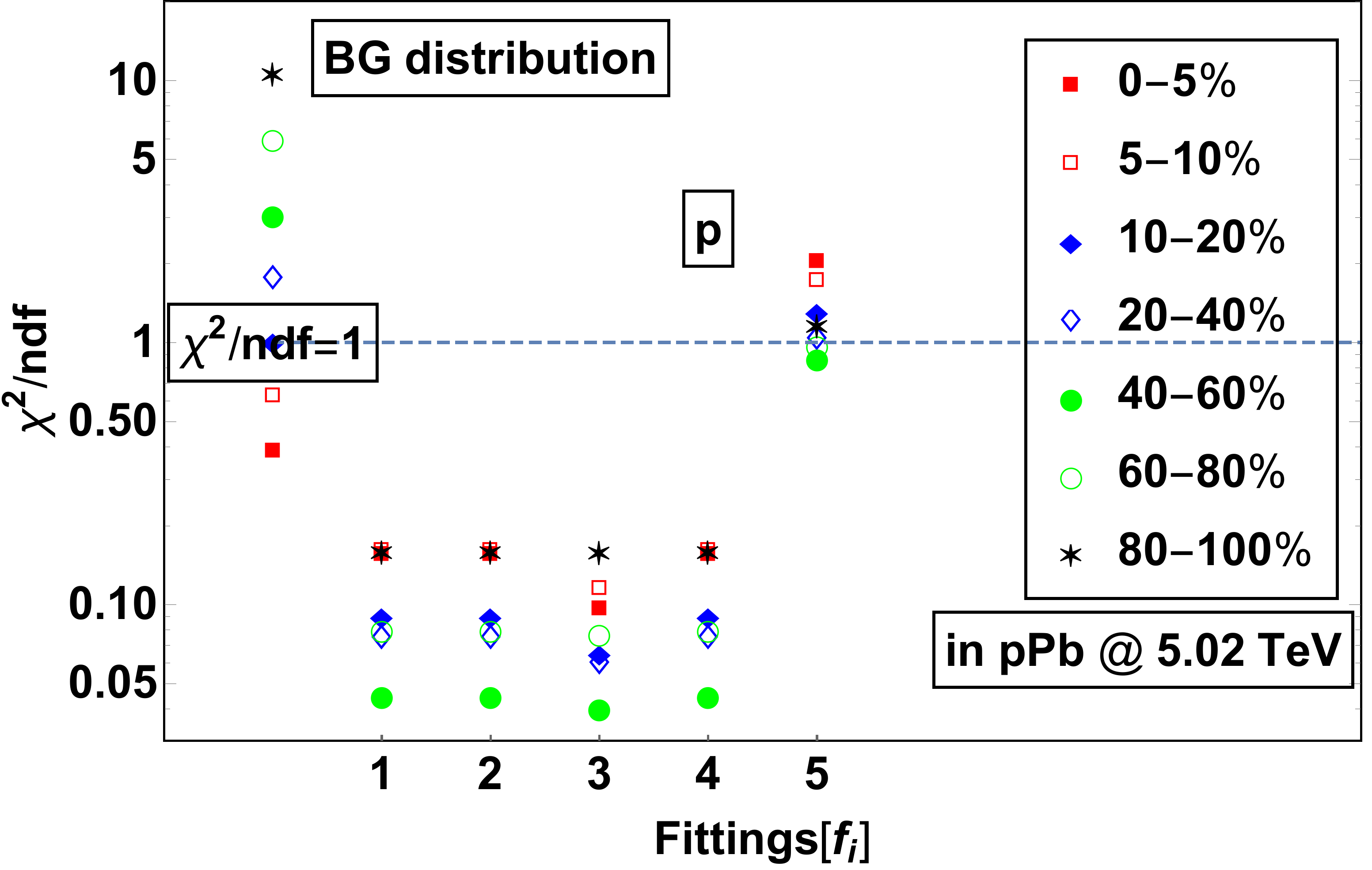}
}
\scalebox{1}[1]{
\includegraphics[width=0.4\linewidth]{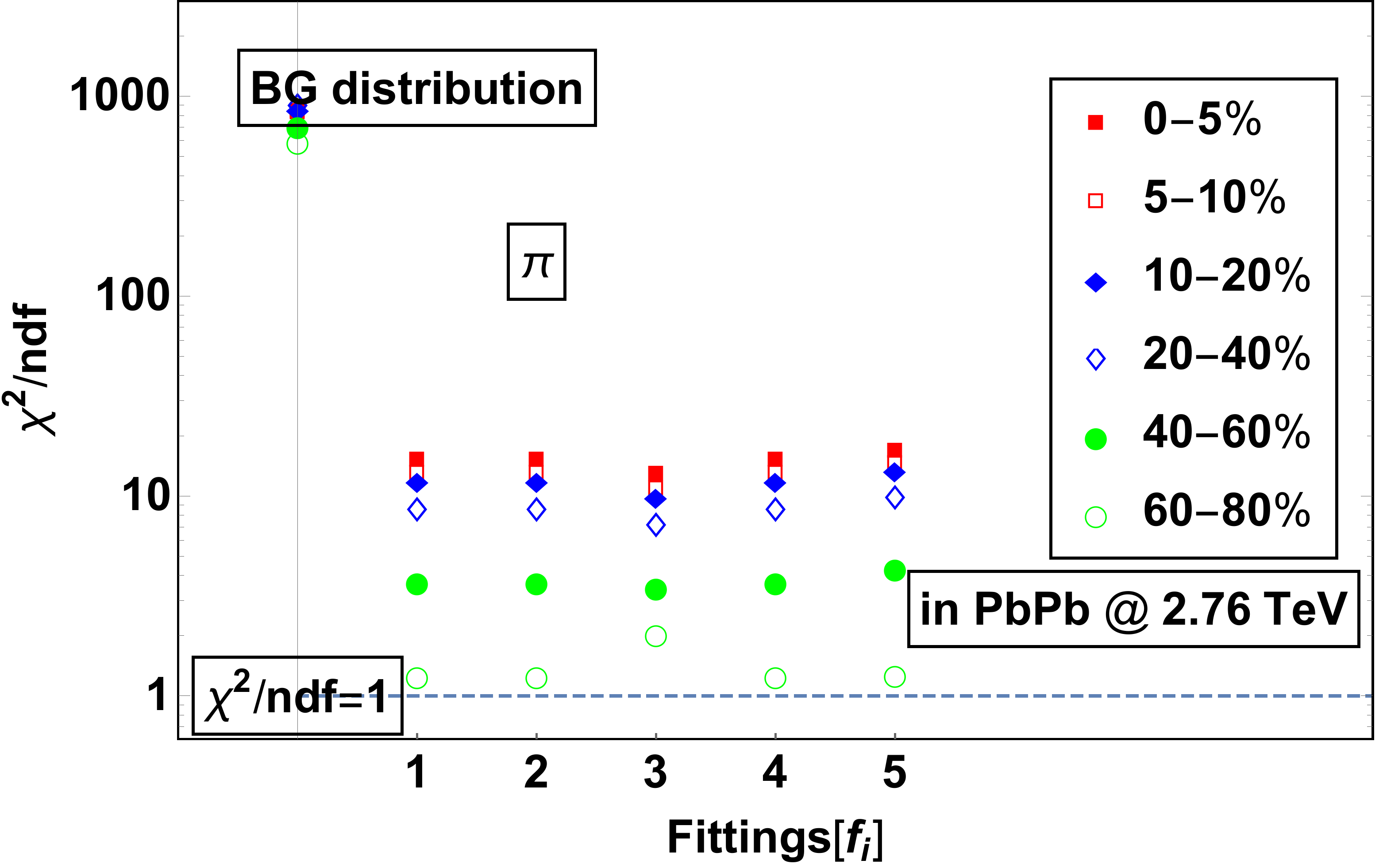}
\includegraphics[width=0.4\linewidth]{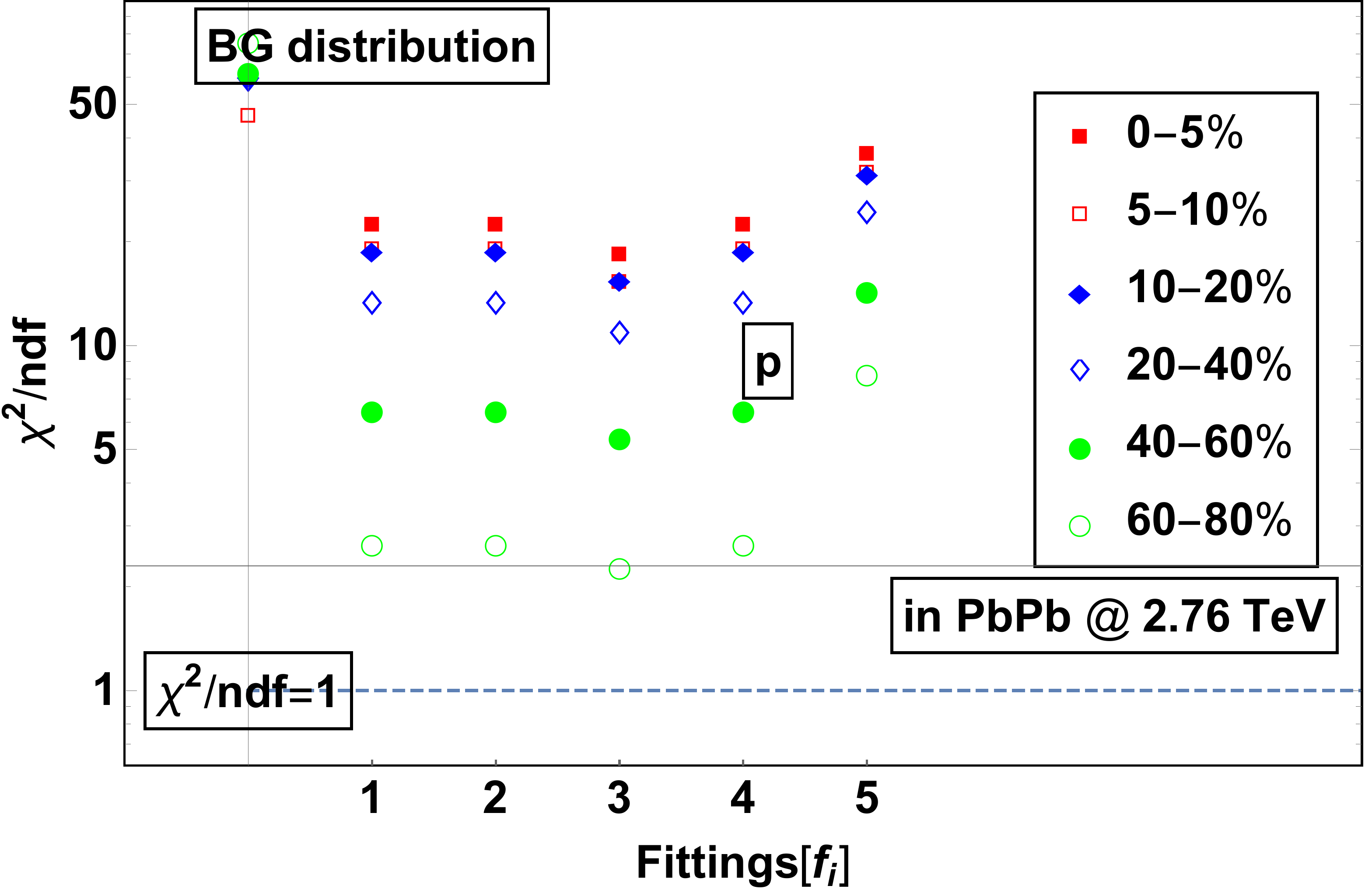}
}
\caption{Plot of $\chi^2/ndf$ of fitting $p_T$ spectra in $pPb$ (upper) at 5.02 TeV and $PbPb$ (lower) collisions at 2.76 TeV for pions (left) and protons (right) within all centralities. The fitting $p_T$ range in each spectrum is listed in Table \ref{tabB1}. The line of $\chi^2/ndf =1$ is also plotted here as a reference.}
\label{figB}
\end{figure*}

\end{document}